\documentclass[a4paper,11pt]{article}
\pdfoutput=1 

\usepackage{jheppub} 

\usepackage[T1]{fontenc} 

\usepackage{upgreek}
\usepackage[titles]{tocloft}
\usepackage{mathabx}
\usepackage[all]{xy}
\usepackage{tikz}
\usepackage{amsthm}
\usepackage{dsfont}
\usepackage{dirtytalk}
\usepackage{xcolor}
\usepackage{tikz-cd}
\usepackage{multirow}
\usepackage{hyperref}
\usepackage{fnpct}

\allowdisplaybreaks

\usepackage[framemethod=tikz]{mdframed}
\definecolor{mycolor}{rgb}{0.122, 0.435, 0.698}
\newmdenv[innerlinewidth=0.5pt, roundcorner=4pt,linecolor=mycolor,innerleftmargin=6pt,
innerrightmargin=6pt,innertopmargin=6pt,innerbottommargin=6pt]{mybox}

\newcommand{\pushright}[1]{\ifmeasuring@#1\else\omit\hfill$\displaystyle#1$\fi\ignorespaces}
\newcommand\gameo{\vcenter{\hbox{\rotatebox{180}{$\omega$}}}}
\newcommand\Gameo{\vcenter{\hbox{\rotatebox{180}{$\Omega$}}}}

\newcommand{\la}{\langle}
\newcommand{\ra}{\rangle}
\newcommand{\braket}[2]{\big\langle{#1}|{#2}\big\rangle}

\newcommand{\Space}{\Cbb^{n_L} \setminus \Tw\cup\Rel}
\newcommand{\Spacedual}{\Cbb^{n_L}\setminus \Tw }

\newcommand{\Cbb}{\mathbb{C}}

\newcommand{\alg}{{\rm alg}}
\newcommand{\dr}{{\rm dR}}
\newcommand{\drc}{{\rm dR,c}}

\renewcommand{\th}{\theta}

\newcommand{\hooklongrightarrow}{\lhook\joinrel\longrightarrow}

\renewcommand{\th}{\theta}
\newcommand{\nn}{\nonumber}
\newcommand{\ed}{\text{d}}
\newcommand{\vep}{\varepsilon}
\newcommand{\vphi}{\varphi}
\newcommand{\bs}[1]{\boldsymbol{#1}}
\newcommand{\Dn}{ {D_{\mathds{N}}} }
\newcommand{\wc}[1]{\widecheck{#1}}
\newcommand{\vth}{\vartheta}

\renewcommand{\vec}[1]{\boldsymbol{#1}}

\newcommand{\Dsf}{\textsf{D}}
\newcommand{\Tw}{\textsf{Tw}}
\newcommand{\Rel}{\textsf{Rel}}

\newcommand\wedgedot[1][1]{
\tikzset{every picture/.style={line width=0.75pt}}      
\begin{tikzpicture}[x=0.75pt,y=0.75pt,yscale=-1,xscale=1]
\draw   (100,130.12) -- (105.06,120) -- (110.12,130.12) ;
\draw  [fill={rgb, 255:red, 0; green, 0; blue, 0 }  ,fill opacity=1 ] (104,127) .. controls (104,126.45) and (104.45,126) .. (105,126) .. controls (105.56,126) and (106.01,126.45) .. (106.01,127) .. controls (106.01,127.56) and (105.56,128.01) .. (105,128.01) .. controls (104.45,128.01) and (104,127.56) .. (104,127) -- cycle ;
\end{tikzpicture}
}

\title{Loop-by-loop Differential
Equations for Dual (Elliptic) Feynman Integrals}

\author[a]{Mathieu Giroux,}
\author[a,b]{Andrzej Pokraka,}

\affiliation[a]{Department of Physics, McGill University, 3600 Rue University, Montr\'eal, QC Canada H3A 2T8}
\affiliation[b]{Department of Physics, Brown University, Providence, RI 02912, USA}

\emailAdd{mathieu.giroux2@mail.mcgill.ca}
\emailAdd{andrzej\_pokraka@brown.edu}

\abstract{We present a loop-by-loop method for computing the differential equations of Feynman integrals using the recently developed dual form formalism. We give explicit prescriptions for the loop-by-loop fibration of multi-loop dual forms. Then, we test our formalism on a simple, but non-trivial, example: the two-loop three-mass elliptic sunrise family of integrals. We obtain an $\varepsilon$-form differential equation within the correct function space in a sequence of relatively simple algebraic steps. In particular, none of these steps relies on the analysis of $q$-series. Then, we discuss interesting properties satisfied by our dual basis as well as its simple relation to the known $\vep$-form basis of Feynman integrands. The underlying K3-geometry of the three-loop four-mass sunrise integral is also discussed. Finally, we speculate on how to construct a ``good'' loop-by-loop basis at three-loop.
}

\begin{document} 
\maketitle

\section{Introduction \label{sec:intro}}
Over the last decade, our ability to compute scattering amplitudes that evaluate to a special class of functions called multiple polylogarithms (MPLs) has increased tremendously. 
In supersymmetric theories, many amplitudes
with two or more loops and/or five or more particles
have been computed
\cite{Caron-Huot:2011dec,Dixon:2013eka,Dixon:2014voa,Henn:2016jdu,Caron-Huot:2016owq,Dixon:2016apl,Bourjaily:2019jrk,Caron-Huot:2018dsv,Drummond:2018caf,Abreu:2018aqd,Chicherin:2018yne,He:2019jee,Li:2021bwg}. 
Substantial progress in non-supersymmetric theories has also been made with the computation of $2\to3$ scattering amplitudes at NNLO \cite{Chicherin:2017dob,Gehrmann:2018yef,Abreu:2018zmy,Bonciani:2020tvf,Abreu:2020cwb,Badger:2021owl,Chawdhry:2021hkp,Gerlach:2021xtb,Czakon:2021mjy,Badger:2021imn,Abreu:2021asb,Badger:2022ncb} 
and beyond \cite{Henn:2016wlm,Dulat:2018bfe,Gehrmann:2018odt,Duhr:2019kwi,Duhr:2020seh,Duhr:2020kzd,Chen:2021isd,Camarda:2021ict,Caola:2022dfa,vonManteuffel:2020vjv,Lee:2021lkc,Lee:2022nhh,He:2022ujv}. 
Much of this progress has been facilitated by a firm mathematical understanding of the MPL function space.
This has led to the development of efficient analytical \cite{Remiddi:1999ew,Goncharov:2010jf,Brown:2011ik,Duhr:2012fh} and numerical \cite{Bauer:2000cp,Gehrmann:2001pz,Vollinga:2004sn,Maitre:2005uu,Panzer:2014caa,Duhr:2019tlz} tools for manipulating and evaluating MPLs.

It is also known that functions beyond the space of MPLs appear in scattering amplitudes at high enough loops or when there are many scales.
Such functions appear in theoretical laboratories such as scalar theories \cite{Caffo:1998du,Adams:2013nia,Bourjaily:2017bsb,Adams:2018yfj,Bogner:2018uus,Bourjaily:2018ycu,Bonisch:2021yfw,Duhr:2022pch,Pozo:2022dox}, $\mathcal{N}=4$ SYM \cite{Caron-Huot:2012awx,Bourjaily:2017bsb,Kristensson:2021ani,Bourjaily:2019hmc}, string theory \cite{Broedel:2014vla,Broedel:2015hia,Broedel:2017jdo,Broedel:2018izr} as well as precision studies of the standard model  \cite{sabry1962fourth,Broadhurst:1987ei,Caola:2022ayt,Abreu:2022vei}. 
Thus, developing a solid mathematical understanding of the special functions beyond MPLs will not only enhance our theoretical grasp of scattering amplitudes but also play an important role in future searches for new physics at colliders \cite{deBlas:2019rxi,EuropeanStrategyforParticlePhysicsPreparatoryGroup:2019qin}.

MPLs are defined by iterated integrals over rational functions. 
The simplest generalization introduces the square root of an irreducible cubic or quartic polynomial, defining an elliptic curve, into the integrand (in addition to rational functions). 
When only one elliptic curve appears in the integrand, it is known how to formulate the elliptic generalization of MPLs called elliptic MPLs (eMPLs). 
Mathematically, eMPLs can be thought of as iterated integrals over the rational functions on the elliptic curve \cite{Adams:2014vja,Remiddi:2017har,Broedel:2017kkb,Broedel:2017siw} or as iterated integrals on a genus-1 Riemann surface \cite{zagier1991periods,levin2007towards,Brown:2011ik,Duhr:2019rrs,Bogner:2019lfa,Weinzierl:2022eaz,Muller:2022gec}.
Much of the MPL technology seems to have elliptic analogues, such as symbol calculus \cite{Broedel:2018iwv,Broedel:2018qkq,Kristensson:2021ani,Wilhelm:2022wow,Forum:2022lpz}.
However, elliptic symbol letters are complicated functions that satisfy non-trivial identities. 
Additional tools such as the symbol prime are needed to manifest such identities \cite{Wilhelm:2022wow}.
Moreover, the transcendental map relating ordinary kinematic variables to torus variables coupled with the complexity of the elliptic letters obfuscates the connection between symbol letters and Landau surfaces. 

Homological and cohomological perspectives have  also played an increasingly important role in our understanding of multi-loop integrals. 
In particular, Feynman integrands can be thought of as elements of a \emph{twisted} cohomology group \cite{Mastrolia:2018uzb,Mizera:2019gea,Frellesvig:2019kgj,Frellesvig:2019uqt,Mizera:2019vvs,Mizera:2019ose,Weinzierl:2020gda,Chestnov:2022xsy}.
Then, the intersection number (an algebro-geometric invariant) defines an inner product on the vector space of Feynman integrands. 
Thus, any Feynman integral can be decomposed into a minimal basis of master integrals without generating and solving a system of integration-by-parts (IBP) identities \cite{Laporta:2000dsw}.
Moreover, whenever there is a vector space and an inner product there exists an associated dual space.
The elements of this dual space, called dual forms, must be localized to generalized unitarity cuts in order for the definition of the intersection number to make sense. 
Mathematically, dual forms are elements of a \emph{relative} twisted cohomology group.
Here, relative simply means that the generalized unitarity cuts are now treated as geometric boundaries \cite{Caron-Huot:p1,Caron-Huot:p2}.
Paired with the intersection number, dual forms  provide a systematic formalism for extracting generalized unitarity coefficients in general dimensions. 
Intuitively, dual forms can be thought of as making the duality between generalized unitarity cuts and Feynman integrals precise. 

In this work, we compute the \emph{dual} differential equations (DEs) associated to the \emph{elliptic} two-loop three-mass sunrise family using a \emph{loop-by-loop} fibration. 
Loop-by-loop fibrations are particularly enticing since they allow us to recycle the lower loop differential equations and integral bases. 
While previous loop-by-loop attempts have had some success \cite{Broadhurst:1987ei, Gluza:2007rt, Frellesvig:2017aai, Marquard:2018rwx}, widespread adoption has not happened. 
However, the authors are optimistic that twisted cohomology can accommodate most loop-by-loop decompositions. 
In fact, we find constraints from the loop-by-loop splitting that help us choose a good basis. 
While no obvious obstruction for the loop-by-loop fibration is observed, generalizing to higher loops or different diagrams is needed to test the generality of our method. 

Since the dual DEs and Feynman DEs are simply related, we do not lose information by working with dual forms. 
In fact, there are several advantages. 
First, dual IBP identities are simpler. 
Since there are no propagators in dual forms, IBP identities cannot introduce square propagators.  
Moreover, all IBP vectors are restricted to generalized unitarity cuts. Second, dual forms do not have to look like Feynman integrands -- we can choose a basis motivated purely by the underlying geometry. 

This paper is organised as follows: in section \ref{SEC:review}, we provide a short review of dual forms. 
Then, the general framework for loop-by-loop fibrations is introduced in section \ref{SEC:loopByLoop}.
In section \ref{SEC:ellipticSunrise}, we define the sunrise and dual sunrise integrals, detail the loop-by-loop splitting and study the elliptic curve on the maximal-cut.
Section \ref{SEC:lblDE} details the loop-by-loop computation of the dual sunrise DEs. 
We also discuss various properties of these DEs and the related basis. 
In section \ref{SEC:feynmanBasis}, we show how the \emph{dual} differential equations simultaneously characterize dual forms and the Feynman integrands of \cite{Bogner:2019lfa}. 
In section \ref{SEC:K3banana}, we show that the geometry for the three-loop four-mass sunrise integral is associated to an elliptically fibred K3-surface and discuss the period domain of such surfaces. 
We close this section by speculating on how to construct modular invariant dual bases for the three-loop four-mass sunrise diagram from our loop-by-loop approach.
\section{Feynman integrands and their duals \label{SEC:review}}

In section \ref{sec:Feynman integrands}, we describe how Feynman integrands fit into the framework of twisted cohomology. 
Then, from the definition of the intersection number, an algebraic invariant that acts as an inner product on the vector space of Feynman integrands, we deduce the definition of dual forms (section \ref{sec:int num and duals}). 
\subsection{Feynman integrands and twisted cohomology \label{sec:Feynman integrands}}

An $n$-point multi-loop Feynman integrand $\mathcal{I}$ is a rational function of $L$ loop momenta $\{\ell_i^\mu\}_{a=1}^L$ and $(n-1)$ external momenta $\{p_i^\mu\}_{a=1}^{n-1}$ multiplied by the volume form
\begin{align}
\label{eq:FeynmanIntegrand}
    \mathcal{I}\left(\{\ell_i^\mu\},\{p_i^\mu\}\right) 
    = \frac{
        \mathcal{N}\left(\{\ell_i^\mu\},\{p_i^\mu\}\right)
    }{
        \prod_a \textsf{D}_a\left(\{\ell_i^\mu\},\{p_i^\mu\}\right)
    }
    \bigwedge_b \frac{\ed^{D}\ell_b}{\pi^{D/2}}
    .
\end{align}
The numerators $\mathcal{N}$ are theory dependent and in general can be quite complicated. 
On the other hand, the denominators are universal -- they must be a product of propagators
\begin{align}
    \textsf{D}_a = q_a^2\left(\{\ell_i^\mu\},\{p_i^\mu\}\right) 
        + m_a^2.
\end{align}
Here, $q_a$ is a linear combination of loop and external momenta and $m_a$ is a mass. 

To regulate divergences that appear in Feynman integrals, we take the spacetime dimension $D=D_{\mathds{N}}-2\vep$ to be near some integer $D_\mathds{N}\in\mathds{N}$ where the dimensional regularization parameter $\vep$ is assumed to be non-integer $\vep\notin\mathds{Z}$ and infinitesimal.\footnote{{Note that only the $\Dn = \text{even}$ and $\Dn = \text{odd}$ cases have to be considered separately due to well known relations between Feynman integrals in $D$-dimensions and Feynman integrals in $(D+2)$-dimensions \cite{Tarasov:1996br}.}}
While Feynman integrands look like single-valued rational differential forms, they become multi-valued when the spacetime dimension is non-integer. 

To see the emergence of this multi-valuedness, consider a one-loop $n$-point integrand. 
While the complexified loop and external momentum are $D$-dimensional $\ell^\mu,p_a^\mu \in \mathds{C}^D$,
the external momenta are physical and therefore constrained to lie in a $\Dn$-dimensional subspace
\begin{align} \label{eq:perp par}
	p^\mu_i = (p_i^0, p_i^1, \dots, p_i^{\Dn}, \bs{0}_\perp), 
	\qquad 
	\ell^\mu = (\ell^0, \ell^1, \dots, \ell^{\Dn},\bs{\ell}_{\perp}).
\end{align}
Thus, the one-loop volume form becomes
\begin{align} \label{eq:1loop measure}
    \frac{\ed^D\ell}{\pi^{D/2}} 
    = \mathcal{C}_{\Dn}\ u\ \frac{\ed^\Dn\ell \wedge d\ell_\perp^2}{\ell_\perp^2},
    \qquad 
    u = (\ell_\perp^2)^{-\varepsilon},
    \qquad 
    \mathcal{C}_{\Dn} = \frac{1}{\pi^{\Dn/2} \Gamma(-\varepsilon)},
\end{align}
where the multi-valued function $u$ is universal to all one-loop integrals and the overall prefactor $\mathcal{C_\Dn}$ can be ignored in many situations.
At higher loops, the radius $\ell_{\perp}^2$ generalizes to the Gram determinant of the $\perp$-space loop momenta. 
The multi-valued function $u$ is often called the \emph{twist}.

Note that dimensionally regulated one-loop integrals are $(\Dn+1)$-forms rather than $\Dn$-forms.
Generalizing to more loops is straightforward (see \cite{Caron-Huot:p1,Caron-Huot:p2}). 
For $L\geq2$, there are $L$ radii $\ell_{i,\perp}^2$ and $\binom{L}{2}$ scalar products $\ell_{i,\perp}\cdot\ell_{j,\perp}$.
Thus, $L$-loop Feynman forms are generically $n_L$-forms where 
\begin{align} \label{eq:nL}
    n_L = L\times(\Dn + 1) + \binom{L}{2}.
\end{align}
As we will see, when the number of external particles $n$ is less than $\Dn+1$ some of the $n_L$ integration variables are spectators and can be integrated out.
For example, the $L=2$ Feynman forms associated to the sunrise topology are four-forms in our parameterization rather than ($n_2 =11$)-forms.

Since the twist $u$ is universal (for a given $L$), we can work with single-valued forms $\vphi$ rather than multi-valued integrands $\mathcal{I}$ by factoring out $u$
\begin{align}
    \mathcal{I} \equiv u \vphi.
\end{align}
Then the presence of $u$ is encoded by a covariant derivative 
\begin{align}
    \ed (u \vphi) 
    = u~(\ed + u^{-1}\ed u \wedge) \vphi
    = u~\nabla \vphi,
\end{align}
where 
$\nabla = \ed + \omega \wedge$ 
and 
$\omega = \ed\log u ~\propto~ \vep$
is a \emph{flat (integrable) Gau\ss-Manin connection}.
In this way, we can ``forget'' about $u$. 
We will refer to the single-valued forms $\vphi$ associated to Feynman integrands $\mathcal{I}$ as \emph{Feynman forms}. 

Feynman forms $\vphi$ can have poles on the locus 
\begin{align}
    \textsf{Tw} = \{ u = 0,\infty \}.
\end{align} 
Such singularities are called \emph{twisted singularities} and are said to be regulated by $\vep$. 
That is, the integral $\int u \vphi$ evaluates to well-defined expressions (via analytic continuation) even when $\vphi$ has poles on $\textsf{Tw}$.
Multiplying a Feynman integrand by integer powers of $u^{-1/\vep}$ relates Feynman forms in different dimensions. Thus, physically, these singularities correspond to dimension shifting.

In addition to twisted singularities, Feynman forms have also unregulated singularities at the on-shell conditions $\textsf{D}_a=0$ called \emph{relative singularities}. 
We denote the collection of all relative singularities by 
\begin{align}
    \textsf{Rel} = \bigcup_a \{ \Dsf_a = 0 \}. 
\end{align}
These singularities are much more dangerous since the integral $\int u \vphi$ does not necessarily evaluate to something well-defined when $\vphi$ has poles on $\textsf{Rel}$. 

Thus, Feynman forms are defined on the manifold of complexified loop momentum space where both twisted \emph{and} relative singularities have been excised
\begin{align}
    X = \mathds{C}^{n_L}
    \setminus (\textsf{Tw} \cup \textsf{Rel}) \, .
\end{align}
In particular, Feynman forms are middle-dimensional twisted de Rham cochains\footnote{
{Here, \say{middle-dimensional} refers to the \emph{complex} dimension of of the space $X$, which is \emph{half} the \emph{real} dimension of $X$. In fact, it was shown by Aomoto that (in generic situations) only middle-dimensional twisted cohomology groups are non-empty \cite{Aomoto1975OnVO}.}}
\begin{align}
    \vphi \in \Omega_\text{dR}^{\text{dim}_\mathds{C}X}(X,\nabla),
\end{align}
where $\text{dim}_\mathds{R} X = 2~\text{dim}_\mathds{C}X = 2{n_L}$ is the real dimension of $X$.

There also exists non-trivial relations between Feynman \emph{integrals called integration-by-parts} (IBP) \emph{identities}.
In the language of differential forms, IBP identities are equivalent to the statement that the total covariant derivatives integrate to zero
\begin{align}
    \sum_{a} c_a \int \mathcal{I}_a = 0
    \iff
    \sum_{a} c_a~\vphi_a = \nabla \phi. 
\end{align}
Equivalently, this means that a Feynman form is unique only up to a covariant derivative
\begin{align}
    \vphi \simeq \vphi + \nabla \phi.
\end{align}
This redundancy can be removed by working with the equivalence classes of Feynman forms modulo IBP identities. 
Mathematically, Feynman forms are representatives of equivalence classes belonging to \emph{twisted cohomology} groups
\begin{align}
	\vphi \in H^p_{\textnormal{dR}}(X;\nabla) 
	\equiv \frac{
		\ker \nabla: \Omega^p_{\textnormal{dR}}(X;\nabla) \to \Omega^{p+1}_{\textnormal{dR}}(X;\nabla)
	}{
		\textnormal{im} \nabla: \Omega^{p-1}_{\textnormal{dR}}(X;\nabla) \to \Omega^{p}_{\textnormal{dR}}(X;\nabla)
	}
        \sim \frac{\text{Feynman forms}}{\text{IBP identities}}.
\end{align}
Here, Feynman forms belong to the kernel of $\nabla$ simply because they are top-dimensional holomorphic forms ($p=\text{dim}_\mathds{C}X$), while IBP identities originate from the image of $\nabla$.

\subsection{The intersection number and dual forms \label{sec:int num and duals}}

\begin{table}
\begin{center}
\begin{tabular}{c|c}
        \textbf{\emph{Constraints}}
        & \textbf{\emph{Dual forms}}
    \\
        \multirow{2}{*}{Feynman forms are middle dimensional}
        & \multirow{2}{*}{Also middle dimensional}
        \\
        &
    \\ \hline
         \multirow{2}{*}{$\braket{\bullet}{\bullet}$ independent of IBPs}
        & \multirow{2}{*}{Belong to a dual cohomology group}
        \\
        &
    \\ \hline
         \multirow{2}{*}{$\braket{\bullet}{\bullet}$ single-valued}
        & Dual twist: $\wc{u} = u^{-1}$
        \\
        & Dual connection: $\wc{\omega} = -\omega$
    \\ \hline
        \multirow{2}{*}{$\braket{\bullet}{\bullet}$ finite}
        & Compactly supported near \emph{all}
    \\
        & singularities of Feynman forms
\end{tabular}
\caption{Table summarizing the properties of dual forms and why they are necessary. \label{tab:dual def}}
\end{center}
\end{table}

The \emph{intersection number} is a pairing between two differential forms. 
The simplest way to get a number from two differential forms is to wedge a $p$-form $\wc\vphi$ with a $(2{n_L} - p)$-form $\vphi$ and integrate over the whole manifold $X$ 
\begin{equation} \label{eq:int num def}
    \braket{\widecheck{\vphi}}{\vphi} 
    \equiv \frac{(-1)^{ \frac{p(p-1)}{2}} }{(2\pi i)^{n_L}} \int_X (\wc{u}\ \widecheck{\vphi}_c) \wedge (u\ \vphi)
    = \frac{1}{(2\pi i)^{n_L}} \int_X (\wc{u}\ \widecheck{\vphi}_c^\top) \wedge (u\ \vphi). 
\end{equation}
Here, the symbol \say{$\top$} instructs us to transpose the order of the differentials in the wedge product.\footnote{While not obvious yet, dual forms have $p=n_L$ and depend on anti-holomorphic variables. Therefore, the transpose ensures that anti-holomorphic and holomorphic pairs of differentials are always adjacent if we integrate one variable at a time: $(-1)^{ \frac{n_L(n_L-1)}{2}} \int (\ed \bar{z}_1 \wedge \cdots \wedge \ed \bar{z}_{n_L})^\top \wedge (\ed z_1 \wedge \cdots \wedge \ed z_{n_L} \to \int \bar{z}_{n_L} \wedge \cdots \wedge (\ed \bar{z}_{1} \wedge \ed z_1) \wedge \cdots \wedge \ed z_{n_L} \to \cdots \to \int \ed \bar{z}_{n_L} \wedge \ed z_{n_L}$. In the end, the transpose and factor of $(-1)^{ \frac{p(p-1)}{2}}$ are just convenient normalization choices for our applications.} 
In practice, \eqref{eq:int num def} evaluates to a sequence of ${n_L}$ residues (hence the normalization $(2\pi i)^{-n_L}$).
When in $n<\Dn+1$ and the counting is not given by \eqref{eq:nL}, $n_L \equiv \text{dim}_\mathds{C}X$.

The space of dual forms can be deduced by requiring that \eqref{eq:int num def} is well-defined (the resulting conditions and their consequences are summarized in table \ref{tab:dual def}). 
Since Feynman forms are middle dimensional, their duals must also be middle dimensional. 
To ensure that the intersection number is independent of IBP identities dual forms must also belong to a dual cohomology group.
Requiring that the intersection number is single-valued, fixes the dual twist $\wc{u}=u^{-1}$ and  the dual covariant derivative
\begin{align}
    \wc\nabla=\ed+\wc\omega\wedge
    ,
\end{align}
where $\wc\omega=-\omega$.
Lastly, the intersection number should return a finite number. 
Thus, dual forms must vanish in the neighbourhood of possible Feynman form singularities.
The compact support of the dual forms ensures that the intersection number reduces to a sequence of residues.

Mathematically, these conditions imply that dual forms are elements of a \emph{compactly supported cohomology} on $X$
\begin{equation} \label{eq:Hc}
    \wc\vphi_c
    \in H^p_\drc(\Space;\wc\nabla) 
    \hspace{2cm} \mbox{(compact)}. 
\end{equation}
Working in the compactly supported cohomology may seem cumbersome since elements must have some anti-holomorphic dependence. 
However, one can always find representatives where the anti-holomorphic dependence simply originates from Heaviside theta functions and their derivatives. 

Since the compact support condition is only required for the computation of the intersection number, it is often convenient to instead work with an \emph{algebraic relative twisted cohomolgy} that is isomorphic to \eqref{eq:Hc} 
\begin{align}
    \wc\vphi
	& \in H^p_\text{alg} (\Spacedual,\Rel;\wc\nabla)
	\hspace{2cm} \mbox{(relative alg./holo. forms)}
 \nn\\
	&\simeq H^p_\dr(\Spacedual,\Rel;\wc\nabla) 
        \hspace{2cm} \mbox{(smooth relative)}
\nn\\
	&\simeq H^p_\drc(\Space;\wc\nabla) \hspace{2cm} \mbox{(compact)}.
\end{align}
While the compactly supported dual forms are defined on the same manifold as Feynman forms $\wc{X}_c = X$, the algebraic dual forms are defined on a manifold where \emph{only} the twisted singularities have been removed $\wc{X}_\alg = \mathds{C}^{n_L} \setminus \textsf{Tw}$.
These algebraic forms are mapped to compactly supported forms through the following sequence of maps
\begin{equation}
	H^p_\alg  
	\hooklongrightarrow
	H^p_\dr
	\overset{c}{\longrightarrow}       
	H^p_\drc,                   \,
\end{equation}
where the first map is the canonical inclusion and the second is called the \emph{c-map}. 
In particular, the c-map produces compactly supported forms whose anti-holomorphic dependence comes exclusively from Heaviside theta functions and their derivatives. An explicit example of the c-map is given in section \ref{SEC:feynmanBasis}.

Intuitively, the compact support condition can be thought of as including small boundaries around the on-shell conditions: $\{\Dsf_a=0\}$. 
One should think of algebraic relative twisted cohomology simply as a bookkeeping scheme that keeps track of the surface terms generated by IBP identities on a manifold with boundaries. 
Hence, while Feynman forms are singular on the loci $\{\Dsf_a=0\}$, dual forms are non-singular. 
On the other hand, IBP identities do not produce surface terms at twisted singularities: $\{u=0,\infty\}$. 
Thus, twisted singularities are not boundaries and dual forms can have also poles at the twisted singularities $\{u=0,\infty\}$.
See table \ref{tab:Feyn-dual compare} for a comparison of Feynman and dual forms.\footnote{For those familiar with the calculation of intersection numbers, note that the difference between how twisted and relative singularities/boundaries are handled is essential for the existence of the c-map. 
Since $\vep\notin\mathds{Z}$, forms with twisted singularities always have well-defined images under the c-map. 
In particular, one can always find local primitives near twisted singularities. 
On the other hand, relative singularities do not have well-defined images under the c-map since local primitives may not exist. 
}

\begin{table}
\begin{center}
\begin{tabular}{c|c}
        \textbf{\emph{Feynman forms}}
        & \textbf{\emph{Dual forms}}
        \\ \\
        \multirow{2}{*}{$H^{n_L}_{\dr}(\Space; \nabla_\omega)$}
        & \multirow{2}{*}{$H^{n_L}_{\alg}(\Spacedual, \{\Dsf=0\}; \nabla_{-\omega})$}
        \\
        &
    \\ \hline
        \multirow{2}{*}{Top-dimensional holomorphic}
        & \multirow{2}{*}{Top-dimensional holomorphic}
        \\
        &
    \\ \hline
        \multicolumn{2}{c}{\multirow{2}{*}{Possible singularities on the locus $\{u=0,\infty\}$}}
        \\
        \multicolumn{2}{c}{}
    \\ \hline
        \multirow{2}{*}{Possible singularities on the loci $\{\Dsf_a=0\}$}
        & \multirow{2}{*}{Non-singular on the loci $\{\Dsf_a=0\}$}
        \\
        &
\end{tabular}
\caption{Table comparing the singularity structure of Feynman and dual forms. \label{tab:Feyn-dual compare}}
\end{center}
\end{table}

Elements of relative twisted cohomology are formal sums where each term has support on one of the boundaries
\begin{equation} \label{eq:delta notation}
        \th\ \wc\vphi_0
 	+ \delta_{1} \wedge \th\ \wc\vphi_1
	+ \delta_{2} \wedge \th\ \wc\vphi_2
 	+ \delta_{12} \wedge \th\ \wc\vphi_{12}
	+ \cdots .
\end{equation}
Here, the first term $\wc\vphi_0$ is a bulk form while the remaining $\wc\vphi_I$ are forms supported on boundaries denoted by the symbol $\delta_I$. 
For example, $\delta_a$ corresponds to the boundary $\{\Dsf_a=0\}$, while $\delta_{ab}$ corresponds to the boundary $\{\Dsf_a=0\}\cap\{\Dsf_b=0\}$. 
It is also important to note that the $\delta_I$ are totally anti-symmetric: $\delta_{ab} = - \delta_{ba}$.
Moreover, each form $\wc\vphi_I$ is also multiplied by another formal symbol $\th$, which reminds us to keep track of boundary terms generated by derivatives 
\begin{equation}
    \label{eq:boundary terms}
    \wc\nabla \left[ \delta_I \wedge (\theta\ \wc\vphi_I) \right]
    = (-1)^{|I|} \delta_I \wedge 
    \left(
        \theta\ \wc\nabla\vert_I\ \wc\vphi_I 
        + \sum_{j \notin I} \delta_j \wedge 
            \theta\ \wc\vphi_I\big|_{j}  
    \right).
\end{equation}
These rules are easy to remember if one views $\th$ as a literal product of step functions each vanishing in the neighbourhood of a boundary. 
Then, the notation naturally suggests
\begin{align} \label{eq: dtheta nabla delta}
    \textnormal{d} \theta  =  \sum_{j} \delta_j\ \theta, 
    \qquad 
    \wc\nabla \delta_{I} \wedge \bullet 
    = (-1)^{|I|} \delta_I \wedge (\wc\nabla\vert_{I}\ \bullet),
\end{align}
where $\wc\nabla\vert_I = \ed + \wc\omega\vert_I \wedge$ is the covariant derivative restricted to the boundary $I$.

When restricting to a boundary, note that we always solve the on-shell conditions for the radial variables $\ell_{i,\perp}^2$ first. 
While it is not obvious at this point, each loop must be cut at least once to produce a non-trivial cohomology class. 
For example, the bulk cohomology class $\vphi_0$ in \eqref{eq:delta notation} actually does not exist.
Moreover, the co-dimension 1 cohomology classes $\vphi_a$ in \eqref{eq:delta notation} do not exist when $L>1$.
Since each loop must be cut and we solve for the $\ell_{i,\perp}$ first, the remaining un-cut propagators are linear. 
In contrast, the twist will generally contain some degree $2L$ polynomial.
One motivation for a loop-by-loop fibration is lowering the degree of the twist polynomial. 

The c-map isomorphism essentially replaces the combinatorial symbols $\th$ and $\delta$ by literal Heaviside and delta functions, which satisfy equivalent rules.
Explicitly, under the c-map, the combinatorial symbols become
\begin{align}
    \label{eq:c-map for delta and theta}
    \th 
    \to \prod_{a} \theta_a
    \, ,
    \qquad
    \th_a = \begin{cases}
        1 & \text{if } |\Dsf_a| > \epsilon
        \\
        0 & \text{if } |\Dsf_a| < \epsilon
    \end{cases}
    \, ,
    \qquad
    \delta_I 
    \to \frac{\wc{u}\vert_I}{\wc{u}} \bigwedge_{a \in I} \ed\th_a
    \, .
\end{align}
The $\ed\th_a$'s are essentially delta functions supported on a circle of radius $\epsilon$ around the on-shell condition $\{\Dsf_a=0\}$. 
Inside the intersection number, the $\ed\th_a$ take residues about the corresponding on-shell conditions.
Furthermore, the factor $u\vert_I/u$ parallel transports the covariant derivative to the boundary $I$ 
\begin{align}
    \wc\nabla \delta_{I} 
    = (-1)^{|I|} \delta_{I} \wedge \wc\nabla\vert_I.
\end{align}
The overall factors of $\th_a$ ensures that the form does not have support on any of the propagators. 

In addition to only having support away from the on-shell conditions, the c-map must also ensure that its image has no support on the twisted singularities. 
This is accomplished by multiplication by $\th_u$. 
The factor $\th_u$ is a product of step functions that has support away from the twisted singularities. 

While straightforwardly applying the rules \eqref{eq:c-map for delta and theta} to an element from the algebraic relative cohomology yields a form that has the correct compact support, it will not be closed. 
When testing for closure the derivative hits the $\theta$ factors and produces extra $\ed \theta$ terms.
By adding local algebraic primitives multiplied by the appropriate $\ed\th$'s, we ``patch up'' this form so that it is closed. 
This is the last step of the c-map.
Physically, the patch up terms can be thought of as subtractions for the box, triangle, and so on. Intersection theory provides an algorithmic way of constructing these subtractions.

The c-map procedure makes it obvious why relative singularities do not have well-defined images: forms with simple poles (i.e., $\textnormal{d}\log$-forms) do not have local \emph{algebraic} primitives.
We also see why the intersection number evaluates to a sequence of residues since the only terms that survive in the wedge product $\wc\vphi_c\wedge\vphi$ contain $n_L$ factors of  $\ed\th$'s. 
These $\ed\th$'s come from either a boundary $\delta_I$ or the ``patch up'' step of the c-map.

Since the c-map will only play a minor role in the following, we direct the reader to \cite{Caron-Huot:p1,Caron-Huot:p2} for further details.

\section{Loop-by-loop approach to differential equations\label{SEC:loopByLoop}}

Naively, it should be possible to evaluate multi-loop Feynman integrals loop-by-loop.
That is, to integrate over $\ell_1$ then over $\ell_2$ and so on. 
This option is attractive because it offers a way of recycling lower loop results to generate new higher loop results.
Loop-by-loop methods can also be used to reduce the number of integration variables. 
Motivated by these observations, we initiate the loop-by-loop study of differential equations for (dual) Feynman integrals.

In section \ref{SEC:lbl review} we review some previous loop-by-loop applications. 
In particular, we comment on some known problems and explain how our perspective differs. 
Then, in section \ref{SEC:lbl math} we summarize the  mathematical formalism underlying the loop-by-loop fibration. 

\subsection{Loop-by-loop in the literature\label{SEC:lbl review}}

For most loop-by-loop approaches, one tries to perform the integration one loop at a time \cite{Broadhurst:1987ei, Gluza:2007rt, Frellesvig:2017aai, Marquard:2018rwx}. 
While this method can be successful, it also fails when (after integrating out a certain number of loops) one is left with an integrand that does not look like a Feynman integrand (as defined in \eqref{eq:FeynmanIntegrand}). 
For instance, when algebraic roots or more complicated special functions are introduced into the integrand, it takes a considerable amount of effort to keep brute force numerical integration under control.

Loop-by-loop approaches can also yield integral representations with less integration variables. 
For example, the package \texttt{AMBRE} \cite{Gluza:2007rt} automatically applies loop-by-loop integration to generate Mellin-Barnes representations with minimal integration variables. 
However, the loop-by-loop approach does not seem to be the most efficient for non-planar integrals \cite[section 8]{Gluza:2007rt}. 
More recently, the leading singularity of various integrals was computed using a loop-by-loop Baikov representation \cite{Frellesvig:2017aai}. 
While the authors succeeded in reducing the number of integration variables, they still needed to preform challenging integrations to arrive at the final representation.

Since Feynman integrals and their duals are complicated transcendental functions of the kinematic data, trying to integrate these functions against additional propagators is extremely difficult. 
In contrast, the differentials of (dual) Feynman integrals are relatively simple.
Therefore, by examining differential equations instead of integrals we can hope to make progress using a loop-by-loop method. Moreover, due to recent advances  in our understanding of twisted cohomologies \cite{Mastrolia:2018uzb,Mizera:2019gea,Frellesvig:2019kgj,Frellesvig:2019uqt,Mizera:2019vvs,Mizera:2019ose,Caron-Huot:p1,Caron-Huot:p2,Chestnov:2022alh,Chestnov:2022xsy}, it is possible to account for the appearance of algebraic roots that appear in the differential equations.

\subsection{Mathematical setup \label{SEC:lbl math}}

{In order to break up a multi-loop problem into smaller more manageable one-loop problems, we need to understand the fibre bundle structure of multi-loop integrals. 
This will allow us to write any two-loop form as a linear combination of the 1-loop basis \begin{align}
    \wc{\vphi}^{(\text{2-loop})}_a(\ell_1, \ell_2, p)
    = \wc{\vphi}^{(\text{1-loop})}_b(\ell_1,\ell_2,p) 
    \wedge \wc{\vphi}^{(\text{left-over})}_{ba}(\ell_1,p).
\end{align}
This splitting will impose constraints on the left-over pieces and inform our choice of ``good'' $\wc{\vphi}^{(\text{left-over})}_{ba}$ when building the two-loop basis. 
Moreover, since the differential equations for the one-loop basis are known, we can commute $\wc{\nabla}$ across the one-loop basis to get a new covariant derivative acting on the left-over part 
\begin{align}
    \wc{\nabla} \left( \wc{\vphi}^{(\text{1-loop})}_b
    \wedge \wc{\vphi}^{(\text{left-over})}_{ba} \right)
    = \wc{\vphi}^{(\text{1-loop})}_b \wedge \wc{\nabla}_{bc}^{(\text{new})}
    \left(\wc{\vphi}^{(\text{left-over})}_{ca} \right). 
\end{align}
Here, the one-loop differential equations provide the connection for the left-over part
\begin{align}
    \wc{\nabla}_{ab}^{(\text{new})} = \delta_{ab} \textnormal{d} + \Omega_{ab}^{(\text{1-loop})} \wedge 
    \quad \text{where} \quad 
     \wc{\nabla} \wc{\vphi}^{(\text{1-loop})}_{a}
    = \wc{\vphi}^{(\text{1-loop})}_{a} 
    \wedge \Omega_{ba}^{(\text{1-loop})} 
    .
\end{align}
Naively, it should be easier to compute the two-loop differential equations from the action of $\wc{\nabla}_{bc}^{(\text{new})}$ on $\wc{\vphi}^{(\text{left-over})}_{ca}$ instead of by the action of $\wc{\nabla}$ on $\wc{\vphi}^{(\text{2-loop})}$.
\\
\\
To avoid notational clutter from keeping track of boundaries, we describe the fibration of Feynman forms below. 
Including the boundaries of dual forms is straightforward and explicitly treated in sections \ref{SEC:ellipticSunrise} and \ref{SEC:lblDE}.
}

Fibre bundles are useful constructions in topology that describe complicated spaces in terms of simpler pieces (see \cite{NakaharaTextbook} for the basics). 
For our purposes, we can restrict the discussion to a special kind of fibration known as a \emph{Serre fibration}.

Loosely speaking, a fibration $\pi: X\twoheadrightarrow B$ is Serre if all fibres $F_b=\pi^{-1}(b)$, for any $b$ in some neighbourhood of $B$, are homotopically equivalent to each other. In that sense, a Serre fibration behaves like a locally trivial fibre bundle \emph{up to deformation}. 
In particular, it can be shown that any fibre (vector/principle) bundle is a Serre fibration \cite{mccleary2001user}. Therefore, they constitute a large class of fibrations including many relevant for physics applications.

\begin{figure}
\centering
\includegraphics[scale=0.4]{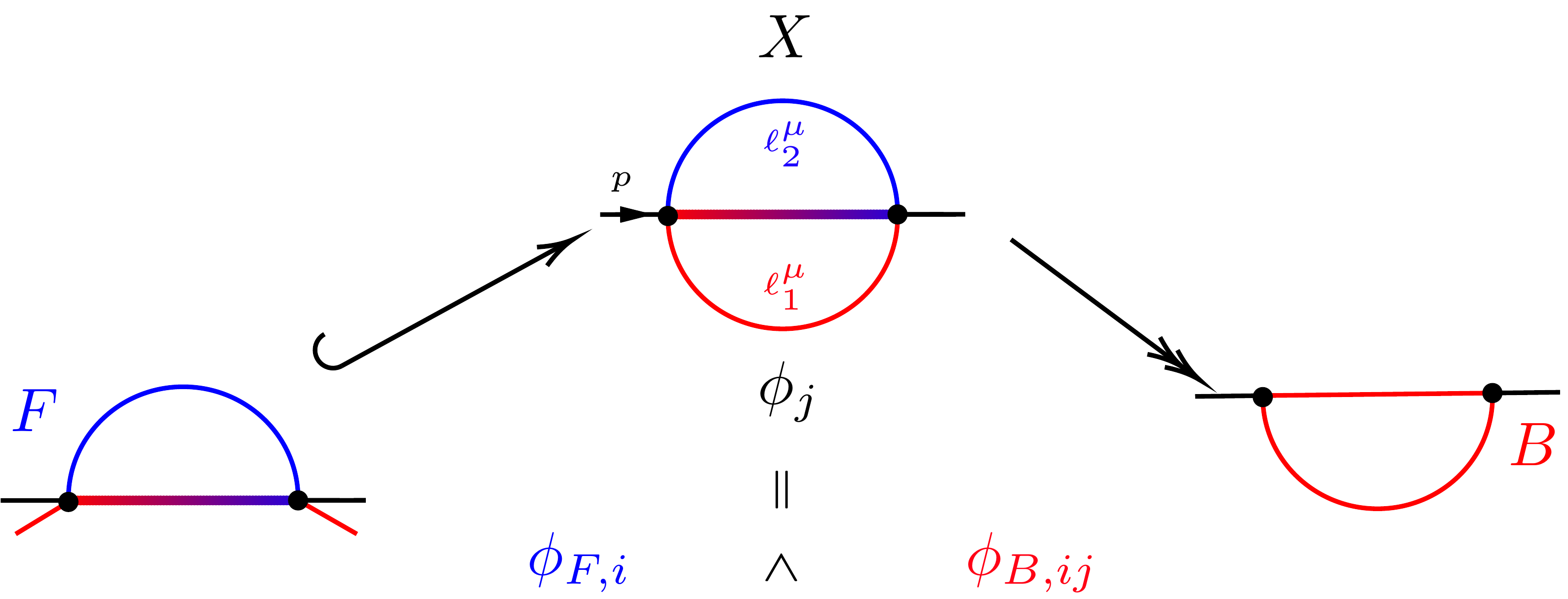}
    \caption{A sketch of a loop-by-loop fibration of the sunrise graph. The fibre space $F$ is defined as the subspace of $X$ spanned by $\ell_2$, which is the loop momentum running into the one-loop sub-bubble shown on the bottom left. For each different value of the base parameter $\ell_1$ we have a different one-loop sub-bubble.}
    \label{fig:sunFib}
\end{figure}

Recall that the total space $X$ is the complexified space of loop momenta where the singularities have been excised. 
For us, the space $F$ is naturally identified with the subspace of $X$ that is spanned by \emph{one} of the loop momentum. 
The orthogonal complement of $F$ in $X$ is called the base $B$. 
Varying the loop momenta spanning the orthogonal complement $B$ of $F$ in $X$ defines a fibre bundle with base $B$. 
Physically, $B$ is interpreted as the subspace of $X$ that \say{looks like} external kinematics from the viewpoint of the internal loop (fibre). This is schematically illustrated in figure \ref{fig:sunFib} for the sunrise diagram.

Associated to a Serre fibration is a \emph{Serre spectral sequence} \cite{serre1951homologie}. 
This spectral sequence can be used to extract information about the twisted (co)homologies of fibered spaces. 
Since the details of Serre spectral sequences are beyond the scope of this paper, we only comment on a few useful results for the purposes of our calculations. 
See \cite{hatcher2004spectral,mccleary2001user} for introductory material. 

Serre spectral sequences provide the following useful decomposition of the total space (co)homologies in terms of the fibre and base (co)homologies
\begin{equation}
\label{isomCohom}
    H^{p+q}(X,\nabla)
    \simeq H^{p}(B, H^q(F;\nabla\vert_F) )
    = H^{p}(B,\boldsymbol{\nabla}).
\end{equation}
 Equation \eqref{isomCohom} states that the information contained in the $(p+q)$ twisted cohomology group on $X$ is equivalently contained in the $p$ twisted cohomology group on the base space $B$ with coefficients in the $q$ cohomology of the fibre $F$.
 In plain language, this simply means that the twisted fibre cohomology generates a matrix-valued connection $\bs\nabla$ on the base cohomology, which is now vector-valued. Incidentally, base forms (on which $\bs\nabla$ acts) are vector-valued.
 
 To understand the origin of this vector-valuedness, let $\{\phi_j\}$ be a basis of $(p+q)$ twisted forms on $X$. 
 Next, let
 $F \hookrightarrow X \twoheadrightarrow B$, be a Serre fibration. Then, for each $j$, we express the elements of the total space  cohomology as a linear combination of the base cohmology basis with coefficients in the fibre cohomology
\begin{equation} 
\label{fibForm}
    \vphi_j=\textcolor{black}{\vphi_{F,i}}\wedge\textcolor{black}{\vphi_{B,ij}}.
\end{equation}
Here, $\vphi_{F,i} \in H^q(F,\nabla)$ is a basis element of the fibre cohomlogy and $\vphi_{B,ij} \in H^p(B,\boldsymbol{\nabla})$ is a basis element of the base cohomology. 
To keep our formulas clean, we denote $\bs\vphi_{F}$ as the vector of fibre basis elements. 
Similarly, $\bs\vphi_{B,j}$ is a vector-valued basis form on the base.
In this notation, equation \eqref{fibForm} becomes
\begin{align}
    \vphi_j = \bs\vphi_{F} ~\wedgedot~ \bs{\vphi}_{B,j},
\end{align}
where $\wedgedot$ is the combination of the standard vector dot product and wedge product. 

Next, we demonstrate how to compute the kinematic connection using the loop-by-loop fibration. 
Explicitly, we wish to find $\bs\Omega$ where $\nabla\bs\vphi = \bs\vphi~\wedgedot~\bs\Omega$. 
In addition to the fibre-base splitting of $\vphi_j$ (equation \eqref{fibForm}), we also split the twist into a piece that is independent of all fibre variables $u_B$ and one that is not $u_F$
\begin{align} \label{fibu}
    u = u_F \ u_B.
\end{align}
Then, the total space covariant derivative becomes $\nabla = \ed + \ed\log u_F \wedge + \ed\log u_B \wedge$.
Taking the covariant derivative of $\vphi_j$ and using equations \eqref{fibForm} and \eqref{fibu} yields 
\begin{equation}
\label{eq:lblDE1}
\begin{split}
    \nabla\vphi_j
    &=(\ed + \ed\log u_F \wedge + \ed\log u_B \wedge)
        \vphi_{F,i} \wedge \vphi_{B,ij}
    \\&=(\ed\vphi_{F,i} + \ed\log u_F \wedge \vphi_{F,i}) 
        \wedge \vphi_{B,ij}
    \\ & \quad \quad +(-1)^q~\vphi_{F,i} \wedge 
        (\ed\vphi_{B,ij}+\ed\log u_B \wedge \vphi_{B,ij})
    .
\end{split}
\end{equation}
Using an IBP reduction, we obtain the differential equation for the fibre basis
\begin{equation}
\label{eq:lblDE2}
    (
        \ed\vphi_{F,i} 
        + \textnormal{d}\log(u_F)\wedge\vphi_{F,i}
    )
        \simeq (-1)^q~\vphi_{F,k} \wedge \Omega_{F,ki},
    \end{equation}
where the symbol \say{$\simeq$} stands for \say{\emph{equivalent modulo IBP identities}}. Plugging \eqref{eq:lblDE2} back into \eqref{eq:lblDE1}, we find
\begin{equation}
\label{eq:lblDE3}
    \begin{split}
        \nabla\vphi_j&\simeq (-1)^q\ \vphi_{F,k}\wedge
        (
            \delta_{ki}~ \ed 
            + \Omega_{F,ki} \wedge
            +\delta_{ki}~ \ed\log u_B  \wedge
        ) \vphi_{B,ij}
        \equiv \bs\vphi_{F} ~\wedgedot~ \bs\nabla_B \bs\vphi_j.
    \end{split}
\end{equation}
From \eqref{eq:lblDE3}, we define the induced matrix-valued covariant derivative $\boldsymbol{\nabla}_B$ acting on the base forms
\begin{equation} \label{eq:nabla base def}
    \nabla_{B,ki}
    := \delta_{ki}~ \ed + \omega_{B,ki} \wedge 
    \quad \textnormal{where} \quad 
    \omega_{B,ki}
    :=\Omega_{F,ki} + \delta_{ki}~ \ed\log u_B.
\end{equation}
Note that the connection on the base $\boldsymbol{\omega}_B$ knows about $H^q(F,\nabla)$ through the kinematic connection on the fibre $\boldsymbol{\Omega}_F$.  
It is not hard to show that $\boldsymbol{\nabla}_B$ is also a flat (integrable) Gau\ss-Manin connection on a vector bundle \cite{kulikov1998mixed} -- i.e., 
\begin{equation}
   0=\boldsymbol{\nabla}_B \circ\boldsymbol{\nabla}_B =\textbf{d}\boldsymbol{\omega}_B
   + \boldsymbol{\omega}_B~\wedgedot~\boldsymbol{\omega}_B.
\end{equation}
Then, after an IBP reduction with respect to the base connection $\bs\nabla_B$, we obtain the kinematic connection  $\boldsymbol{\Omega}$ on $X$
\begin{equation}
    \begin{split}
        \nabla\vphi_j
        \simeq \vphi_{F,k} 
            \wedge \nabla_{B,ki}~ \vphi_{B,ij}
        \simeq \vphi_{F,k} 
            \wedge \vphi_{B,ki} 
            \wedge \Omega_{ij}
        = \vphi_i \wedge \Omega_{ij}
        . 
    \end{split}
\end{equation}

Up to some minor technicalities, the material covered in this section straightforwardly generalizes to the \emph{relative} twisted cohomologies discussed in section \ref{SEC:review}. 
The only added complication is that one has to keep track of boundary terms.
See \cite[p.17]{hatcher2004spectral}, \cite[Ex 5.5-6]{mccleary2001user} or \cite{Caron-Huot:p2} for the formal details on this extension.
\section{The three-mass (elliptic) sunrise family}
\label{SEC:ellipticSunrise}
\begin{figure}
    \centering
    \includegraphics[scale=.3]{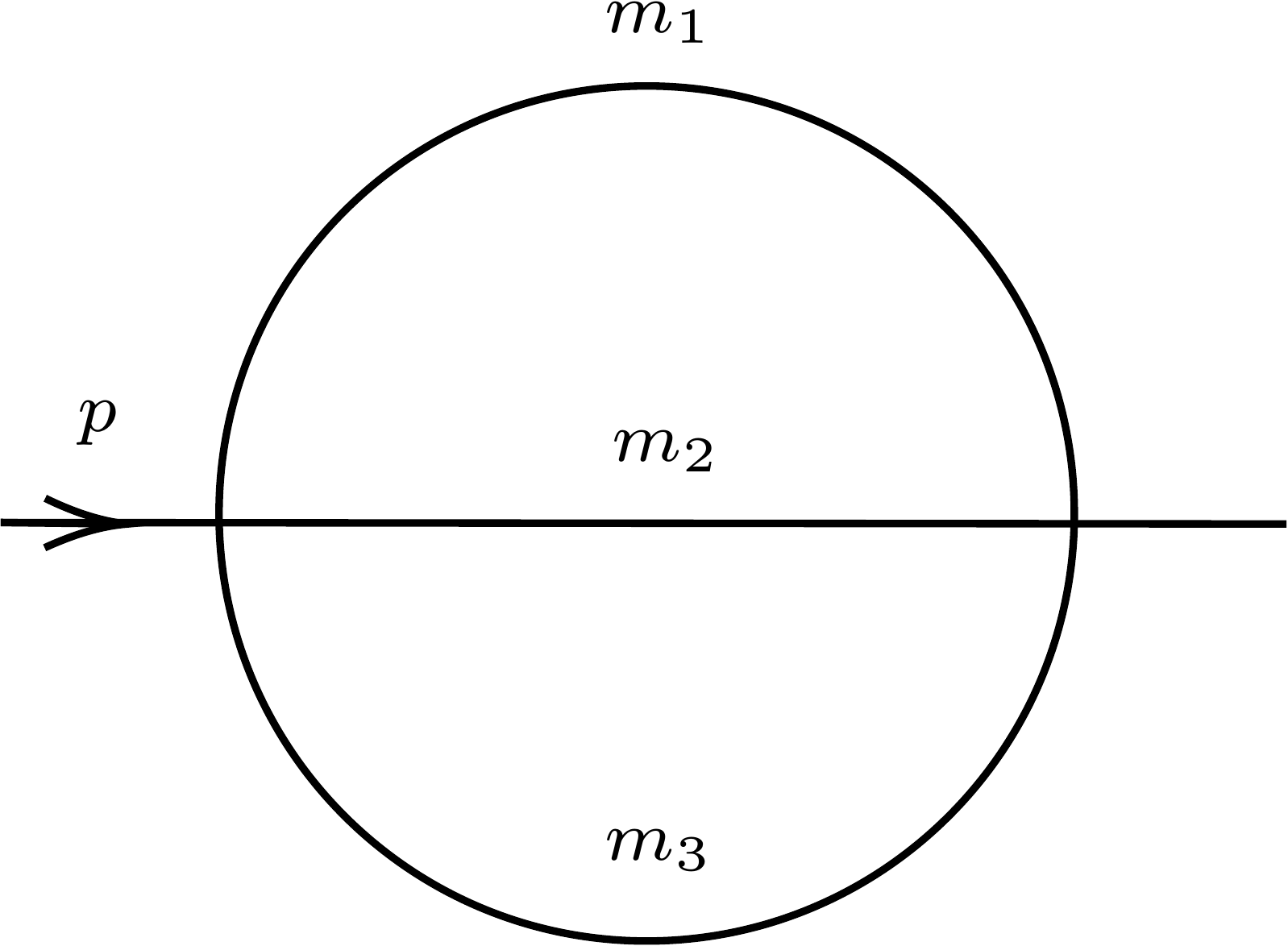}
    \caption{The three-mass sunrise graph.}
    \label{sunGraph}
\end{figure}
As a proof of concept, we apply the machinery discussed in section \ref{SEC:lbl math} to the simplest family of integrals that \emph{cannot} be expressed in term of MPLs: The \emph{two-loop sunrise} family (see figure \ref{sunGraph}), where 
\begin{align} \label{eq:3mass propagators}
    \Dsf_1 = \ell_1^2+m_1^2,
    \qquad
    \Dsf_2 = \ell_2^2+m_2^2,
    \qquad
    \Dsf_3 = (\ell_1+\ell_2+p)^2+m_3^2,
\end{align} 
are the inverse propagators.
Along with the twist, the propagators \eqref{eq:3mass propagators} dictate the geometry for both Feynman and dual forms.
Unless specified otherwise, we work near four-dimension ($D=4-2\vep$) and normalize our forms to be dimensionless once multiplied by the twist.

{While most of the literature on the sunset integral assumes $D=2-2\vep$, we prefer to work near the physical number of dimensions $D=4-2\vep$. However, in some sense, we are secretly studying $D=2-2\vep$ integrals. Our basis forms are defined with additional powers of $\ell_{i,\perp}^2$ that corresponds to the embedding of $D=2-2\vep$ integrals in $D=4-2\vep$. Without these dimension shifting factors, we could not achieve an $\vep$-form basis.
We find it more satisfying to keep the dimension fixed near the physical value and modify the forms to have nice mathematical properties.
}

Our construction of the basis of dual forms is intimately tied to the loop-by-loop procedure, for which some of the details are postponed to section \ref{SEC:lblDE}. Summarizing, the dual basis
\begin{align}
\label{eq:sunriseBasis}
    \wc\vphi_a 
    = \wc{\bs{\vphi}}_{F} ~\wedgedot~ \wc{\bs{\vphi}}_{B,a}, 
\end{align}
has the form \eqref{fibForm}, where the fibre basis $\{\wc\vphi_{F,a}\}$ is given by \eqref{fibBasis} and yields an $\vep$-form connection on the base (c.f., $\widecheck{\boldsymbol{\omega}}_B$ in \eqref{2loopBaseCD}).
Our starting base basis $\{\wc{\bs{\vphi}}_{B,a}\}$ (given in \eqref{baseTads} and 
 \eqref{mcBasis2New}) does not yield an $\vep$-form differential equation, but rather a differential equation \eqref{eq:linearForm} in \emph{linear form} \cite{Ekta:2019dwc}.
Our final base basis $\{\wc{\bs{\vartheta}}_{B,a}\}$ (given in \eqref{2loopBaseBasis}) satisfies an $\vep$-form differential equation \eqref{epsCon} with simple poles. It is related to $\{\wc{\bs{\vphi}}_{B,a}\}$ by two gauge transformations \eqref{eq:Ugauge} and \eqref{eq:Vgauge} fixed \emph{uniquely} by the modular properties of the differential equation in linear form.

In section \ref{sec:lbl param}, we define the loop-by-loop parameterization of the loop variables, while in section \ref{sec:2loopfibBasis}, we define the fibre basis.
Then we examine the maximal-cut corresponding to the 123-boundary in section \ref{sec:ecurve and marked points} uncovering the underlying elliptic curve.

\subsection{Loop-by-loop parameterization \label{sec:lbl param}}

Since the number of independent external momenta is less than $\Dn=4$, the perpendicular space (c.f., equation \eqref{eq:perp par}) can be enlarged to include the physical directions perpendicular to the single external momentum $p$.
Explicitly, we set 
\begin{align} \label{paramLoop1}
    \ell_{2}^\mu
    =\ell_{2,\perp}^\mu+\ell_{2,\parallel}^\mu, 
    & \qquad\ell_{2,\parallel}^\mu
        \in\text{span}\left\{ q^\mu \equiv \ell_{1}^\mu+p^\mu \right\}
    \implies \ell_{2,\parallel}^{\mu}
    = x_2\ q^\mu,
    \\
    \ell_{1}^\mu
    =\ell_{1,\perp}^\mu+\ell_{1,\parallel}^\mu, 
    & \qquad\ell_{1,\parallel}^\mu
        \in \text{span}\left\{p^\mu\right\} \implies\ell_{1,\parallel}^{\mu}
    = x_1\ p^{\mu},
\end{align}
such that 
\begin{align}
    \ell_{2,\perp} \cdot \ell_{2,\parallel} 
    =\ell_{2,\perp} \cdot q 
    = 0,
    \qquad
    \ell_{1,\perp} \cdot \ell_{1,\parallel}
    =\ell_{1,\perp} \cdot p 
    = 0.
\end{align}
When viewed from the perspective of the $\ell_2$ sub-bubble, $q=\ell_1+p$ is the external momentum.
Note that this parameterization implicitly fixes an integration order: $\ell_2$ is integrated over first. 

In the loop-by-loop parameterization \eqref{paramLoop1}, the boundaries (propagators) become
\begin{gather} 
    \textsf{D}_{1}  = \ell_{1,\perp}^{2} + x_1^{2}\ p^{2} + m_{1}^{2},
    \nn \\
    \textsf{D}_{2} 
     =\ell_{2,\perp}^{2} + x_2^{2}\ \ell_{1,\perp}^{2}
        + x_2^{2} \left(x_1+1\right)^{2}p^{2} + m_{2}^{2},
    \label{eq:2loop props} \\
    \textsf{D}_{3} 
     =\ell_{2,\perp}^{2} + \left(x_2+1\right)^{2}\ell_{1,\perp}^{2} 
        + \left(x_2+1\right)^{2}\left(x_1+1\right)^{2}p^{2}+m_{3}^{2}.
    \nn
\end{gather}
Also, the $\ell_2$ and $\ell_1$ integration measures follow straightforwardly from the one-loop measure \eqref{eq:1loop measure} 
\begin{align}
    \frac{\ed^{D}\ell_{2}}{\pi^{D/2}}
    &= \frac{\ed\Omega_{D-2}}{2\pi^{D/2}} \wedge
        \ed\ell_{2,\perp}^{2} \wedge 
        \ed x_2 \
        \sqrt{q^{2}}\ 
        \left(\ell_{2,\perp}^{2}\right)^{\frac{D-3}{2}}
        ,
    \\
    \frac{\ed^{D}\ell_{1}}{\pi^{D/2}}
    &=\frac{\ed\Omega_{D-2}}{2\pi^{D/2}} \wedge
        \ed\ell_{1,\perp}^{2} \wedge 
        \ed x_1 \
        \sqrt{p^{2}}\ 
        \left(\ell_{1,\perp}^{2}\right)^{\frac{D-3}{2}}
        .
\end{align}
Next, the angular integration can be performed independently of the $\ell_{i,\perp}$ and $x_i$ integrals
\begin{align}
\label{eq:hyperSphereArea}
    f(\vep)
    = \int\frac{\ed\Omega_{D-2}}{2\pi^{D/2}}
    =\frac{1}{\sqrt{\pi}~\Gamma\left(\frac{D-1}{2}\right)},
\end{align}
where we recall that $D=4-2\vep$.

For the purposes of computing differential equations, factors of $f(\vep)$ are irrelevant and will be dropped from now on.
Putting things together, we find that the full integration measure is a four-form
\begin{align}
    \frac{\ed\ell_2^D}{\pi^{D/2}} \wedge 
    \frac{\ed\ell_1^D}{\pi^{D/2}}
    \sim u\ \ed\ell_{2,\perp} \wedge \ed x_2
    \wedge \ed\ell_{1,\perp} \wedge \ed x_1,
\end{align}
multiplied by the multi-valued twist
\begin{align}
\label{eq:uTwist}
    u 
    &= \sqrt{q^{2}}\ 
    \left(\ell_{2,\perp}^{2}\right)^{\frac{D-3}{2}}
    \sqrt{p^2}
    \left(\ell_{1,\perp}^{2}\right)^{\frac{D-3}{2}}
    \nn \\
    &= \sqrt{\ell_{1,\perp}^2 + (x_1+1)^2 p^2}\ 
    \left(\ell_{2,\perp}^{2}\right)^{\frac{1}{2}-\vep}
    \sqrt{p^2}
    \left(\ell_{1,\perp}^{2}\right)^{\frac{1}{2}-\vep}
    .
\end{align}
We define the dual twist by the replacement $\vep\to-\vep$
\begin{align} \label{eq:lbl udual}
    \wc{u} = u\vert_{\vep\to -\vep}
    =\sqrt{\ell_{1,\perp}^2 + (x_1+1)^2 p^2}\ 
    \left(\ell_{2,\perp}^{2}\right)^{\frac{1}{2}+\vep}
    \sqrt{p^2}
    \left(\ell_{1,\perp}^{2}\right)^{\frac{1}{2}+\vep}
    . 
\end{align}
While \eqref{eq:lbl udual} is not the inverse of $u$ as defined in section \ref{sec:int num and duals}, the product $u \wc{u}$ is single-valued. 

Due to the square root factor in \eqref{eq:lbl udual},
not all twisted singularities are regulated by $\vep$ in the loop-by-loop parameterization.  
Still, the square root twisting is generic in the sense that the c-map is well-defined (see section \ref{SEC:feynmanBasis}).
Moreover, in section \ref{sec:ecurve and marked points}, we will see that these square roots define an elliptic curve. 

Next, we break up the space $\wc{X}_\alg$ ($\wc{X}_c$) into the fibre and base. 
The dual twist associated to the base is the part of $\wc{u}$ that does not depend on the $\ell_2$ variables $\ell_{2,\perp}^2$ and $x_2$
\begin{equation}
\label{eq:2loopBtwist}
    \wc{u}_B 
    := \sqrt{p^{2}} \left(
        \ell_{1,\perp}^{2}
    \right)^{\frac{1}{2}+\vep}.
\end{equation}
The remaining part of $\wc{u}$ is then the dual twist associated to the fibre 
\begin{equation}
    \wc{u}_F
    := \sqrt{q^2} \left(
        \ell_{2,\perp}^{2}
    \right)^{\frac{1}{2}+\vep}
    = \sqrt{\ell_{1,\perp}^2 + (x_1+1)^2 p^2}\ 
    \left(\ell_{2,\perp}^{2}\right)^{\frac{1}{2}-\vep}
    .
\end{equation}
Thus, the twisted singularities on the fibre and base are
\begin{align}
    \textsf{Tw}_F &= \left\{
        (\ell_{2,\perp}^2,x_2) \in \mathds{C}^2
        \setminus \{ \wc{u}_F = 0,\infty \}
    \right\},
    \\
    \textsf{Tw}_B &= \left\{
        (\ell_{1,\perp}^2,x_1) \in \mathds{C}^2
        \setminus \{ \wc{u}_B = 0,\infty \}
    \right\}
    .
\end{align}
On the base, there is only one boundary
\begin{align}
    \textsf{Rel}_B &= \left\{
        (\ell_{2,\perp}^2,x_2) \in \mathds{C}^2
        \setminus \{ \Dsf_1 = 0 \} 
    \right\},
\end{align}
since $D_1$ is the only boundary independent of $\ell_{2,\perp}^2$ and $x_2$.
On the other hand, the boundaries on the fibre are
\begin{align}
    \textsf{Rel}_F &= \left\{
        (\ell_{2,\perp}^2,x_2) \in \mathds{C}^2
        \setminus \{ \Dsf_2 = 0 \} 
        \cup \{ \Dsf_3 = 0 \}
    \right\}.
\end{align}
Note that in addition to the codimension-1 boundaries $\{\Dsf_2=0\}$ and $\{\Dsf_3=0\}$, there is also the codimesion-2 boundary $\Dsf_{23} = \{\Dsf_2=0\}\cap\{\Dsf_3=0\}$ on the fibre.

\subsection{The fibre dual basis and base connection \label{sec:2loopfibBasis}}

Having understood the singularity and boundary structure of the fibre and base, we can define a basis for the fibre cohomology $H^2_\alg(\mathds{C}^2\setminus\Tw,\Rel;\wc{\nabla}_F)$ where we 
recall that $\widecheck{\nabla}_F=\ed+\widecheck{\omega}_F\wedge$ and $\widecheck{\omega}_F=\textnormal{d}\log(\widecheck{u}_{F})$. 

Since a canonical ($\vep$-form) dual basis at one-loop is known \cite{Caron-Huot:p1}, we can reuse this result for the dual tadpole and bubble basis forms with only slight modification. 
We normalize the forms in \cite{Caron-Huot:p1} by an additional factor of $\wc{u}_B^{-1}\vert_{\vep=0} = (p^2 \ell_{1,\perp}^2)^{-1/2}$ 
to cancel the non-$\vep$ terms generated by the $\ed\log \wc{u}_B$ in the base connection \eqref{eq:nabla base def}
\begin{align} \label{fibBasis}
    &\textbf{\emph{Tadpoles}}: 
    \quad 
    \begin{cases}
        \wc{\varphi}_{F,1}
        = \textcolor{red}{ \frac{1}{ (\wc{u}_B\vert_{\vep=0}) } }\ 
        \frac{2\vep\ \th\ \delta_2 \wedge \ed \ell_{2,\parallel} }{(\ell_{2,\perp}^{2}\vert_{_2})\ q}
        =\textcolor{red}{ \frac{1}{ \sqrt{p^2\ell_{1,\perp}^2}} }\ 
        \frac{2\vep\ \th\ \delta_2 \wedge \ed x_2 }{\ell_{2,\perp}^{2}\vert_{_2}},
    \\
        \widecheck{\varphi}_{F,2}
        =\textcolor{red}{ \frac{1}{ (\wc{u}_B\vert_{\vep=0}) } }\ 
        \frac{2\varepsilon\ \th\ \delta_3 \wedge \ed\ell_{2,\parallel}}{(\ell_{2,\perp}^{2}\vert_{_3})\ q}
        =\textcolor{red}{ \frac{1}{\sqrt{p^2\ell_{1,\perp}^2}} }\ 
        \frac{2\varepsilon\ \th\ \delta_{3} \wedge \ed x_2}{\ell_{2,\perp}^{2}\vert_{_3}}
        ,
    \end{cases} 
\\
    &\textbf{\emph{Bubble}}: 
    \quad 
    \begin{cases}
        \widecheck{\varphi}_{F,3}
        =\textcolor{red}{ \frac{1}{ (\wc{u}_B\vert_{\vep=0}) } }\ 
        \frac{\delta_{23}}{ 
            \textcolor{blue}{
                \sqrt{
                    q^2 \ell_{2,\perp}^{2} \vert_{_{23}}
                }
            }
        }
        =\textcolor{red}{ \frac{1}{\sqrt{p^2\ell_{1,\perp}^2}} }\ 
        \frac{\delta_{23}}{ 
            \textcolor{blue}{
                \sqrt{q^2 \ell_{2,\perp}^{2} \vert_{_{23}}
                }
            }
        }
        .
    \end{cases}
    \nn
\end{align}
Explicitly, the above restrictions are 
\begin{align}
    -\ell_{2,\perp}^{2}\vert_{_2} 
    	&=  q^2 x_2^2 + m_2^2  
    ,
    \\
    -\ell_{2,\perp}^{2}\vert_{_3} 
    	&= q^2 x_2^2 + 2q^2 x_2 + m_3^2 + q^2
    ,
    \\
    -\ell_{2,\perp}^{2}\vert_{_{23}} 
    &= \frac{q_+^2 q_-^2}{4q^2},
\end{align}
where $q_\pm^2 = \left( q^2 + (m_2 \pm m_3)^2 \right)$.
The \textcolor{blue}{blue} factor in $\wc\vphi_{F,23}$ is the familiar ``kinematic'' square root normalization for bubble Feynman forms \cite{Caron-Huot:p1}, while the \textcolor{red}{red} square roots are new.
They are needed to ensure that the fibre basis yields a connection proportional to $\vep$ on the base. 

Recall that dual forms $\wc{\vphi}_{F,\bullet}$ are supposed to be single-valued.
Yet, we have added square root normalizations in the fibre basis \eqref{fibBasis}.
These square roots are allowed since from the perspective of the fibre, the variables $\ell_{1,\perp}^2$ and $x_1$ are ``external kinematic''.
However, due to the single-valued constraint on the wedge product $\wc{\bs{\vphi}}_F ~\wedgedot~ \wc{\bs{\vphi}}_{B,a}$, these square root factors place constraints on the choice of base basis. 

The differential equation for the basis \eqref{fibBasis} is straightforwardly obtained by recycling the canonical DE from \cite{Caron-Huot:p1} and keeping track of the new normalization factor in \textcolor{red}{red}.
Explicitly, 
\begin{equation}
     \widecheck{\nabla}_F~\widecheck{\boldsymbol{\varphi}}_F\simeq \widecheck{\boldsymbol{\varphi}}_F~\wedgedot~\widecheck{\boldsymbol{\Omega}}_F,
\end{equation}
where
\begin{small}
\begin{equation} \label{omegaF}
\small
\begin{split}
    \widecheck{\boldsymbol{\Omega}}_F
    &= \textcolor{red}{
        -\frac{1}{2}\
        \textbf{Id}\ 
        \ed\log( p^2 \ell_{1,\perp}^2)
    }
    + \varepsilon
    \underset{ \text{given in  \cite{Caron-Huot:p1}} }{\underbrace{
        \begin{pmatrix}
            {\textnormal{d}}\log\left(m_{2}^{2}\right) & 0 & 0
            \\
            0 & {\textnormal{d}}\log\left(m_{3}^{2}\right) & 0
            \\
            \widecheck{\Omega}_{F,31} 
                & \widecheck{\Omega}_{F,32} 
                & {\textnormal{d}} \log \left(
                    q_+^2 q_-^2 / q^{2}
                \right)
        \end{pmatrix}
    }}
    .
\end{split}
\end{equation}
\end{small} 
Here, the off-diagonal terms are given by
\begin{align}
    \widecheck{\Omega}_{F,31} &= \textnormal{d}\log\left[
        \frac{
            m_2^2 \left(q_-q_++2m_3^2\right)
            -\left(m_3^2+q^2\right)
            \left(q_-q_++m_3^2+q^2\right)
            -m_2^4
        }{2m_2^2 q^2}
    \right],
   \\
   \widecheck{\Omega}_{F,32} &=   \widecheck{\Omega}_{F,31}|_{m_2\leftrightarrow m_3}.
\end{align}
While $\wc{\bs{\Omega}}_F$ is not in $\vep$-form, the base connection $\wc{\bs{\omega}}_B$ (dual analogue of \eqref{eq:nabla base def}) is
\begin{align} \label{eq:nabla base}
    \wc{\bs{\omega}}_B 
    &= \wc{\bs{\Omega}}_{F} + \textbf{Id}\ \ed \log u_{B}
    \nn \\
    &= \vep\ \textbf{Id}\ \ed \log \ell_{1,\perp}^2
    + \varepsilon
        \begin{pmatrix}
            {\textnormal{d}}\log\left(m_{2}^{2}\right) & 0 & 0
            \\
            0 & {\textnormal{d}}\log\left(m_{3}^{2}\right) & 0
            \\
            \widecheck{\Omega}_{F,31} 
                & \widecheck{\Omega}_{F,32} 
                & {\textnormal{d}} \log \left(
                    q_+^2 q_-^2 / q^{2}
                \right)
        \end{pmatrix}
    .    
\end{align}
As a quick sanity check, we can verify if the base cohomology $H^2_\alg(\mathds{C}^2\setminus\Tw_B;\wc{\bs{\nabla}}_B)$ has the correct dimension. 

Counting the critical points corresponding to the diagonal elements of equation \eqref{eq:nabla base} yields an upper bound on the size of the base cohomology \cite{Frellesvig:2019uqt}.
Without restricting $\wc\omega_{ii}$ we find only one critical point coming from the $\wc\omega_{33}$ component. 
Evidently, this corresponds to the 23-double-tadpole. 
On the other hand, restricting $\wc\omega_{ii}$ to the $1$-boundary we find that $\wc\omega_{11}\vert_1$ and $\wc\omega_{22}\vert_{1}$ each have one critical point. 
These must correspond to the 12- and 13- double-tadpoles. 
Lastly, we find that $\wc\omega_{33}\vert_1$ has 4 critical points corresponding to the different maximal-cut integrals. 
Comparing to the known basis of master integrals for this family \cite{Bogner:2019lfa}, we find an exact match. 
There are 7 forms: 3 are double-tadpoles while 4 are maximal-cuts. 
Here, we stress that when \eqref{eq:nabla base} is not in $\vep$-form such counting is not as straightforward due to possible relations between the diagonal and off-diagonal cohomologies.

\subsection{The elliptic curve and its marked points \label{sec:ecurve and marked points}}

In order to construct a fibre basis that admits an $\vep$-form connection, we had to normalize the fibre basis by square roots that depend on the base variables. 
To ensure that $\wc{\bs{\vphi}}_F ~\wedgedot~ \wc{\bs{\vphi}}_{B,a}$ is single-valued, we have to undo the square root normalization in \eqref{fibBasis} when defining base forms. 
That is, the elements of the base cohomology need to contain one of the following factors
\begin{align}
    \label{eq:undo roots}
    \textbf{\emph{Tadpoles}:}\quad
    \sqrt{p^2~\ell_{1,\perp}^2 },
    \qquad\qquad
    \textbf{\emph{Bubble}:}\quad
     \sqrt{p^2~\ell_{1,\perp}^2 } \sqrt{q^2~\ell_{2,\perp}^2 \vert_{23} }.
\end{align}
We refer to this condition as the \emph{single-valued constraint} (sv-constraint).

Depending on whether or not there is a $\delta_1$ in the base form, these factors may be restricted to the $1$-boundary. 
Such a restriction is not problematic for the single-valuedness of $\wc{\bs{\vphi}}_F ~\wedgedot~ \wc{\bs{\vphi}}_{B,a}$ since the product/ratio of the restricted/un-restricted factors have a single-valued Taylor expansion near any of the twisted singularities. For example, the product
\begin{equation}
   \left(q \sqrt{\ell_{1,\perp}^2 } \sqrt{\ell_{2,\perp}^2 \vert_{23} }\right)\left(q\vert_{1} \sqrt{\ell_{1,\perp}^2\vert_{1}} \sqrt{ \ell_{2,\perp}^2 \vert_{123} }\right)
   = (q^2 \ell_{1,\perp}^2)\vert_{1}\ \ell_{2,\perp}^2 \vert_{123} 
   + \mathcal{O}(\textsf{D}_1),
\end{equation}
has a single-valued Taylor expansion near the maximal-cut boundary.

When the second factor in \eqref{eq:undo roots} is restricted to the $1$-boundary something special happens: the restriction defines a \emph{quartic elliptic curve} (reviewed in appendix \ref{SEC:GenericEllCurve})
\begin{align}
\label{eq:ecurve def}
    E_Y(\mathds{C}) &= \{ (x_1, \pm Y) \
        \vert x_1 \in \mathds{C} \},
    \\
    Y^2 &\equiv 
    \left.\left(
        q \sqrt{\ell_{1,\perp}^2 } \sqrt{\ell_{2,\perp}^2 \vert_{23} } 
    \right)^2\right\vert_{1}
    = \mathcal{N} \prod_{a=1}^4 (x_1 - r_a).
\end{align}
Here,  $\mathcal{N} = p^6$ and 
\begin{align} \label{2loopRoots}
    r_1 = \frac{- 1 - X_0^2 + (X_2+X_3)^2 }{ 2 X_0^2 } ,
    & \qquad
    r_2 = \frac{1}{X_0} ,
    \\
    r_4 = \frac{- 1 - X_0^2 + (X_2-X_3)^2 }{ 2 X_0^2 } , 
    & \qquad
    r_3 = - \frac{1}{X_0}  ,
\end{align}
where we have defined the dimensionless variables
\begin{equation}
    X_0=\frac{p}{m_1}, \quad X_2=\frac{m_2}{m_1} \quad \textnormal{and} \quad X_3=\frac{m_3}{m_1}.
\end{equation}
We also  work in a kinematic regime where $r_4<r_3<0<r_2<r_1$ are real. 
One can also see this elliptic curve by examining the maximal-cut of the twist in the $\vep\to0$ limit
\begin{equation}
    \label{eq:u max cut}
    Y^2 = \frac{1}{p^2}
    \left( \lim_{\vep \to 0} \wc{u}\vert_{123} \right)^2
    = \left.\left(
        q^{2}\ 
        \ell_{2,\perp}^{2} \
        \ell_{1,\perp}^{2}
    \right)\right\vert_{123}. 
\end{equation}
Thus, to undo the square root normalization of \eqref{fibBasis}, we have to include a factor of the elliptic curve $Y$ in all maximal-cut dual forms. 
\begin{figure}
    \centering
    \includegraphics[scale=0.30]{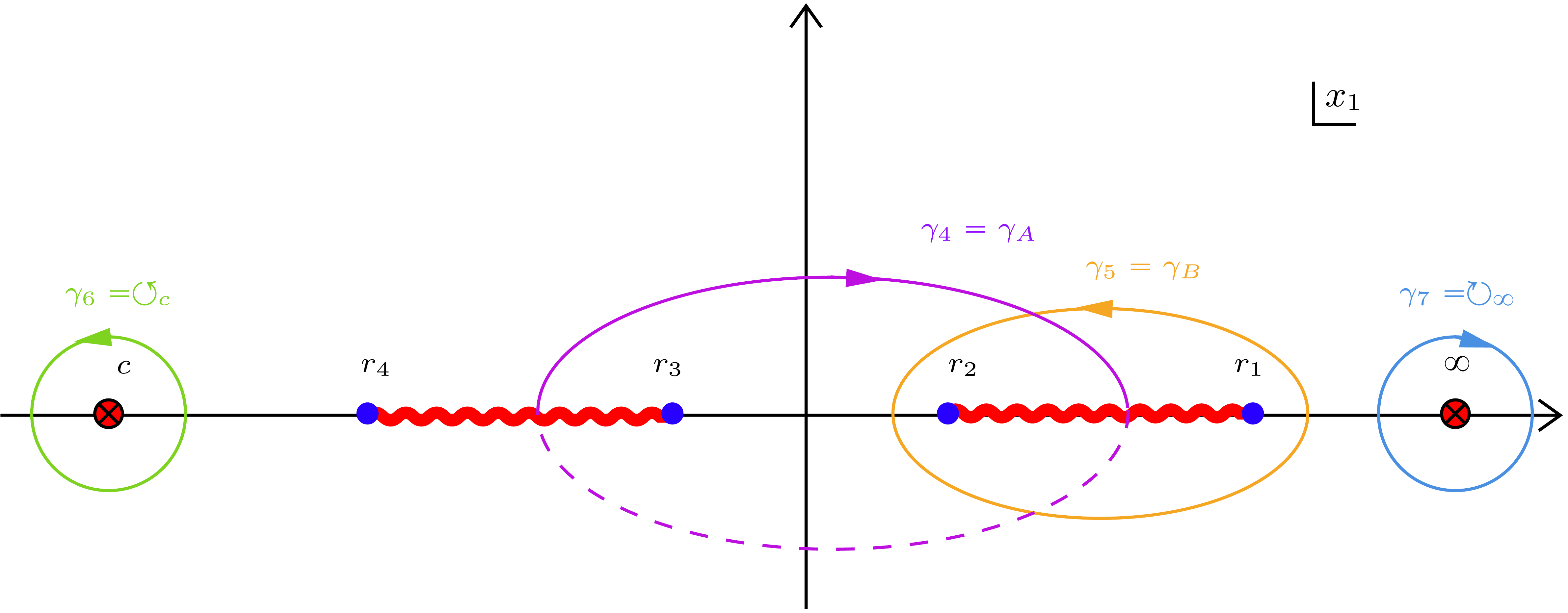}
    \caption{The homology basis in $D=4$ ($\vep=0$). Following standard conventions, the residue contour at finite point is oriented counterclockwise, while the residue contour at infinity is oriented clockwise. The orientations of the non-algebraic $A$- and $B$-cycles are such that the intersection number $[\widecheck{\gamma}_5|\widecheck{\gamma}_4]=+1$.
    }
    \label{fig:homologyCycles}
\end{figure}

The roots of $Y^2$ are always branch points for any $\vep$ (see figure \ref{fig:homologyCycles}). 
When $\vep\neq0$, the covering space requires infinitely many sheets. 
On the other hand, the covering space is only a double cover when $\vep=0$.  
Moreover, this double cover is conveniently described by a torus, $\mathds{T}$, that is constructed by gluing two copies of the complex plane pictured in figure \ref{fig:homologyCycles} along the branch cuts.
Then, the contours $\wc{\gamma}_{4}$ and $\wc{\gamma}_5$ form the $A$- and $B$-cycles on the torus.

Examining the maximal-cut restriction of the dual twist for generic $\vep$ 
\begin{equation}
    \label{2loopIntegrand}
    \widecheck{u}\vert_{123} 
    = \sqrt{p^2}\ 
    \left[ 2 p^2 (x_1-c) \right]^{-\vep}\
    \left(Y^2\right)^{ \frac{1}{2}+\vep },
\end{equation}
or the maximal-cut component of the base connection \eqref{eq:nabla base}
\begin{equation}
   \wc{\omega}_{B,33}\vert_1=\varepsilon \left(\textnormal{d}\log
   \left(Y^2\right)-\textnormal{d}\log \left(x_1-c\right)-\textnormal{d}\log \left(p^2\right)\right),
\end{equation}
reveals that there are also twisted singular points at $x_1 = c, \infty$  where 
\begin{equation}
    \label{eq:defC}
    c=-\frac{m_1^2+p^2}{2p^2}=-\frac{1+X_0^2}{2X_0^2}.
\end{equation}
However, when $\vep=0$ these points correspond to poles. 
Thus, to complete the $\vep=0$ homology basis, we have to also include residue contours about $x_1=c,\infty$ (see figure \ref{fig:homologyCycles}).

\begin{figure}
    \centering
\includegraphics[scale=0.33]{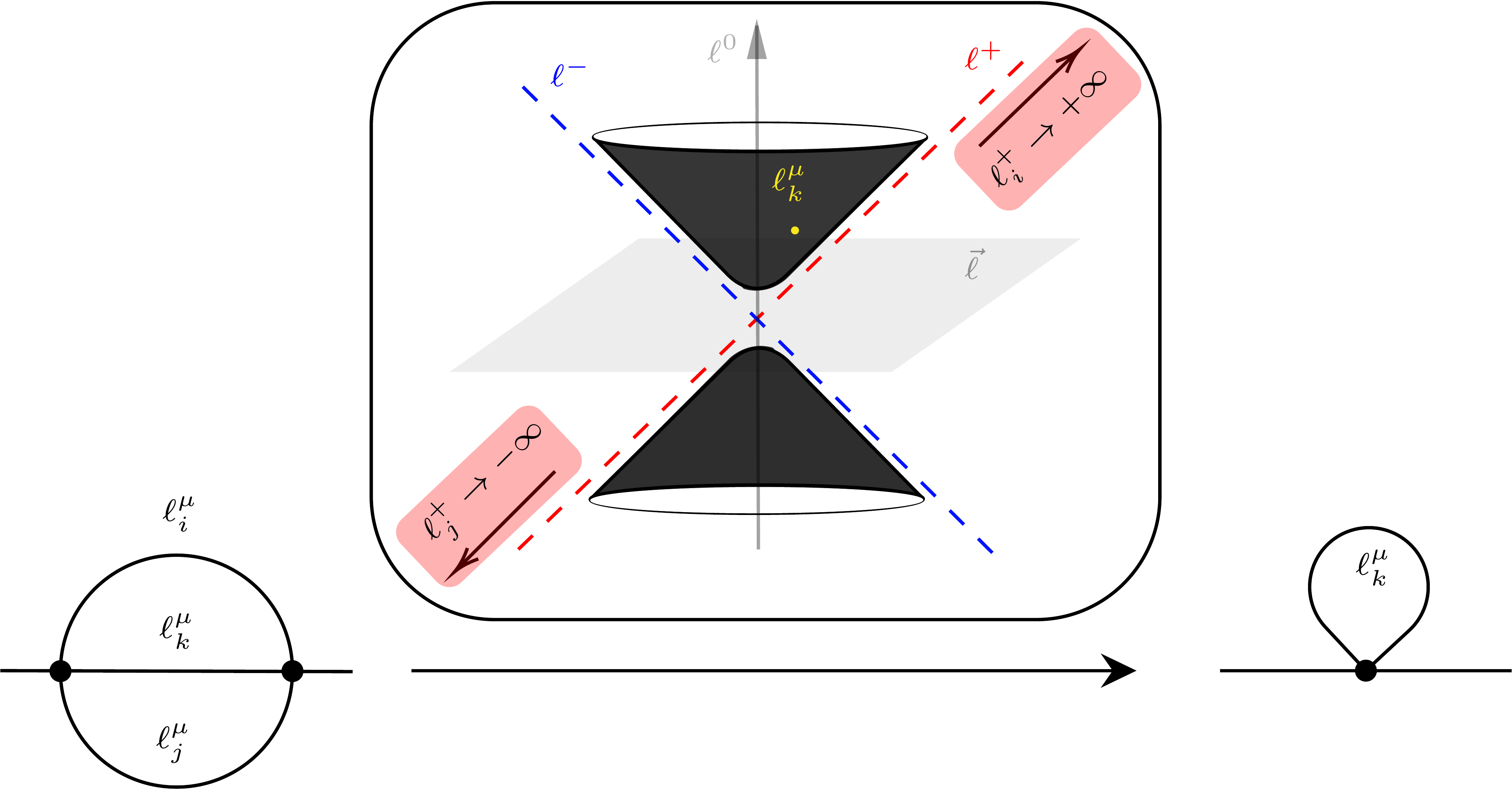}
    \caption{Special configurations of the sunrise graph through loop momenta going to infinity along opposite directions on the light-cone (either $+$ or $-$). The ambiguity in choosing the direction ($(\ell_j^+,\ell_j^+)\to (-\infty,+\infty)$ or $(\ell_j^-,\ell_j^-)\to (-\infty,+\infty)$) is transparent at the diagrammatic level since the sunrise graph contracts to the same one-loop tadpole.
    By momentum conservation, $\ell_k^\mu$ labels a finite point on the momentum shell. There are $\binom{3}{2}=3$ distinct pairs of $i,j$. The result of two virtual particles propagating infinitely fast in a sunrise diagram (left most Feynman diagram) is a one-loop tadpole (right most Feynman diagram).
    }
    \label{fig:LCmom}
\end{figure}

The marked point at $x_1=\infty$ corresponds to a degenerate limit where the sunrise graph degenerates into a one-loop tadpole.
Physically, this corresponds to loop momentum configurations where the momentum flowing through two of the propagators diverge in opposite directions along the light-cone (see figure \ref{fig:LCmom}). 
While we only detect the degeneration compatible with our choice of loop integration order, there are three distinct ways the sunrise diagram can degenerate into a tadpole with mass $m_i$. 
This is visible by a complimentary analysis where one does not use a loop-by-loop parameterization. 
For example, see \cite{Bogner:2019lfa,Weinzierl:2022eaz} where the analysis is performed in parametric space rather than loop momentum space.\footnote{From the parametric space analysis, the marked points are given by the intersection points of the domain of integration in Feynman parameter space with the zero set of the second graph polynomial: $(\alpha_i,\alpha_j)=(0,0)$ for $i\neq j$. Since $\alpha_i\to0$ implies that the corresponding $i^\textnormal{th}$-edge is contracted to a point \cite{schwartz2014quantum}, imposing the $(\alpha_i,\alpha_j)=(0,0)$ for $i\neq j$ produces three distinct one-loop tadpoles from the sunrise graph.}
It turns out that the marked point $x_1=c$ is a symptom of the loop-by-loop parameterization. 
On the maximal-cut, $x_1=c$ corresponds to a singularity of the sub-bubble $q^2\vert_{123}=0$ and not to a degeneration of the sunrise. 
While $x_1=c$ will play an important role in the loop-by-loop construction of our basis and differential equations, we choose to express final results in terms of quantities that are visible outside the loop-by-loop parameterization.

Since understanding the $\vep=0$ geometry is essential to constructing an $\vep$-form basis, it is important that we use variables and functions adapted to the torus. 
Functions on the torus must be doubly periodic with periods $\psi_{1,2}$. 
These periods define a lattice that, in turn, defines the torus as a quotient space
\begin{align}
    \mathds{T} &= \mathds{C} / \Lambda_{(\psi_1,\psi_2)},
    \\
    \Lambda_{(\psi_1,\psi_2)} &= \{m \psi_1 + n \psi_2 \vert m,n \in \mathds{Z} \}
    .
\end{align}
The periods are obtained by the integrating the holomorphic form on the torus
\begin{align}
    \ed Z = \frac{\ed x_1}{Y}
    \qquad
    \big(
        Z \in \mathds{T}
        \text{ and }
        (x_1,Y)\in E_Y(\mathds{C})
    \big),
\end{align}
over the $A$- and $B$-cycles
\begin{equation}
\label{periodsEllCurve}
 \psi_1 
    = 2c_4 \int_{r_3}^{r_2} \frac{\ed x_1}{Y} 
    = 2K(k^2), 
    \quad \quad
    \psi_2 
    = 2 c_4 \int_{r_1}^{r_2}
        \frac{\ed x_1}{Y} 
      = 2 i K(1-k^2).
\end{equation}
Here, $K$ is the elliptic integral of the first kind and 
\begin{align}
\label{eq:ellQUantities}
    c_4 = \frac{1}{2}\sqrt{r_{13} \ r_{24}} >0,
    \qquad
    0 <
    k^2 = \frac{ r_{23}\ r_{14} }{ r_{13}\ r_{24} }
    < 1,
    \qquad
    r_{ij} &= r_i - r_j.
\end{align} 
It is convenient to normalize one period to unity such that the other lies in the upper half plane $\tau = \psi_2/\psi_1 \in \mathds{H}$.\footnote{There is no loss of generality in assuming $\tau \in \mathds{H}$. If $\text{Im}\tau<0$, then one simply swaps the periods: $\psi_1\leftrightarrow\psi_2$.}
Then, the torus becomes 
\begin{align}
    \mathds{T} &= \mathds{C} / \Lambda_{(1,\tau)},
    \\
    \Lambda_{(1,\tau)} &= \{m  + n \tau \vert m,n \in \mathds{Z} \}
    .
\end{align}
The map between elliptic curve and torus is given by \emph{Abel's map}
\begin{equation}
\label{eq:goodAbel1}
     Z_{x_1^\prime}
     \equiv \frac{c_4}{\psi_1}
        \int_{x_1^\prime}^{r_2}\frac{\textnormal{d}x_1}{Y} 
     \quad \textnormal{mod} \ \Lambda_{(1,\tau)}
     \equiv\frac{F(\sin^{-1}(u_{x_1})|k)}{2K(k)}
     \quad \textnormal{mod} \ \Lambda_{(1,\tau)}
     .
\end{equation}
where 
\begin{equation}
    u_{x_1}=\sqrt{\frac{(x_1-r_1)(r_2-r_4)}{(x_1-r_2)(r_1-r_4)}}.
\end{equation}
In particular, the marked points $x_1 = c,\infty$ on the elliptic curve are mapped to the marked points $Z_c$ and $Z_\infty$ on the torus via \eqref{eq:goodAbel1}.
The inverse is given in equation \eqref{elllambv2}.
While it is mathematically important to use variables adapted to the torus, the physical meaning of the torus variables is harder to understand due to the transcendental nature of Abel's map.

Thanks to the translation symmetry of the torus, only distances (i.e., differences of $Z$'s) appear in final formulae. 
To this end, we define the distance $z_3$ such that 
\begin{align} \label{eq:z3DEF}
    z_3-1 
    \equiv Z_\infty - (-Z_\infty) \text{ mod } \Lambda_{(1,\tau)}
    &= 2 Z_\infty \text{ mod } \Lambda_{(1,\tau)}.
\end{align}
The \say{$-1$} above is conventional and chosen to make the equal mass limit simple.
Recall that physically, $Z_\infty$ and hence $z_3$ corresponds to the configuration where the sunrise degenerates to the $m_1$-loop tadpole (see figure \ref{fig:LCmom}).
The other degenerate configurations are related by mass permutations: $z_1 = 2 Z_\infty\vert_{m_1 \leftrightarrow m_2}$ and $z_2 = 2 Z_\infty\vert_{m_1 \leftrightarrow m_3}$.
Note that these quantities are true distances in the sense that  
\begin{align}
    z_2 = (-Z_\infty) - Z_c \text{ mod } \Lambda_{(1,\tau)},
    \quad\text{and}\quad
    z_1 = 1 - z_2 - z_3 \text{ mod } \Lambda_{(1,\tau)}.
\end{align}
Further note that the translation symmetry of the torus allowed us to fix $z_3$ such that 
\begin{equation}
    z_3=1-z_1-z_2.
\end{equation}
This way, the moduli degenerate to the same rational cusp $z_i = 1/3$ in equal mass limit.  
This translation symmetry also explains why there are no active moduli in the equal mass limit \cite{Adams:2017ejb}.

Before moving onto the construction of the base basis, it is important to note that any two lattices $\Lambda_{(\psi_1,\psi_2)}$ and $\Lambda_{(\psi_1^\prime,\psi_2^\prime)}$ are equivalent under $\textnormal{SL}(2,\mathds{Z})$ transformations
\begin{align}
    \begin{pmatrix}
        \psi_2
        \\
        \psi_1
    \end{pmatrix} 
    \to 
    \begin{pmatrix}
        \psi_2^\prime 
        \\
        \psi_1^\prime
    \end{pmatrix}
    =
    \begin{pmatrix}
        a & b
        \\
        c & d
    \end{pmatrix}
    \cdot
    \begin{pmatrix}
        \psi_2
        \\
        \psi_1
    \end{pmatrix}
    ,
    \qquad 
    \begin{pmatrix}
        a & b
        \\
        c & d
    \end{pmatrix}
    \in \textnormal{SL}(2,\mathds{Z})
    .
\end{align}
Similarly, any two normalized lattices $\Lambda_{(1,\tau)}$ and $\Lambda_{(1,\tau^\prime)}$ are equivalent under modular transformations
\begin{align}
    \tau \to 
    \tau^\prime = \frac{a\tau+b}{c\tau+d}
    ,
    \qquad 
    \begin{pmatrix}
        a & b
        \\
        c & d
    \end{pmatrix}
    \in \textnormal{SL}(2,\mathds{Z})
    .
\end{align}
Understanding the modular transformation properties of differential equation will play an essential role in writing our kinematic connection on the torus. 

\section{Constructing an $\vep$-form dual basis \label{SEC:lblDE}}

In this section, we construct a basis of dual forms that admits an $\varepsilon$-differential equation spanned by modular and Kronnecker forms, which are both reviewed in appendix \ref{SEC:DiffFormOnTorus}. 
Starting from a natural seed basis, we show how the gauge transformations bringing the kinematic connection into $\vep$-form are fixed from its modular properties (see section \ref{SEC:epsForm}). In section \ref{SEC:perMAT}, we discuss the period matrix of our $\varepsilon$-form basis.
Then, in section \ref{SEC:readyToIntegrateDE}, we write the kinematic connection on the torus in a form that is \say{ready-to-integrate}. 
We close this section by comparing the resulting elliptic symbol alphabet with the alphabet of \cite{Wilhelm:2022wow}.

\subsection{A seed basis for the base \label{SEC:baseBasis}}

Recall (from equation \eqref{eq:nabla base}) that the covariant derivative on the base is 
\begin{equation}
\label{2loopBaseCD}
    \widecheck{\boldsymbol{\nabla}}_B
    := \textbf{d}
        + \wc{\boldsymbol{\omega}}_B ~\wedgedot, 
    \quad \text{where}  \ 
    \wc{\bs{\omega}}_B =
        \wc{\bs{\Omega}}_F 
        + \textbf{Id}\ \ed\log(\wc{u}_B\vert_{\vep=0})
         \xrightarrow{\varepsilon\to 0}0.
\end{equation}
We know from the critical points of $\wc\omega_{B,11}\vert_1$, $\wc\omega_{B,22}\vert_1$ and $\wc\omega_{B,33}$ that there should be 3 double-tadpoles. 
Two of these should lie on the  1-boundary while one should be a bulk form. 

A general guiding principle for constructing a ``nice'' basis is  to use, as much as possible, $\ed{}\log$-forms where the argument of the $\log$'s vanish on the twisted singularities.
However, in our case, these $\ed\log$-forms must also cancel the square root normalizations  \eqref{eq:undo roots}. 
This constraint is rather strong and suggests the following basis for the double-tadpoles
\begin{equation}
\label{baseTads}
\small
   \textbf{Double-tadpoles}: \
   \begin{cases}
        \wc{\bs{\varphi}}_{B,1}
        = m_1^{-4\vep}
        \ed\log\left[
            \frac{
                1-\frac{ix_1}{\sqrt{\ell_{1,\perp}^2\vert_1/m_1^2}}
            }{
                1+\frac{ix_1}{\sqrt{\ell_{1,\perp}^2\vert_1/m_1^2}}
            }
            \right]
        \wedge \delta_{1}\begin{tiny}
        \begin{pmatrix}1\\0\\0\end{pmatrix}\end{tiny}, 
        \qquad 
        \\
    \widecheck{\boldsymbol{\varphi}}_{B,2}=m_1^{-4\vep}\ed\log\left[
            \frac{
                1-\frac{ix_1}{\sqrt{\ell_{1,\perp}^2\vert_1/m_1^2}}
            }{
                1+\frac{ix_1}{\sqrt{\ell_{1,\perp}^2\vert_1/m_1^2}}
            }
            \right]\wedge \delta_{1}\begin{tiny}\begin{pmatrix}0\\1\\0\end{pmatrix}\end{tiny},\\ \widecheck{\boldsymbol{\varphi}}_{B,3}=i~\varepsilon~m_1^{-4\vep}~\theta~ \textnormal{d}\log\left[\frac{p~\left(x_1+1\right)+\sqrt{\ell_{1,\perp}^{2}}}{p~\left(x_1+1\right)-\sqrt{\ell_{1,\perp}^{2}}}\right]\wedge\textnormal{d}\log\left[\frac{q_{+}-q_{-}}{q_{+}+q_{-}}\right]\begin{tiny}\begin{pmatrix}0\\0\\1\end{pmatrix}\end{tiny}.
    \end{cases}
\end{equation}
The $\varepsilon$-normalization of these forms were chosen such that the diagonal components of the kinematic connection corresponding to the double-tadpoles are in $\vep$-form, while the kinematic-dependent normalization factor is such that the products $\widecheck{u}~\widecheck{\varphi}_{1\le a\le3}$ (c.f., \eqref{eq:lbl udual} and \eqref{eq:sunriseBasis}) are dimensionless.\footnote{From \eqref{eq:lbl udual}, the mass dimension of $\widecheck{u}$ is $4+4\varepsilon$, while from \eqref{fibBasis} and \eqref{baseTads} the mass dimension of the double-tadpoles on $\widecheck{X}_{\textnormal{alg}}$ is $-4-4\varepsilon$.}

 Next, we need to construct 4 independent maximal-cut forms corresponding to the critical points of $\wc\omega_{B,33}\vert_{123}$.
The elliptic curve \eqref{eq:u max cut} in the denominator of $\widecheck{\varphi}_{F,3}|_1$ prevents us from using $\ed \log$-forms as basis elements. 
Instead, we use the following guiding principles to help choose a ``nice'' basis on the maximal-cut:
\begin{itemize}  \itemsep0em 
    \item The sv-constraint is satisfied.
    
    \item One basis element is the standard holomorphic form and one is proportional to its partial derivative in some kinematic variable.\footnote{We can motivate this requirement by recalling that the cohomology group of an elliptic curve is two-dimensional and spanned by such forms. {In particular, the \emph{simplest} basis choice is given by this specific pair. Furthermore, as pointed out in \eqref{eq:derAper}, the derivative form has vanishing $A$-cycle period at leading order in $\vep$. Section \ref{SEC:perMAT} makes clear that having a triangular period matrix simplifies the construction of a good basis. Moreover, the derivative also signals that one should normalize this form by a factor of $1/\vep$. Amazingly, the $1/\vep$ terms in the kinematic connection will cancel rather non-trivially.}}
    
    \item The holomorphic form has constant $A$-period and $B$-period proportional to $\tau$ at leading order in $\varepsilon$.
    
    \item The derivative form has vanishing $A$-period and constant $B$-period at leading order in $\varepsilon$.
    
    \item One basis element has a simple pole at infinity and nowhere else.
    
    \item One basis element has a simple pole at $c$ and nowhere else.
    

\end{itemize}
One possible choice that satisfies all of the above constraints is 
\begin{equation}
\label{mcBasis2New}
\small
   \textbf{\emph{Maximal-cut}}: 
   \begin{cases}
       \widecheck{\boldsymbol{\varphi}}_{B,4}
        = \frac{\psi_1^2}{\pi~ \varepsilon~W_{0}}
        \widecheck{\nabla}_{0}~ \widecheck{\boldsymbol{\varphi}}_{B,7}, 
        \quad   
        \widecheck{\boldsymbol{\varphi}}_{B,5}
        = \delta_{1} \wedge 
            m_1^{-4\vep} (x_1-r_1)
            \frac{\ed x_1}{Y_\textnormal{r}} 
            \begin{scriptsize}
                \begin{pmatrix}
                    0\\0\\1
                \end{pmatrix}
            \end{scriptsize},
        \\ 
        \widecheck{\boldsymbol{\varphi}}_{B,6}
        = \delta_{1} \wedge 
            \frac{m_1^{-4\vep} Y_\textnormal{r}(c)}{(x_1-c)}
            \frac{\ed x_1}{Y_\textnormal{r}}
            \begin{scriptsize}
                \begin{pmatrix}
                    0\\0\\1
                \end{pmatrix}
            \end{scriptsize}, 
        \quad \widecheck{\boldsymbol{\varphi}}_{B,7}
        = \delta_{1} \wedge 
            \frac{m_1^{-4\vep} \pi c_4}{\psi_1}
            \frac{\ed x_1}{Y_\textnormal{r}} 
            \begin{scriptsize}
                \begin{pmatrix}
                    0\\0\\1
                \end{pmatrix}
            \end{scriptsize}.
    \end{cases}
\end{equation}
Here, $Y_\textnormal{r}$ is the rescaled elliptic curve $Y_\textnormal{r}^2=Y^2/\mathcal{N}$, $\widecheck{\nabla}_{0} = \bs{\partial}_{X_0} + \wc{\bs{\omega}}_{B}\vert_{\ed X_{a=2,3} = 0}$ and   $W_0$ is the Wronksian in the $X_0$ variable
\begin{equation}
\begin{split}
     W_{0}&=
     \psi_1\partial_0 \psi_2-\psi_2\partial_0\psi_1
    \\&=-i \pi  
    \left(
        \frac{1}{(-+-)}
        +\frac{1}{(--+)}
        -\frac{1}{(+--)}
        -\frac{1}{(+++)}
        +\frac{1}{X_0}
    \right),
\end{split}
\end{equation}
where $(\pm\pm\pm) = \pm X_0 \pm X_2 \pm X_3 + 1$. The dimensionful kinematic normalization is, once again, such that the products $\widecheck{u}~\widecheck{\varphi}_{4\le a\le 7}$ are dimensionless. 
Note that we have put a factor of $1/\vep$ in the normalization of $\wc{\bs{\vphi}}_{B,4}$. 
This factor was chosen to cancel the \say{weight} of the derivative.
While the concepts of elliptic weights and purity are not fully understood yet, we would like our starting basis to be as close as possible to having uniform weights.

It is easy to check that $\widecheck{\boldsymbol{\varphi}}_{B,7}$ and $\widecheck{\boldsymbol{\varphi}}_{B,4}$ have the desired periods. Indeed, at leading order in $\vep$ we have 
\begin{equation}
    \oint_A \widecheck{\boldsymbol{\varphi}}_{B,7}\sim1 \quad \textnormal{and} \quad \oint_B \widecheck{\boldsymbol{\varphi}}_{B,7}\sim \tau,
\end{equation}
Then, it is easy to verify that the $A$-period of $\wc{\bs{\vphi}}_{B,4}$ vanishes at leading order
\begin{equation}
\label{eq:derAper}
    \oint_A \widecheck{\boldsymbol{\varphi}}_{B,4}=\frac{\psi_1^2}{\pi~ \varepsilon~W_{0}}\partial_{0} \oint_A \widecheck{\boldsymbol{\varphi}}_{B,7}=0,
\end{equation}
and, similarly, that the $B$-period of $\wc{\bs{\vphi}}_{B,4}$ is constant at leading order
\begin{equation}
\begin{split}
    \oint_B \widecheck{\boldsymbol{\varphi}}_{B,4}&=\frac{\psi_1^2}{\pi~ \varepsilon~W_{0}}\partial_{0} \oint_B \widecheck{\boldsymbol{\varphi}}_{B,7}\sim\frac{\psi_1^2}{\varepsilon~W_{0}}\partial_{0}\tau=\varepsilon^{-1}.
\end{split}
\end{equation}

Now, using dual IBP identities \cite{Caron-Huot:p1}, we compute the kinematic connection $\widecheck{\boldsymbol{\Omega}}$ associated to the seed basis \eqref{fibBasis} and \eqref{mcBasis2New}  
\begin{align}
    \wc{\bs{\nabla}} \wc{\bs{\vphi}}_{B,a} 
    = \wc{\bs{\vphi}}_{B,b} \wedge \wc{\Omega}_{b,a}. 
\end{align}
In particular, this basis has a kinematic connection in linear form (e.g., see \cite{Ekta:2019dwc})
\begin{equation}
\label{eq:linearForm}
    \widecheck{\boldsymbol{\Omega}}
    =\widecheck{\boldsymbol{\Omega}}^{(0)}+\varepsilon~ {\widecheck{\boldsymbol{\Omega}}}^{(1)},
\end{equation}
where the leading term is lower triangular
\begin{equation}
\label{eq:linearFormO0}
   \widecheck{\boldsymbol{\Omega}}^{(0)}
   = \begin{small}
   \left(\begin{array}{ccccccc}
        0& 0 & 0 & 0 & 0 & 0 & 0\\
        0 & 0& 0 & 0& 0 & 0 & 0\\
        0 & 0 & 0& 0 & 0 & 0 & 0\\
        0 & 0 & 0 & 0 & 0 & 0 & 0\\
        0 & 0 & 0 & \bullet & 0 & 0 & 0\\
        0 & 0 & 0 & \bullet & 0 & 0 & 0\\
        0& 0 & 0 & \star & \bullet & \bullet & 0
    \end{array}\right).
    \end{small}
\end{equation}
Moreover, $\widecheck{\boldsymbol{\Omega}}$ is independent of $a$ and $b$ under an arbitrary modular transformations. This is a smoking gun the pullback of the kinematic connection on the torus is spanned by \emph{only} modular and Kronnecker forms, as shown later in section \ref{sec:pullback}.

\subsection{The $\vep$-form basis\label{SEC:epsForm}}

Now that we have a starting seed basis and the associated kinematic connection is linear form, we would like to make a gauge transformation that removes the $\mathcal{O}(\vep^0)$ term in \eqref{eq:linearForm} (i.e., \eqref{eq:linearFormO0}).

There are two kinds of components in $\wc{\bs{\Omega}}^{(0)}$: those that are closed in the $\vep\to0$ limit and those that are not. 
Explicitly, 
\begin{align}
 \ed \widecheck{\Omega}^{(0)}_{54}   
 = \ed \widecheck{\Omega}^{(0)}_{64}
 = \ed \widecheck{\Omega}^{(0)}_{75}
 = \ed \widecheck{\Omega}^{(0)}_{76}
 = 0,
\end{align}
while 
\begin{align}
    \ed \widecheck{\Omega}^{(0)}_{74} \neq 0 .
\end{align}
Unfortunately, the non-vanishing of  $\ed \widecheck{\Omega}^{(0)}_{74}$
prevents us from integrating out the $\mathcal{O}(\varepsilon^0)$-piece in a way that is independent of the choice of integration contour.\footnote{The integral of closed one-forms is invariant under path-homotopy \cite[Thm. 16.26]{lee2013smooth}.} 
In order to remove the $\mathcal{O}(\vep^0)$ term in $\widecheck{\bs{\Omega}}^{(0)}$ and obtain a connection proportional to $\vep$, we make two gauge transformations. 
The first ensures the $\mathcal{O}(\vep^0)$ term of the new connection is closed in the $\vep\to0$ limit. 
The second removes the $\mathcal{O}(\vep^0)$ term.

\subsubsection*{The first gauge transformation}

The first gauge transform is simple and only has two non-trivial components
\begin{equation}
\label{eq:Ugauge}
    \boldsymbol{\mathcal{U}}= \textbf{Id} +
    \begin{small}\begin{pmatrix}
        0&0&0&0&0&0&0\\
        0&0&0&0&0&0&0\\
        0&0&0&0&0&0&0\\
        0&0&0&0&0&0&0\\
        0&0&0&0&0&0&0\\
        0&0&0&0&0&0&0\\
        0&0&0&0&u_{75}&u_{76}&0\\
    \end{pmatrix}\end{small}.
\end{equation}
We denote the corresponding kinematic connection by 
\begin{equation}
\tilde{\boldsymbol{\Omega}}=\boldsymbol{\mathcal{U}}^{-1}\cdot\widecheck{\boldsymbol{\Omega}}\cdot\boldsymbol{\mathcal{U}}+\boldsymbol{\mathcal{U}}^{-1}\cdot\textnormal{d}\boldsymbol{\mathcal{U}}.
\end{equation}
Then, assuming that $\boldsymbol{\mathcal{U}}$ is independent of $\varepsilon$, both 
$\widecheck{\bs{\Omega}}^{(0)}$ 
and 
$\tilde{\bs{\Omega}}^{(0)}$
satisfy the usual integrability conditions
\begin{equation}
\label{intCondU}
    \ed\widecheck{\bs{\Omega}}^{(0)}
    + \widecheck{\bs{\Omega}}^{(0)}
         ~\wedgedot~ \widecheck{\bs{\Omega}}^{(0)}
    = 0 = 
    \ed\tilde{\bs{\Omega}}^{(0)}
    + \tilde{\bs{\Omega}}^{(0)}
        ~\wedgedot~ \tilde{\bs{\Omega}}^{(0)}
    .
\end{equation}
If we can find $u_\bullet$'s that simultaneously solve
$\ed\tilde{\bs{\Omega}}^{(0)} = 0$ 
and 
$\tilde{\bs{\Omega}}^{(0)}~\wedgedot~\tilde{\bs{\Omega}}^{(0)}=0$, 
the integrability conditions will be satisfied and the $\mathcal{O}(\vep^0)$ term of $\tilde{\Omega}$ will be closed in the $\vep\to0$ limit: 
$\ed\tilde{\bs{\Omega}}^{(0)}=0$. The vanishing of 
$\tilde{\bs{\Omega}}^{(0)}~\wedgedot~\tilde{\bs{\Omega}}^{(0)}$
yields the following constraints on the $u_\bullet$'s
\begin{equation} \label{ucon0}
    (\ed u_{75} + \widecheck{\Omega}^{(0)}_{75})
        \wedge \widecheck{\Omega}^{(0)}_{54}
    +
    (\ed u_{76} +\widecheck{\Omega}^{(0)}_{76})
        \wedge \widecheck{\Omega}^{(0)}_{64}
    = 0.
\end{equation}
Next, by using the modular transformation properties of $\widecheck{\Omega}^{(0)}_{7j}$, we show there exists some $u_{7j}$ and $\lambda_{7j}$ such that 
\begin{align} \label{ucon1}
    \ed u_{7j} + \widecheck{\Omega}^{(0)}_{7j} - \lambda_{7j} \widecheck{\Omega}^{(0)}_{j4}
        = 0 
    \qquad (j=5,6).
\end{align}
inferring constraint \eqref{ucon0} is satisfied.

To do this, we note that $\widecheck{\Omega}^{(0)}_{75}$, $\widecheck{\Omega}^{(0)}_{76}$, $\widecheck{\Omega}^{(0)}_{54}$ and $\widecheck{\Omega}^{(0)}_{64}$ transform like the derivative of a weight one modular form. For example, $\widecheck{\Omega}^{(0)}_{7j}$ takes the form
\begin{equation}
\label{structureOmegaU}
    \widecheck{\Omega}^{(0)}_{7j}
    = \alpha_{7j}~\psi_1 
        + \beta_{7j}\ \partial_{X_0}\psi_1,
\end{equation}
where $\alpha_\bullet$ and $\beta_\bullet$ are rational differential forms in the kinematic variables (i.e., modular invariant).
Then, making a modular transformation, one finds that
\begin{align} \label{eq:OmegaCheckModTrans}
    \widecheck{\Omega}^{(0)}_{7j}
    \to 
    (c\tau+d)\widecheck{\Omega}^{(0)}_{7j}
    +c\psi_1 (\partial_0\tau)\ \beta_{7j}.
\end{align}
If \eqref{ucon1} is to hold, $u_{75}$ and $u_{76}$ must be weight one modular forms
\begin{align} \label{eq:uModTrans}
    \ed u_{7j} \to
    (c\tau+d)\ \ed u_{7j} + c u_{7j}\ \ed \tau.
\end{align}
Putting everything together, one finds under a modular transformation of \eqref{ucon1}
\begin{align}
    &0 
    =  (c\tau +d) 
        \underset{0}{\underbrace{
        [
            \ed u_{7j} 
            + \widecheck{\Omega}^{(0)}_{7j} - \lambda_{7j} \widecheck{\Omega}^{(0)}_{j4}
        ] 
        }}
    + c [u_{7j}\ \ed \tau + \psi_1 (\partial_0 \tau)\ (\beta_{7j} - \lambda_{7j} \beta_{j4})], \notag
    \\
    & \label{master1DEu7j}\implies u_{7j}\ \ed \tau + \psi_1 (\partial_0 \tau)\ (\beta_{7j} - \lambda_{7j} \beta_{j4}) = 0.
\end{align}
Expanding \eqref{master1DEu7j} in terms of the $\ed X_a$, yields the following system of equations
\begin{align}
    \label{eq:u7j X3}
    u_{7j}  &= - \psi_1 (\beta_{7j,X_0}-\lambda_{7j}\beta_{j4,X_0}),
    \\ 
    \label{eq:u7j Xa}
    u_{7j} &= - \psi_1 \frac{W_0}{W_a} 
        ( \beta_{7j,X_a} - \lambda_{7j}\beta_{j4,X_a}), 
\end{align}
where we have used $\frac{\partial \tau}{\partial X_a}=\psi_1^{-2}W_a$.
Since $\beta_{j4,X_0}=0$,\footnote{
{Since the expressions involved are algebraically complicated and contain elliptic functions, we checked this equality only numerically. The main obstruction for analytical checks is the existence of many non-trivial relations between elliptic functions that do not simplify automatically in \textsc{Mathematica}.}} equation \eqref{eq:u7j X3} sets 
\begin{align} \label{eq:u7j}
    u_{7j} = - \psi_1 \beta_{7j,X_0}. 
\end{align}
Feeding \eqref{eq:u7j} into \eqref{eq:u7j Xa}, fixes $\lambda_{7j} = 8\pi$. 
Explicit expressions for the $u$'s are given in appendix \ref{SEC:details}.

\subsubsection*{The second gauge transformation}

Our remaining task is to gauge away the non-trivial entries of $\tilde{\boldsymbol{\Omega}}^{(0)}$. 
To this end, we introduce the gauge transformation 
\begin{equation} \label{eq:Vgauge}
    \boldsymbol{\mathcal{V}}= \textbf{Id} + 
    \begin{small}\begin{pmatrix}
        0&0&0&0&0&0&0\\
        0&0&0&0&0&0&0\\
        0&0&0&0&0&0&0\\
        0&0&0&0&0&0&0\\
        0&0&0&v_{54}&0&0&0\\
        0&0&0&v_{64}&0&0&0\\
        0&0&0&v_{74}&v_{75}&v_{76}&0\\
    \end{pmatrix}\end{small}.
\end{equation}
where the $v_\bullet$ are independent of $\vep$.
Next, we denote the corresponding gauge transformed connection by 
\begin{align}
\label{gammaCheck}
    \widecheck{\bs{\Gamma}}
    = \boldsymbol{\mathcal{V}}^{-1} \cdot \tilde{\bs{\Omega}} \cdot \boldsymbol{\mathcal{V}} + \boldsymbol{\mathcal{V}}^{-1} \cdot \ed \boldsymbol{\mathcal{V}}.
\end{align}
Then, requiring that $\widecheck{\bs{\Gamma}}^{(0)} = 0$, yields the following equations for the components of $\boldsymbol{\mathcal{V}}$
\begin{align}
\label{eq:vCV1}
   \textnormal{d}v_{54} + \tilde{\Omega}^{(0)}_{54}=0, 
   & \quad 
   \ed v_{64} + \tilde{\Omega}^{(0)}_{64}=0, \notag\\ 
   \textnormal{d}v_{75} + \tilde{\Omega}^{(0)}_{75}=0, 
   & \quad \textnormal{d}v_{76} + \tilde{\Omega}^{(0)}_{76}=0,
\end{align}
and
\begin{equation}
\label{eq:vCV2}
        \textnormal{d}v_{74}
        {+} \tilde{\Omega}^{(0)}_{74}+
        {-} v_{75} \ed v_{54}
        {-} v_{76} \ed v_{64}
        {+} v_{54} \tilde{\Omega}^{(0)}_{75} 
        {+} v_{64} \tilde{\Omega}^{(0)}_{76}
        {-} v_{75} \tilde{\Omega}^{(0)}_{54}
        {-} v_{76} \tilde{\Omega}^{(0)}_{64}
   = 0.
\end{equation}
Note that if the system \eqref{eq:vCV1} is satisfied, \eqref{eq:vCV2} reduces to 
\begin{equation}
\label{eq:nEq1}
\textnormal{d}v_{74}+\tilde{\Omega}^{(0)}_{74}+v_{54}\tilde{\Omega}^{(0)}_{75}+v_{64}\tilde{\Omega}^{(0)}_{76}=0.
\end{equation}
{It is convenient to set $v_{74}=v_{74}^{\textnormal{(I)}}+v_{74}^{\textnormal{(II)}}$. Here, the job of $v_{74}^{\textnormal{(I)}}$ is to kill  $\tilde{\boldsymbol{\Omega}}_{74}^{(0)}$. 
This is possible since we know that $\tilde{\boldsymbol{\Omega}}_{74}^{(0)}$ is $\ed$-closed.
Then, the role of $v_{74}^{\textnormal{(II)}}$ is to neutralize the remaining terms in $\widecheck{\bs{\Gamma}}_{74}^{(0)}$. That is, we solve \eqref{eq:nEq1} by solving } 
\begin{equation}
\label{eq:vCV3}
\begin{cases}
  \textnormal{d}v_{74}^{\textnormal{(I)}}+\tilde{\Omega}^{(0)}_{74}=0,\\
  \textnormal{d}v_{74}^{\textnormal{(II)}}+v_{54}\tilde{\Omega}^{(0)}_{75}+v_{64}\tilde{\Omega}^{(0)}_{76}=0.
\end{cases}
\end{equation}
{To proceed, we first solve \eqref{eq:vCV1} since it appears as source terms in \eqref{eq:vCV3}. Once $v_{54}$ and $v_{64}$ are known, we solve \eqref{eq:vCV3}. }

We start by rewriting the matrix elements in such a way that their modular properties are manifest
\begin{equation}
\label{allBut74}
    \tilde{\Omega}^{(0)}_{ij}
    = \psi_1\ \sigma_{ij}
        + (\partial_0\psi_1)\ \rho_{ij} \qquad \textnormal{(for $(i,j)\neq(7,4)$)},
\end{equation}
and 
\begin{equation}
\label{just74}
    \tilde{\Omega}^{(0)}_{74} 
    = \psi_1^2\ \sigma_{74} 
    + \psi_1 (\partial_0\psi_1)\ \rho_{74}.
\end{equation}
Here, the $\sigma_\bullet$'s and $\rho_\bullet$'s  are modular invariant differential one-forms in the kinematic variables analogous to the $\alpha$'s and $\beta$'s. 
By making a modular transformation on equation \eqref{eq:vCV1}, we find the following constraint
\begin{align}
\label{eq:vij tot}
    v_{ij}\ \ed \tau 
    + (\partial_0 \psi_1)\ \rho_{ij}  = 0
    \quad \text{for} \quad 
    (i,j) \in \{(5,4),(6,4),(7,5),(7,6)\},
\end{align}
for $(i,j) \in \{(5,4),(6,4),(7,5),(7,6)\}$.
However, equating the components of the above (as done for $\boldsymbol{\mathcal{U}}$) does not help us find a solution since $\rho_{ij,X_0}=0$.\footnote{{Again, since the expressions involved are algebraically complicated and depend on non-trivial combinations of elliptic functions, we checked that equality only numerically.}} Instead, it can be shown that
\begin{align} \label{solVsNot74}
    v_{ij} = -\frac{W_0}{\psi_1} \frac{\rho_{ij}}{\ed\tau}
    \equiv -\frac{W_0}{\psi_1} \sum_{a=0,2,3} \rho_{ij,X_a} \frac{\partial X_a}{\partial \tau},
\end{align}
satisfies equation \eqref{eq:vCV1}.\footnote{This was cross-checked numerically.} 

It is possible to see from the inverse function theorem that the quantities $\frac{\partial X_a}{\partial \tau}$ evaluate to non-trivial combinations of incomplete elliptic integrals. The modular properties of $\frac{\partial X_a}{\partial \tau}$ are given in appendix \ref{SEC:DiffFormOnTorus}. In particular, it is clear from the  transformation rule in \eqref{partialDerMod} that \eqref{solVsNot74} transforms like a quasi-modular form.
This already justifies the appearance of $\omega_{\textnormal{Kr}}^{(k\ge 3)}$'s in the final version of the kinematic connection \eqref{epsCon}. 

Next, we fix $v_{74}^{\textnormal{(I)}}$. From \eqref{just74}, it is clear that $ \tilde{\Omega}^{(0)}_{74}$ transforms like the derivative of a weight two modular form. 
The function $v_{74}^{(\textnormal{I})}$ must therefore be a modular form of weight two 
\begin{equation}
\textnormal{d}v_{74}^{\textnormal{(I)}}\to (c\tau+d)^2\textnormal{d}v_{74}^{\textnormal{(I)}}+2c(c\tau+d)v_{74}^{(\textnormal{I})}\textnormal{d}\tau,    
\end{equation}
if
\begin{equation}
\label{v741Cons}
    \textnormal{d}v_{74}^{(\textnormal{I})}+\tilde{\Omega}^{(0)}_{74}=0,
\end{equation}
is to hold.
Making a modular transformation on \eqref{v741Cons} one finds that
\begin{equation}
    \begin{split}
        2v_{74}^{(\textnormal{I})} \ed\tau 
        + \psi_1^2 (\partial_0\tau) \rho_{74} = 0.
    \end{split}
\end{equation}
Since $\rho_{74,X_3}\neq0$, we can equate the $\ed X_a$ components to find 
\begin{align}
    v_{74}^{(\textnormal{I})}
    = -\frac12 \psi_1^2 \rho_{74,X_0}
    = -\frac12 \psi_1^2 \frac{W_0}{W_a} \rho_{74,X_a}.
\end{align}

Our final task is to fix $v_{74}^{(\textnormal{II})}$ such that
\begin{equation}
\label{v742Cons}
    \textnormal{d}v_{74}^{(\textnormal{II})}+v_{54}\tilde{\Omega}^{(0)}_{75}+v_{64}\tilde{\Omega}^{(0)}_{76}=0.
\end{equation}
From \eqref{allBut74}, \eqref{solVsNot74} and \eqref{v742Cons}, we see that $v_{74}^{(\textnormal{II})}$ needs to transform as a modular form of weight two. 
Using \eqref{eq:vCV2}, one finds that the transformation rule of \eqref{v742Cons} 
\begin{equation}
    0=(c\tau+d)^2\left[\textnormal{d}v_{74}^{(\textnormal{II})}-v_{54}\textnormal{d}v_{75}-v_{64}\textnormal{d}v_{76}\right]+c(c\tau+d)\textnormal{d}\tau\left[2v_{74}^{(\textnormal{II})}-v_{54}v_{75}-v_{64}v_{76}\right],
\end{equation}
uniquely fixes $v_{74}^{(\textnormal{II})}$. 
Since both terms above need to vanish independently, we find that 
\begin{equation}
    v_{74}^{(\textnormal{II})}=\frac{1}{2}\left[v_{54}v_{75}+v_{64}v_{76}\right]=-v_{64}^2-\frac{1}{3}v_{54}^2.
\end{equation}
Once again, the explicit expressions for the $v$'s are listed in appendix \ref{SEC:details}.

\subsubsection*{The $\vep$-form basis}

The base basis that yields an $\vep$-form differential equation is given by applying the gauge transformations 
\eqref{eq:Ugauge} and \eqref{eq:Vgauge} to \eqref{mcBasis2New}
\begin{align}
    \widecheck{\boldsymbol{\nabla}}_B \widecheck{\bs{\vartheta}}_B
    = \widecheck{\bs{\vartheta}}_B\ \wedgedot \
    \widecheck{\bs{\Gamma}}
    \qquad \text{for} \qquad
    \widecheck{\bs{\vartheta}}_B 
    = \widecheck{\boldsymbol{\varphi}}_{B} 
    \cdot \boldsymbol{\mathcal{U}} \cdot \boldsymbol{\mathcal{V}}. 
\end{align}
where only $\widecheck{\bs{\Gamma}}^{(1)} \neq 0$.
More explicitly, the $\vep$-form basis reads
\begin{align*} 
   \widecheck{ \boldsymbol{\vartheta}}_{B,1}=\widecheck{\boldsymbol{\varphi}}_{B,1}, & \quad\widecheck{ \boldsymbol{\vartheta}}_{B,2}=\widecheck{\boldsymbol{\varphi}}_{B,2}, \quad\widecheck{ \boldsymbol{\vartheta}}_{B,3}=\widecheck{\boldsymbol{\varphi}}_{B,3},
\end{align*}
\vspace{-0.7cm}
\begin{align}
\label{2loopBaseBasis}
   \widecheck{ \boldsymbol{\vartheta}}_{B,4}=\widecheck{\boldsymbol{\varphi}}_{B,4}+v_{54}~\widecheck{\boldsymbol{\varphi}}_{B,5}+v_{64}~\widecheck{\boldsymbol{\varphi}}_{B,6}+\left(v_{74}^{\textnormal{(I)}}+v_{74}^{\textnormal{(II)}}+u_{75}v_{54}+u_{76}v_{64}\right)~\widecheck{\boldsymbol{\varphi}}_{B,7},
\end{align}
\vspace{-0.7cm}
\begin{align*}
   \widecheck{ \boldsymbol{\vartheta}}_{B,5}=\widecheck{\boldsymbol{\varphi}}_{B,5}+(u_{75}+v_{75})\widecheck{\boldsymbol{\varphi}}_{B,7}, & \quad\widecheck{ \boldsymbol{\vartheta}}_{B,6}=\widecheck{\boldsymbol{\varphi}}_{B,6}+(u_{76}+v_{76})\widecheck{\boldsymbol{\varphi}}_{B,7},
\end{align*}
\vspace{-0.7cm}
\begin{align*}
   \widecheck{ \boldsymbol{\vartheta}}_{B,7}=\widecheck{\boldsymbol{\varphi}}_{B,7},
\end{align*}
where the explicit expressions for the $v$'s and $u$'s can be found in appendix \ref{SEC:details}. We recall that, by construction, the dual sunrise basis
\begin{align}
\label{eq:2loopSunBasis}
    \widecheck{\vartheta}_a 
    = \widecheck{\bs{\vphi}}_F 
    ~\wedgedot~ \widecheck{\bs{\vartheta}}_{B,a},
\end{align}
also satisfies \eqref{gammaCheck} ({we refer to \eqref{epsCon} for the explicit differential equation}).

Although \eqref{gammaCheck} is not yet the evaluated sunrise integral, a great deal of information about the sunrise family of integrals is contained in it. In particular, its pole structure determines the singularities of the Feynman integrals and thus their analytic properties. A singularity analysis reveals the poles of \eqref{gammaCheck} are all simple and in one-to-one correspondence with the
\begin{equation}
\label{landau1}
  \textbf{\emph{First type Landau singularities}:}\quad \{p^2+(\pm m_1^2\pm....\pm m_{L+1})^2=0\}|_{L=2},
\end{equation}
given in \cite{Mizera:2021icv}, up to additional
\begin{equation}
\label{landau2}
    \textbf{\emph{IR and second type Landau singularities}:}\quad \{m_{1\le i\le 3}^2=0,p^2=0\}.
\end{equation}
Singularities in \eqref{landau2} should be considered spurious, since they do not correspond to (anomalous) thresholds \cite{Eden:1966dnq}. However, they can be used to derive an initial condition for the differential equation (see e.g., \cite{Bogner:2019lfa}).

\subsection{The period matrix and $\varepsilon$-form \label{SEC:perMAT}}

Based on recent observations {\cite{Primo:2017ipr,Frellesvig:2021hkr,Frellesvig:2023iwr}}, one may expect the period matrix of our $\varepsilon$-form basis \eqref{2loopBaseBasis} to be constant -- i.e., the basis has constant leading singularity on all of the spanning homology cycles (see figure \ref{fig:homologyCycles}). 
While this would indeed be correct for polylogarithmic integral, it is slightly more complicated for elliptic integrals. 

We have to first be specific about what we mean by the period matrix.
The period map is simply the pairing between a cycle and cocycle -- i.e., the integral of a cocycle over a cycle.
The period matrix tabulates this pairing for specific basis of cycles and cocycles.
There are two kinds of periods we need to discuss: \emph{twisted} periods and \emph{non-twisted} periods.
A twisted period is the pairing between a twisted cycle and a twisted cocycle.
Dimensionally regulated Feynman integrals and dual Feynman integrals are examples of twisted periods. 
On the other hand, a non-twisted period or period is the pairing between a $\mathds{Z}$-valued cycle and cocycle. 
In integer dimensions, finite Feynman integrals and dual Feynman integrals are periods. 
More common examples of periods include transcendental numbers such as $\pi$ or 
Riemann-zeta's $\zeta_n$.
Intuitively, the $\vep\to0$ limit of a twisted period is a period.

To express the twisted periods in terms of the base cohomology, we need to know how the loop-by-loop fibration breaks up the total space cycles and cocycles
\begin{align}
    \wc\vth_{a}
    = \wc\vphi_{F,b} \wedge \wc\vth_{B,ba},
    \quad
    \wc\gamma_{\alpha} 
    = \wc\gamma_{B,\alpha\beta} \wedge \wc\gamma_{F,\beta}.
\end{align}
Using the above decomposition, the dual Feynman integrals or dual twisted periods become
\begin{align} \label{eq:tpmat def}
    \wc{\mathcal{P}}_{\alpha a}
    = \oint_{\wc{\gamma}_\alpha} \wc{u}\ \wc{\vphi}_a 
    = \oint_{\wc{\gamma}_{B,\alpha\beta}} \left( \oint_{\wc{\gamma}_{F,\beta}} \wc{u}\ \wc{\vphi}_{F,b} \right) \wc{\vphi}_{B,ba} 
    = \oint_{\wc{\gamma}_{B,\alpha\beta}} \wc{u}_{B,\beta b}\ \wc{\vphi}_{B,ba} 
\end{align}
where the matrix-valued twist on the base is simply the twisted period matrix of the fibre 
\begin{align}
\label{eq:fibTwist1}
    \wc{u}_{\alpha a} 
    = \oint_{\wc\gamma_{F,\alpha}} u\ \wc\vphi_{F,a}.
\end{align}
One can also define $\wc{u}_{\alpha a}$ by\footnote{Both the path ordered exponential solution to \eqref{eq:fibTwist1} and the integral over contours in \eqref{eq:tpmat def} solve the same differential equation. Therefore, these definitions can only differ by a constant boundary term.}
\begin{align}
\label{eq:fibTwist2}
    \wc\omega_{B,ab} 
    = \left(\wc{u}^{-1}_{B}\right)_{a \alpha}\ \ed \wc{u}_{B,\alpha b}.
\end{align}
Note that $\wc{u}_{\alpha a} \sim \textbf{Id} + \mathcal{O}(\vep)$ since $\wc{\omega}_{B,ab}$ is directly proportional to $\vep$.

We define the \emph{period matrix} following \cite{Frellesvig:2021hkr} by taking the $\vep\to0$ limit of the twist in \eqref{eq:tpmat def} and set
\begin{align} \label{eq:pmat def}
    \wc{P}_{\alpha a}
    = \oint_{\wc{\gamma}_{B,\alpha b}} \wc{\vphi}_{B,ba}. 
\end{align}
In particular, we are interested in the 4-by-4 block of $\wc{\bs{P}}$ corresponding the maximal-cut. 
That is, the pairing between $\wc{\bs{\vphi}}_{B,a\geq4}$ and 
\begin{align}
    \wc{\bs{\gamma}}_{\alpha\geq4} = 
    \begin{pmatrix}
        0 & 0 & 1   
    \end{pmatrix}
    \wc{\gamma}_a,
\end{align}
where $\wc{\gamma}_4 = \wc{\gamma}_A$ is the $A$-cycle, 
$\wc{\gamma}_5 = \wc{\gamma}_B$ is the $B$-cycle,
$\wc{\gamma}_6 = \wc{\gamma}_{\circlearrowleft_c}$ (c.f., \eqref{eq:defC}) is the residue contour centered at $c$,
and $\wc{\gamma}_7 = \wc{\gamma}_{\circlearrowright_\infty}$ is the residue contour centered at $\infty$.

To motivate further our choice of seed basis \eqref{mcBasis2New} and why our approach differs form \cite{Frellesvig:2021hkr}, 
consider an alternative seed basis $\wc{\bs{\vphi}}_{B,a}^\text{alt}$ where 
\begin{align}
    \wc{\bs{\vphi}}_{B,4}^\text{alt} 
        &= -m_1^{-4\vep}\frac{\psi_1}{4c_4}
        \left(x_1^2-\frac{s_1}{2}x_1+\frac{s_2}{6}\right)\frac{\ed x_1}{Y}
        \begin{scriptsize}\begin{pmatrix}0\\0\\1\end{pmatrix}\end{scriptsize}
        - \phi_1^\prime \wc{\bs{\vphi}}_{B,7},
    \\
    \wc{\bs{\vphi}}_{B,a}^\text{alt} 
        &= \wc{\bs{\vphi}}_{B,a}
        \quad \text{for} \quad a=5,6,7.
\end{align}
By design, the form $\wc{\bs{\vphi}}_{B,4}^\text{alt}$ vanishes on the $A$-cycle and is constant on the $B$-cycle. 
This form is closely related to the quasi-period form \eqref{phiHat} whose 
$A$-cycle integral is the normalized quasi-period \eqref{eq:normalized quasi-period 1}.
Consequently, the alternative basis has a simple period matrix
\begin{align} \label{eq:pmat}
    \wc{P}_{ab}^\text{alt}
    &= \begin{pmatrix}
        1 
        & \tau 
        & 0 
        & 0 
    \\
        0 
        & -i \pi 
        & 0 
        & 0
    \\
        2\pi i {+} \tilde{g}^{(1)}(Z_c,Z_\infty;\tau)
        & 4\pi i Z_c + \tau \tilde{g}^{(1)}(Z_c,Z_\infty;\tau)
        & 2\pi i
        & 0
    \\
        -2\pi i -2  g^{(1)}(Z_\infty;\tau)
        & -4\pi i Z_\infty - 2 \tau g^{(1)}(Z_\infty;\tau)
        & 0 
        & 2\pi i
    \end{pmatrix},
\end{align}
where $\tilde{g}^{(1)}(\xi_1,\xi_2;\tau) = g^{(1)}(\xi_1-\xi_2;\tau) + g^{(1)}(\xi_1+\xi_2;\tau)$ and $a,b \geq 4$.

The differential equations for the alternative seed basis are also in linear form 
\begin{align}
    \wc{\bs\nabla}_B\ \wc{\bs{\vphi}}_{B,a} 
    = \wc{\bs{\vphi}}_{B,b} \wedge 
        \left( \wc{\Omega}_{B,ba}^{\text{alt }(0)} 
            + \vep\ \wc{\Omega}_{B,ba}^{\text{alt }(1)}
        \right).
\end{align}
Following \cite{Frellesvig:2021hkr}, the $\mathcal{O}(\vep^0)$ term above can be removed by gauge transforming by the inverse period matrix
\begin{align} \label{eq:alt ep form deqs}
    \wc{\bs\nabla}_B\ 
    \wc{\bs{\vth}}^\text{alt}_{B,a}
    = \wc{\bs{\vth}}^\text{alt}_{B,b}
    \wedge
    \left[
        \vep\
        \wc{\bs{P}}^\text{alt}
        \cdot
        \wc{\bs{\Omega}}_{B}^{\text{alt }(1)}
        \cdot 
        \left(\wc{\bs{P}}^\text{alt}\right)^{-1}
    \right]_{ba},
\end{align}
where
\begin{align} \label{eq:alt ep form basis}
    \wc{\bs{\vth}}^\text{alt}_{B,a} 
    =\wc{\bs{\vphi}}_{B,b}\cdot  \left(\wc{P}^\text{alt}\right)^{-1}_{ba}.
\end{align}
In general, the $\mathcal{O}(\vep^0)$ term can be removed in this way to produce an $\vep$-form connection since 
\begin{align}
    \ed \wc{\bs{P}}^\text{alt} 
    = \wc{\bs{P}}^\text{alt} 
    \cdot \wc{\bs\Omega}_{B}^{\text{alt }(0)} 
    \implies 
    \ed \left(\wc{\bs{P}}^\text{alt}\right)^{-1}
    = - \wc{\bs\Omega}_{B}^{\text{alt }(0)} 
    \cdot \left(\wc{\bs{P}}^\text{alt}\right)^{-1}.
\end{align}
The corresponding period matrix for the basis $\wc{\bs{\vth}}^\text{alt}_a$ is identity and thus, the basis $\wc{\bs{\vth}}^\text{alt}_a$ has constant \say{leading singularity}.\footnote{{Here, the term \say{leading singularity} is understood as follows. The \say{leading singularity} associated to a basis element is the kinematic function such that its quotient with the basis element is constant in the $\varepsilon\to0$ limit. We note that the definition of what a leading singularity is for integrals outside the MPL function space is still unclear. See for example \cite{Primo:2017ipr,Broedel:2018qkq,Frellesvig:2021hkr,Bourjaily:2020hjv,Bourjaily:2021vyj,Frellesvig:2021vdl,Bourjaily:2022tep,Frellesvig:2023iwr}}} 

While ensuring that $\wc{\bs\vth}_B^\text{alt}$ has constant leading singularity produced an $\vep$-form connection,\footnote{We checked numerically.} our ability to  integrate this connection is hampered by the explicit dependence in $\tau$ in the inverse period matrix.
The connection in \eqref{eq:alt ep form deqs} cannot be written only in terms of Kronnecker and modular forms since the factors of $\tau$ transform differently. 
In order to write the connection on the torus in terms of Kronnecker and modular forms, we have to ensure that under a modular transformation the connection only depends on the modular parameters $c$ and $d$. 
This means that while $\ed\tau$ is allowed, factors of $\tau$ are not. 
The choice $\wc{\bs{\vphi}}_{B,4}$ in the seed basis \eqref{mcBasis2New} ensures that.

The period matrix of \eqref{2loopBaseBasis} is also more subtle than \eqref{eq:pmat} due to the factor of $1/\vep$ in $\wc{\bs{\vartheta}}_{B,4}$. 
While \eqref{eq:alt ep form basis} has constant period matrix by construction, we checked numerically that \eqref{2loopBaseBasis} does not. Yet, they both satisfy an $\varepsilon$-form differential equation. This is a concrete example that although choosing a basis with constant leading singularity is \emph{sufficient} to get an $\varepsilon$-form differential equation, it is not \emph{necessary}. This statement raises interesting tensions with the prescriptive unitarity program, which {focuses on} bases with unit leading singularities for integrals beyond MPLs \cite{Primo:2017ipr,Frellesvig:2021hkr,Bourjaily:2020hjv,Bourjaily:2021vyj,Bourjaily:2022tep}. 

Another intriguing finding is that the determinant of the period matrix of both \eqref{eq:alt ep form basis} and \eqref{2loopBaseBasis} are constant.\footnote{This computation also includes subleading terms in $\varepsilon$ coming from the different normalizations in $\varepsilon$ of the forms.} This observation was recently made in \cite{hjalteTalk} for the non-planar (elliptic) two-loop triangle. 
It would be interesting to investigate more on the precise role constant determinant period matrices play in deriving $\varepsilon$-form differential equations.

\subsection{A ready-to-integrate differential equation on the torus\label{SEC:readyToIntegrateDE}}

We present the general strategy used to pullback $\widecheck{\boldsymbol{\Gamma}}$ on the torus in section \ref{sec:pullback}. 
Then, in section \ref{sec:relations}, we describe relations among the elements of $\wc{\bs{\Gamma}}$ reducing the amount of work needed to pullback $\wc{\bs{\Gamma}}$ componentwise.
Lastly, in section \ref{sec:final kin connection}, we present a compact ready-to-integrate formula for $\wc{\bs{\Gamma}}$. 
\subsubsection{Pullback to the torus: modular bootstrap \label{sec:pullback}}

Our general strategy is to leverage the knowledge of the modular properties of $\wc{\bs{\Gamma}}$ in order to construct an ansatz for it on the torus, built out of suitable building blocks.
Once we have a good ansatz, the coefficients are easily fixed numerically. 
As an illustrative example, we will detail the pullback procedure for one of the simplest components: $\wc{\Gamma}_{11}$.

Under the action of the congruence subgroup of $\textnormal{SL}(2,\mathbb{Z})$
\begin{equation}
    \Gamma(4)=\left\{\left.\begin{pmatrix}
        a &b\\c&d
    \end{pmatrix}\in\textnormal{SL}(2,\mathbb{Z})\right|\begin{pmatrix}
        a &b\\c&d
    \end{pmatrix}\equiv \textbf{Id} \quad \textnormal{mod}~4\right\},
\end{equation}
where the reduction modulo four is regarded entry-wise, all rational differential forms in the $X_i$ are left invariant. We recall that this is because the action does not permute the roots. Therefore, under $\Gamma(4)$, the modular properties of $\widecheck{\boldsymbol{\Gamma}}$ follows exclusively from the factors of $\psi_1$ and the $\partial_\tau X_i$.
Using the modular transformations of the periods and the $\partial_\tau X_\bullet$'s worked out in appendix \ref{SEC:DiffFormOnTorus}, we find the elements of $\wc{\bs{\Gamma}}$ to transform either like modular forms or quasi-modular forms.
Moreover, as mentioned earlier, the components of $\wc{\bs{\Gamma}}$ have also at most simple poles on the Landau surfaces \eqref{landau1} and \eqref{landau2}.

On the torus, we look for building blocks with similar properties. 
The simplest such objects are modular forms depending only on $\tau$: $\eta_n(\tau)$ (see appendix \ref{SEC:DiffFormOnTorus}).
However, these are obviously not sufficient since we also require $z_i$-dependence. 
Therefore, we also consider Kronnecker forms $\omega_n^\text{Kr}(z\vert\tau)$ (see appendix \ref{SEC:DiffFormOnTorus}) in our ansatz. 
Kronnecker forms have at most simple poles in $z$ at the lattice points and also transform like quasi-modular forms \cite{Bogner:2019lfa}. 
Thus, we expect that the components of $\wc{\bs{\Gamma}}$ can be written in terms of $\omega_{n}^\text{Kr}$ and $\eta_n$
\begin{equation}
    \label{eq:ansatz}
    \widecheck{\Gamma}_{ij}
    = \sum_{k=1}^3 c_{ij;k}\ 
    \omega^{\textnormal{Kr}}_{n_{ij}}(z_k-\lambda_{ij;k})
    + c_0 \eta_{n_{ij}}(\tau),
\end{equation}
where $c_{ij;k} \in \mathds{Q}$ and $z_3 = 1-z_1-z_2$ as defined in equation \eqref{eq:z3DEF}.
In particular, the modular properties of $\widecheck{\Gamma}_{ij}$ fix $n_{ij}$ such that $n_{ij}\in\{0,1,2,3,4\}$.


In terms of the kinematic variables $\{X_0,X_2,X_3\}$, the only allowed poles lie on the Landau surfaces described in \eqref{landau1} and \eqref{landau2}. 
Thus, the arguments of the Kronnecker forms must be correlated to these surfaces. 
By examining how the torus shape changes as we approach a Landau surface, the arguments of the Kronnecker forms can be guessed.  
When a Landau surface is approached, a subset of the moduli $\{z_i\}$ on the torus collide with lattice points fixing the set of allowed $\lambda_{ij;k}$.

\begin{figure}
    \centering
    \includegraphics[scale=0.29]{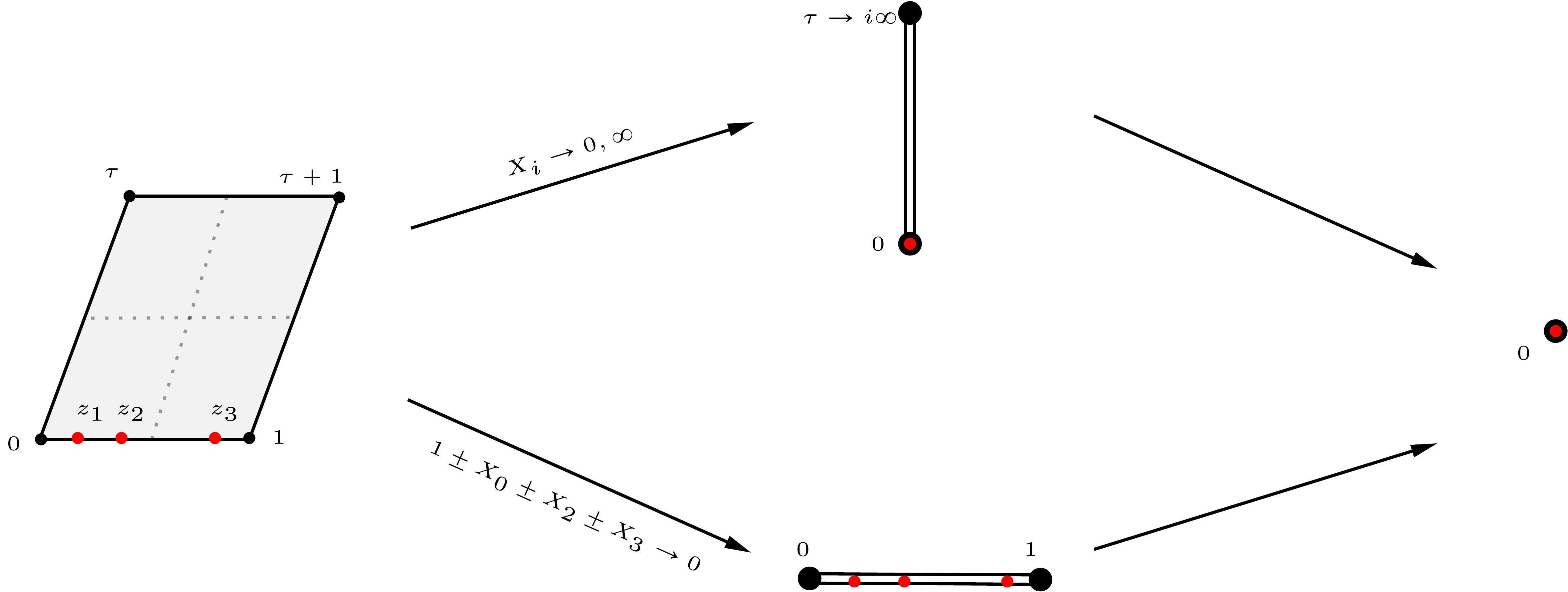}
    \caption{Degenerations of the toric geometry near the differential equation simple poles.}
    \label{funCell}
\end{figure}

To see how this works in detail, we need to understand how the torus shape is correlated with the Landau singularities \eqref{landau1} and \eqref{landau2}, which in our dimensionless coordinates read
\begin{align} \label{eq:dimless Landau}
	(\pm\pm\pm) = 1 \pm X_0 \pm X_2 \pm X_3 = 0
	\quad \text{and} \quad 
	X_i = 0,\infty
	.
\end{align}
When the roots of the elliptic curve collide, a subset of the Landau equations are satisfied. 
With our root order fixed ($r_{i}<r_{i+1}$), we have the following correspondence
\begin{align}
\label{rootsCollide}
    r_1 = r_2 &\implies  (+--), (+++) \to 0,
    \\
    r_2 = r_3 &\implies	X_0 \to \infty,
	\\
    r_3 = r_4 &\implies (-+-), (--+) \to 0, 
    \\
   	r_4 = r_1 &\implies X_2, X_3 \to 0.
 \end{align}
Then, the equations in \eqref{periodsEllCurve} yield
\begin{equation}
\label{tauSing}
    \tau=\frac{\psi_2}{\psi_1}\to\begin{cases}
      i\infty & \textnormal{as} \ X_2, X_3 \to 0,\ X_0 \to \infty,\\
      0 & \textnormal{as} \ (+--), (+++), (-+-), (--+) \to 0. 
    \end{cases}
\end{equation}
To examine the collision of non-adjacent roots one first needs to perform a modular transformation outside of the root preserving subgroup $\Gamma(4)$. 
Still, one finds that $\tau$ either approaches $0$ or $i\infty$. 
Thus, we only need to consider the two degenerations of the torus pictured in figure \ref{funCell}. 
When $\tau \to i\infty$, the geometry degenerates to one associated to polylogarithms. 
This makes sense since the sunrise diagram is known to be polylogarithmic when one internal edge becomes massless \cite{Adams:2013nia}. 
Geometrically, the $A$- or $B$-cycle shrinks to a point when $\tau \to 0$.

It is not hard to show (given our choice of root orderings) that the $z_i$ lie on the real interval $[0,1)$ of the fundamental cell (see the left side of figure \ref{funCell}). Equation \eqref{tauSing} shows that the only accessible (finite) lattice points as the Landau surfaces are approached are either $0 \ \textnormal{or} \ 1$.
This is sketched in the middle panel of figure \ref{funCell}. 
Thus, we expect the following arguments in the Kronnecker forms 
\begin{equation}
    z_j \quad \textnormal{or} \quad z_j-1 \quad \textnormal{for} \quad j\in\{1,2,3\}.
\end{equation}
However, since Kronnecker forms are invariant under shifts by integers, arguments of the form $z_j-1$ are ruled out.  
This particularly simple set of arguments is special to the sunrise example. 
For more complicated examples (e.g., see \cite{Muller:2022gec}), one should expect to find additional (rational) cusps in $\tau$ near the differential equation simple poles. 
Indeed, in general, arguments of the form $z_j-a$, where $a\in \mathbb{Q}$ cannot be ruled out that easy. 

To see how the above \say{bootstrap method} works in practice, we consider the pullback of $\wc{\Gamma}_{11}$. 
In terms of the standard kinematic variables (so before we pullback on the torus) we have
\begin{equation}
\label{eq:gamma11Ex}
    \widecheck{\Gamma}_{11}=2\varepsilon~\textnormal{d}\log(X_2).
\end{equation}
On the kinematic side, this form has two simple poles
\begin{equation}
    X_2=0 \quad \textnormal{and} \quad X_2=\infty.
\end{equation}
Near both of these poles, we see from the middle-top panel of figure \ref{funCell} that the moduli $z$'s collide with the lattice origin, causing simple poles to develop there. More concretely, as we approach $ X_2=0 \quad \textnormal{and} \quad X_2=\infty$ (respectively) the pullback form on the torus has simple poles at $(\tau,z_i)=(i\infty,0)$ and $(\tau,z_j)=(i\infty,0)$ for $i\neq j$ (respectively). To determine which $i,j\in\{1,2,3\}$ to consider, we plug the change of variables \eqref{eq:cvIso} into \eqref{eq:gamma11Ex} and check numerically if any of the $\alpha$'s in
\begin{equation}
\label{eq:GAMMA11alpha}
    \widecheck{\Gamma}_{11}=\alpha_\tau(z_1,z_2)\textnormal{d}\tau+\alpha_{z_1}(z_1,z_2)\textnormal{d}z_1+\alpha_{z_2}(z_1,z_2)\textnormal{d}z_2,
\end{equation}
 vanish. For $\widecheck{\Gamma}_{11}$, we see that $\alpha_{z_2}=0$, while $\alpha_{\tau},\alpha_{z_1}\neq0$. This fixes $i=1$ and $j=3$. Yet, this is not enough information to make a definite ansatz for the pullback of $\widecheck{\Gamma}_{11}$ on the
torus. The extra information needed is the modular transformation rule for  $\widecheck{\Gamma}_{11}$. In this particular example, $\widecheck{\Gamma}_{11}$ is $\Gamma(4)$-invariant.

The simplest expression spanned by Kronnecker forms satisfying these properties is 
\begin{align}
    \widecheck{\Gamma}_{11}
    &= c_1 \left(
         \omega_2^\text{Kr}(z_1 \vert \tau)
         - 2 \omega_2^\text{Kr}(z_1 \vert 2\tau)
    \right)
    + c_3 \left(
        \omega_2^\text{Kr}(z_3 \vert \tau)
         - 2 \omega_2^\text{Kr}(z_3 \vert 2\tau)
    \right)
    \\
    &\to \begin{cases}
      c_1\textnormal{d}\log(z_1) & \textnormal{as} \ (\tau,z_1)\to(i\infty,0)\iff X_2\to 0,\\
      c_3\textnormal{d}\log(z_3) & \textnormal{as} \ (\tau,z_3)\to(i\infty,0)\iff X_2\to \infty.
    \end{cases}
\end{align}
The constants $c_1$ and $c_3$ are fixed by comparing both sides of the last equation numerically against a large set of generic points. For each point, we find the same solution: $c_3=-c_1=2$. 
Thus, 
\begin{align}
    \widecheck{\Gamma}_{11}
    &= 2 \left(
         \omega_2^\text{Kr}(z_1 \vert \tau)
         - 2 \omega_2^\text{Kr}(z_1 \vert 2\tau)
    \right)
    + 2 \left(
        \omega_2^\text{Kr}(z_3 \vert \tau)
         - 2 \omega_2^\text{Kr}(z_3 \vert 2\tau)
    \right).
\end{align}

While not required in the above example, sometimes one needs to also consider modular forms, $\eta_n(\tau)$ \eqref{eq:ansatz}.
This is needed when the  $\ed z_1$ and $\ed z_2$ coefficients of $\wc{\Gamma}_{ij}$ and our ansatz match numerically, but the $\ed \tau$ component does not. 

Using this procedure for the remaining entries of $\widecheck{\boldsymbol{\Gamma}}$ yields \eqref{epsCon}.
However, note that due to some relations between components of $\wc{\bs{\Gamma}}$ (section \ref{sec:relations}), only a subset of the $\wc{\Gamma}_{ij}$ needs to be computed.
\subsubsection{Relations from modular properties \label{sec:relations}}
Based on the behaviour of $\widecheck{\boldsymbol{\Gamma}}$ under a transformation of the root preserving subgroup $\Gamma(4)$ of modular transformations, we observe that 
\begin{equation}
    \widecheck{\Gamma}_{45}=\frac{2}{3}~\widecheck{\Gamma}_{57}, \quad  \widecheck{\Gamma}_{46}=2~\widecheck{\Gamma}_{67},  \quad  \widecheck{\Gamma}_{56}=3~\widecheck{\Gamma}_{65},  \quad  \widecheck{\Gamma}_{33}=\widecheck{\Gamma}_{11}+\widecheck{\Gamma}_{22},
\end{equation}
\begin{equation}
    \widecheck{\Gamma}_{53}=-\widecheck{\Gamma}_{11}+3\widecheck{\Gamma}_{22}-i\widecheck{\Gamma}_{52}, \quad \widecheck{\Gamma}_{52}=\widecheck{\Gamma}_{51}+4i\widecheck{\Gamma}_{11}-4i\widecheck{\Gamma}_{22},
\end{equation}
\begin{equation}
    \widecheck{\Gamma}_{61}=S~(4i~\widecheck{\Gamma}_{11}+\widecheck{\Gamma}_{51}), \quad \widecheck{\Gamma}_{62}=-S~(4i~\widecheck{\Gamma}_{11}+\widecheck{\Gamma}_{51}), \quad \widecheck{\Gamma}_{63}=S~(\widecheck{\Gamma}_{22}-\widecheck{\Gamma}_{11}),
\end{equation}
\begin{equation}
    \widecheck{\Gamma}_{44}=\widecheck{\Gamma}_{77}, \quad  \widecheck{\Gamma}_{75}=\frac{2}{3}\widecheck{\Gamma}_{54}, \quad  \widecheck{\Gamma}_{76}=2\widecheck{\Gamma}_{64}.
\end{equation}
where $S$ is the sign factor defined in appendix \ref{SEC:details}. Below, we work in a kinematic range where $S=+1$. The knowledge of these relations will reduce the work needed to pullback $\widecheck{\boldsymbol{\Gamma}}$ on the torus.

\subsubsection{Kinematic connection on the torus \label{sec:final kin connection}}

Introducing the functions
\begin{align}
    \gameo_n^{(K)}\hspace{-0.1cm}
        \left[\substack{c_1\\c_2\\c_3}\right]
    &:=\sum_{i=1}^3c_i~\omega^{\textnormal{Kr}}_n(z_i|K~\tau), \quad K\in\mathbb{N},
\end{align}
and
\begin{align}
    \Gameo_n^{(K,m)}\hspace{-0.1cm}
        \left[\substack{c_1\\c_2\\c_3}\right]
    &:=\gameo_n^{(K)}\hspace{-0.1cm}
        \left[\substack{c_1\\c_2\\c_3}\right]
    + \gameo_n^{(2K)}\hspace{-0.1cm}
        \left[\substack{-m~c_1\\-m~c_2\\-m~c_3}\right],
\end{align}
the kinematic connection takes the following compact form 
\begin{align}
\label{epsCon}
    &\wc{\bs{\Gamma}} =
    \left(
    \begin{array}{ccccccc}
    0 & 0 & 0 & 0 & 0 & 0 & 0 
    \\
    0 & 0 & 0 & 0 & 0 & 0 & 0 
    \\
    0 & 0 & 0 & 0 & 0 & 0 & 0 
    \\
    0 & 0 & 0 & -6 \eta_{\textcolor{teal}{2}}(\tau ) & 0 & 0 & 0
    \\
    0 & 0 & 0 & 0 & -6 \eta_{\textcolor{teal}{2}}(\tau ) & 0 & 0
    \\
    0 & 0 & 0 & 0 & 0 & -6 \eta_{\textcolor{teal}{2}}(\tau ) & 0
    \\
    0 & 0 & 0 & -288 \eta_{\textcolor{violet}{4}}(\tau ) & 0 & 0 & -6 \eta_{\textcolor{teal}{2}}(\tau ) 
    \end{array}\right) 
    + \\&
    \begin{tiny}
    \left(
\begin{array}{ccccccc}
    \Gameo_{\textcolor{teal}{2}}^{(1,2)}\hspace{-0.1cm}\left[\substack{-2\\0\\2}\right] 
    & 0 & 0 & 0 & 0 & 0 & 0 
\\
    0 
    & \Gameo_{\textcolor{teal}{2}}^{(1,2)}\hspace{-0.1cm}\left[\substack{0\\-2\\2}\right] 
    & 0 & 0 & 0 & 0 & 0 
\\
    0 & 0 
    & \Gameo_{\textcolor{teal}{2}}^{(1,2)}\hspace{-0.1cm}\left[\substack{-2\\-2\\4}\right] 
    & 0 & 0 & 0 & 0 
\\
    0 & 0 & 0 
    & \gameo_{\textcolor{teal}{2}}^{(1)}\hspace{-0.1cm}\left[\substack{0\\0\\4}\right]
        + \gameo_{\textcolor{teal}{2}}^{(2)}\hspace{-0.1cm}\left[\substack{2\\2\\-6}\right] 
    &  \gameo_{\textcolor{orange}{1}}^{(1)}\hspace{-0.1cm}\left[\substack{\frac{1}{2i}\\ \frac{1}{2i}\\0}\right] 
    &
    \gameo_{\textcolor{orange}{1}}^{(1)}\hspace{-0.1cm}\left[\substack{-\frac{1}{2i}\\ \frac{1}{2i}\\0}\right] 
    & \frac{1}{4} \omega_{\textcolor{red}{0}}^{\textnormal{Kr}}(\tau) 
\\
    \Gameo_{\textcolor{teal}{2}}^{(1,2)}\hspace{-0.1cm}\left[\substack{4i\\ -4i\\-8i}\right] 
    &  \Gameo_{\textcolor{teal}{2}}^{(1,2)}\hspace{-0.1cm}\left[\substack{-4i\\ 4i\\-8i}\right] 
    &  \Gameo_{\textcolor{teal}{2}}^{(1,2)}\hspace{-0.1cm}\left[\substack{-2\\ -2\\-4}\right] 
    &  \gameo_{\textcolor{blue}{3}}^{(1)}\hspace{-0.1cm}\left[\substack{-6i\\ -6i\\12i}\right] 
    &  \gameo_{\textcolor{teal}{2}}^{(1)}\hspace{-0.1cm}\left[\substack{-1\\ -1\\0}\right]
        + \gameo_{\textcolor{teal}{2}}^{(2)}\hspace{-0.1cm}\left[\substack{2\\2\\-6}\right] 
    &  \gameo_{\textcolor{teal}{2}}^{(1)}\hspace{-0.1cm}\left[\substack{3\\ -3\\0}\right] 
    &  \gameo_{\textcolor{orange}{1}}^{(1)}\hspace{-0.1cm}\left[\substack{ \frac{3}{4i}\\ \frac{3}{4i}\\0}\right] 
\\
    \Gameo_{\textcolor{teal}{2}}^{(1,2)}\hspace{-0.1cm}\left[\substack{-4i\\ -4i\\0}\right] 
    &  \Gameo_{\textcolor{teal}{2}}^{(1,2)}\hspace{-0.1cm}\left[\substack{4i\\ 4i\\0}\right] 
    &  \Gameo_{\textcolor{teal}{2}}^{(1,2)}\hspace{-0.1cm}\left[\substack{2\\ -2\\0}\right] 
    &  \gameo_{\textcolor{blue}{3}}^{(1)}\hspace{-0.1cm}\left[\substack{-6i\\ 6i\\0}\right] 
    &
    \gameo_{\textcolor{teal}{2}}^{(1)}\hspace{-0.1cm}\left[\substack{1\\ -1\\0}\right] 
    &  \gameo_{\textcolor{teal}{2}}^{(1)}\hspace{-0.1cm}\left[\substack{-3\\ -3\\4}\right]
        + \gameo_{\textcolor{teal}{2}}^{(2)}\hspace{-0.1cm}\left[\substack{2\\2\\-6}\right] 
    &
    \gameo_{\textcolor{orange}{1}}^{(1)}\hspace{-0.1cm}\left[\substack{-\frac{1}{4i}\\ \frac{1}{4i}\\0}\right] 
\\
    \Gameo_{\textcolor{blue}{3}}^{(1,4)}\hspace{-0.1cm}\left[\substack{16i\\ -16i\\16i}\right] 
    &  \Gameo_{\textcolor{blue}{3}}^{(1,4)}\hspace{-0.1cm}\left[\substack{-16i\\ 16i\\16i}\right] 
    &  \Gameo_{\textcolor{blue}{3}}^{(1,4)}\hspace{-0.1cm}\left[\substack{-8i\\ -8i\\8i}\right] 
    &
    \gameo_{\textcolor{violet}{4}}^{(1)}\hspace{-0.1cm}\left[\substack{48\\ 48\\48}\right] 
    &  \gameo_{\textcolor{blue}{3}}^{(1)}\hspace{-0.1cm}\left[\substack{-4i\\ -4i\\8i}\right] 
    &  \gameo_{\textcolor{blue}{3}}^{(1)}\hspace{-0.1cm}\left[\substack{-12i\\ 12i\\0}\right] 
    &
    \gameo_{\textcolor{teal}{2}}^{(1)}\hspace{-0.1cm}\left[\substack{0\\0\\4}\right] 
        + \gameo_{\textcolor{teal}{2}}^{(2)}\hspace{-0.1cm}\left[\substack{2\\2\\-6}\right]
\end{array}
\right).
    \end{tiny}
    \nonumber
\end{align}

\subsection{Elements of the sunrise symbol alphabet\label{sec:symbol}}
This subsection is an invitation for future analysis of the symbol for the full seven-dimensional two-loop sunrise basis. For reasons discussed below, a complete analysis seems beyond the scope of this paper.

Once an $\varepsilon$-form differential equation with simple poles (call it $\varepsilon\boldsymbol{\Gamma}$) is known, the associated master integrals can be evaluated order by order in $\vep$ by iteratively integrating the differential equation along a path that starts at a suitable boundary point in the kinematic space. Both $\tau=i\infty$ (for a path with constant $z_1$ and $z_2$) and the limits where one or more masses vanish (for paths with constant $\tau$) would be good examples of boundary conditions \cite{Bogner:2019lfa}. More precisely,
a Laurent series in $\varepsilon$ for the master integrals is obtained by multiplying the matrix
\begin{equation}
\label{eq:S}
    \boldsymbol{S}=\textbf{Id}+\varepsilon~[\boldsymbol{\Gamma}]+\varepsilon^2~[\boldsymbol{\Gamma}|\boldsymbol{\Gamma}]+\varepsilon^3~[\boldsymbol{\Gamma}|\boldsymbol{\Gamma}|\boldsymbol{\Gamma}]+\cdots,
\end{equation}
with an initial condition vector and then collect terms at each order in $\varepsilon$. In \eqref{eq:S}, each square bracket is computed by multiplying the matrices and concatenating the one-forms appearing in their entries into words (see \cite{Forum:2022lpz}). In that language, a word of length-$k$ then corresponds to an iterated integral of the same length with kernels specified by the concatenations. For differential equations with kernels written in terms of Kronnecker forms (e.g., if $\tau$ is constant in \eqref{epsCon}), we expect the result to be written in terms of \emph{elliptic multiple polylogarithms} (eMPLs). In \cite{Broedel:2017kkb}, eMPLs were defined 
as iterated integrals on an elliptic curve
with fixed modular parameter $\tau$
\begin{equation}
\label{eq:gamT}
    \tilde{\Gamma}\left(\left.\substack{n_1\\ w_1}\substack{n_2\\ w_2}\substack{...\\ ...}\substack{n_k\\ w_k};w\right|\tau\right)=
    \int_{0}^{w} \textnormal{d} w'\,g^{(n_{1})}(w'{-}w_{1},\tau)\tilde{\Gamma}\left(\left.\substack{n_2\\ w_2}\substack{...\\ ...}\substack{n_k\\ w_k};w'\right|\tau\right).
\end{equation}
The numerical evaluation of eMPLs is discussed in detail in \cite{Walden:2020odh}. 

Recent progress on elliptic Feynman integrals suggests that the notion of \say{symbol} is still helpful beyond polylogarithms \cite{Broedel:2018iwv,Wilhelm:2022wow, Forum:2022lpz}. For example, in \cite{Wilhelm:2022wow}, the symbol (prime) was used to bootstrap the 10-point elliptic double-box. Roughly, the symbol of a Feynman integral is a coarse version of its iterated integral expansion, which knows everything about the iterated integral structure but forgets about constants (they are mapped to zero). In particular, one can define the symbol $\mathcal{S}$ of $\tilde{\Gamma}^{(n)}_k$ of weight $n=\sum n_i$ and length $k$ from the differential equation
\begin{align} \nonumber\label{devoftG}
    &\quad \textnormal{d}\tilde{\Gamma}(A_{1},\ldots,A_{k};w)  \\
    &= \sum_{p=1}^{k-1}(-1)^{n_{p+1}}\tilde{\Gamma}(A_{1},\ldots,A_{p-1},\vec{0},A_{p+2},\ldots,A_{k};w) 
    \times\omega^{(n_{p}+n_{p+1})}(w_{p+1,p})
      \\
    &\quad+ \sum_{p=1}^{k}\sum_{r=0}^{n_{p}+1} \Biggl[ 
        \binom{n_{p-1}{+}r{-}1}{n_{p-1}{-}1}\tilde{\Gamma}(A_{1},\ldots, A_{p-1}^{[r]},A_{p+1},\ldots,A_{k};w) 
         \times\omega^{(n_{p}-r)}(w_{p-1,p}) \nonumber \\
    &\qquad \qquad  - \binom{n_{p+1}{+}r{-}1}{n_{p+1}{-}1} \tilde{\Gamma}(A_{1},\ldots, A_{p-1},A_{p+1}^{[r]},\ldots,A_{k};w) 
  \times \omega^{(n_{p}-r)}(w_{p+1,p})
    \Biggr],\nonumber
\end{align}
satisfied by \eqref{eq:gamT} \cite{Broedel:2018iwv}. In \eqref{devoftG}, we have $w_{i,j}=w_i-w_j$ as well as
\begin{equation}
    A^{[r]}_i=\substack{n_i+r\\ w_i} \quad \textnormal{and} \quad  A^{[0]}_i=A_i.
\end{equation}
Schematically, the differential of the \emph{renormalized} eMPL $\tilde{\underline{\Gamma}}_{k}^{(n)}:=(2\pi i)^{k-n}\tilde{\Gamma}_{k}^{(n)}$ takes the form 
\begin{equation}
\label{eq:schemDE}
     \textnormal{d} \tilde{\underline{\Gamma}}_{k}^{(n)}=\sum_i   \tilde{\underline{\Gamma}}^{(n-j_{i})}_{k-1} \textnormal{d} \Omega^{(j_i)}(w_i),
\end{equation}
 where $\Omega^{(\bullet)}$ are primitives for Kronnecker forms. From \eqref{eq:schemDE}, it is then natural to define the symbol as in \cite{Wilhelm:2022wow}
\begin{equation}
\label{eq: elliptic symbol}
     \mathcal{S}(\tilde{\underline{\Gamma}}_{k}^{(n)})=\sum_i \mathcal{S}(\tilde{\underline{\Gamma}}_{k-1}^{(n-j_{i})})\otimes \Omega^{(j_i)}(w_i) \,.
\end{equation}
Fixing, say $z_2$ and $\tau$ constant, the relevant iterated integrals for the computation of the symbol from our differential equation \eqref{epsCon} are obtained at each order in $\varepsilon$ from 
\begin{equation}
\label{eq:Shat}
    \widecheck{\boldsymbol{S}}=\textbf{Id}+\varepsilon~[\widecheck{\boldsymbol{\Gamma}}]+\varepsilon^2~[\widecheck{\boldsymbol{\Gamma}}|\widecheck{\boldsymbol{\Gamma}}]+\varepsilon^3~[\widecheck{\boldsymbol{\Gamma}}|\widecheck{\boldsymbol{\Gamma}}|\widecheck{\boldsymbol{\Gamma}}]+\cdots.
\end{equation}
Looking back at \eqref{epsCon}, we immediately note that one would obtain a solution in terms of eMPLs if terms with $g^{(k)}(z_j, 2 \tau)$ were missing. In order to convert the result to eMPLs we have to express the functions $g^{(1)}(z_j, 2 \tau)$ in terms of functions $g^{(1)}(Z_j, \tau)$. We checked explicitly up to $\mathcal{O}(\varepsilon^2)$ that using the identity (c.f., \cite{Weinzierl:2022eaz})
\begin{equation}
     g^{(k)}\left(z,2\tau\right) =
 \frac{1}{2}
 \left[
 g^{(k)}\left(\frac{z}{2},\tau\right)
 +
 g^{(k)}\left(\frac{z}{2}+\frac{1}{2},\tau\right)
 \right],
\end{equation} the rescaling $z\to Z=z/2$ as well as the identities\footnote{These identities were derived from the relation 
$$\Tilde{\Gamma}\left(\left.\substack{1\\ \sigma}; z\right|\tau\right)=i\pi+\log \frac{\theta_1\left(\pi(z-\sigma)|q\right)}{\theta_1\left(\pi\sigma|q\right)},$$ together with the well known double-angle formula for the odd Jacobi $\theta$-function $\theta_1$.}
\begin{align}
    &\int_{0}^{z}\textnormal{d}z_{1}^{(1)}g^{(1)}\left(\frac{1}{2}z_{1}^{(1)}-\alpha,\tau\right)\int_{0}^{z_{1}^{(1)}}\textnormal{d}z_{1}^{(2)}g^{(1)}\left(\frac{1}{2}z_{1}^{(2)}-\beta,\tau\right)=4\tilde{\Gamma}\left(\left.\substack{1\\ \\ \alpha}~\substack{1\\ \\ \beta}~;~\frac{z}{2}\right|\tau\right),
    \\
    &\int_{0}^{z}\textnormal{d}z_{1}^{(1)}g^{(1)}\left(\frac{1}{2}z_{1}^{(1)}-\alpha,\tau\right)\int_{0}^{z_{1}^{(1)}}\textnormal{d}z_{1}^{(2)}g^{(1)}\left(z_{1}^{(2)}-\beta,\tau\right)
    = 2\left[
        -3\pi i~\tilde{\Gamma}\left(\left.\substack{1\\ \\ \alpha}~;~\frac{z}{2}\right|\tau\right)
    \right.
    \\&\qquad
    \left.
        +\tilde{\Gamma}\left(\left.\substack{1\\ \\ \alpha}~\substack{1\\ \\ \beta/2}~;~\frac{z}{2}\right|\tau\right)
        +\tilde{\Gamma}\left(\left.\substack{1\\ \\ \alpha}~\substack{1\\ \\ (\beta-1)/2}~;~\frac{z}{2}\right|\tau\right)
        +\tilde{\Gamma}\left(\left.\substack{1\\ \\ \alpha}~\substack{1\\ \\ (\beta-\tau)/2}~;~\frac{z}{2}\right|\tau\right)
        +\tilde{\Gamma}\left(\left.\substack{1\\ \\ \alpha}~\substack{1\\ \\ (\beta+1+\tau)/2}~;~\frac{z}{2}\right|\tau\right)
    \right],
    \nn
    \\
    &\int_{0}^{z}\textnormal{d}z_{1}^{(1)}g^{(1)}\left(z_{1}^{(1)}-\alpha,\tau\right)\int_{0}^{z_{1}^{(1)}}\textnormal{d}z_{1}^{(2)}g^{(1)}\left(\frac{1}{2}z_{1}^{(2)}-\beta,\tau\right)
    = 4\left[
        \tilde{\Gamma}\left(\left.\substack{1\\ \\ \alpha/2}~\substack{1\\ \\ \beta}~;~\frac{z}{2}\right|\tau\right)
    \right.
    \nn\\&\qquad
    \left.
        +\tilde{\Gamma}\left(\left.\substack{1\\ \\ (\alpha-1)/2}~\substack{1\\ \\ \beta}~;~\frac{z}{2}\right|\tau\right)
        +\tilde{\Gamma}\left(\left.\substack{1\\ \\ (\alpha-\tau)/2}~\substack{1\\ \\ \beta}~;~\frac{z}{2}\right|\tau\right)
        +\tilde{\Gamma}\left(\left.\substack{1\\ \\ (\alpha+1+\tau)/2}~\substack{1\\ \\ \beta}~;~\frac{z}{2}\right|\tau\right)
    \right],
\end{align}
is enough to express the sunrise components $\widecheck{S}_{4j}$ in \eqref{eq:Shat} in terms of eMPLs. From $\widecheck{S}_{4j}^\textnormal{R}$,\footnote{The R-operation
reverses words -- i.e., $[\omega_1|...|\omega_k]^\textnormal{R}=[\omega_k|...|\omega_1]$.} it is then possible to write the symbol \eqref{eq: elliptic symbol} for the sunrise integral. Up to $\mathcal{O}(\varepsilon^2)$,\footnote{ Near four-dimension ($\varepsilon\to 0$) higher order terms in $\varepsilon$ are suppressed \cite{Adams:2015pya}, so we are mainly interested in the symbol up to $\mathcal{O}(\varepsilon^2)$.} we observe that the only letters showing up are $\Omega^{(0\le i\le 3)}$. The dependence in $\Omega^{(3)}$ is through a unique linear combination of $\Omega^{(3)}(\xi)\otimes\Omega^{(-1)}$ with arguments 
\begin{equation}
\begin{split}
 2~\xi\in&\left\{0,~1,z,~1+z,~ z_2-2,~z_2-1,~z_2,~z_2-\tau-2,~ z_2+\tau-1,\right.\\&\left. z+z_2-2,~ z+z_2-1,~z+z_2,~z+z_2-\tau-2,~
   z+z_2+\tau-1\right\},
\end{split}\end{equation}
at $\mathcal{O}(\varepsilon^2)$. It is therefore not immediate that our expression for the symbol is consistent with that of  \cite{Wilhelm:2022wow}.

That said, there are a few important caveats in the above discussion. 
Firstly, the initial conditions are important for producing functions with the correct physical properties. For example, the correct initial condition could mean that the maximal-cut integrals are suppressed by  additional power(s) of $\vep$ compared to the double-tadpoles. In principle, an intersection calculation is needed to get the linear combination of Feynman integrals our basis corresponds to in order to extract this initial condition vector. 
Secondly, in \cite{Wilhelm:2022wow}, the authors only computed the symbol of the simplest maximal-cut integral, namely the (two-dimensional) sunrise. Perhaps there is no $\Omega^{(3)}$ in this element, once a proper initial condition is taken into account. 
Lastly, if the $\Omega^{(3)}$'s are indeed present, one may need to construct the symbol prime-prime to see cancellations.  However, 
without the input of initial conditions, it is hard to make a direct comparison with \cite{Wilhelm:2022wow}. We can \emph{only} confirm that their alphabet is a subset of our larger alphabet.

\section{Connection to Feynman integrands and loop-by-loop intersections  \label{SEC:feynmanBasis}}

So far, we have constructed an $\vep$-form basis of dual forms and the associated differential equation. 
However, for physics applications, we need the differential equation for Feynman forms \emph{not} dual forms.
Fortunately, the dual and Feynman differential equations are simply related
\begin{align}
\label{eq:dualFromDE}
    \nabla\bs{\vphi} = -\vep\ \bs{\Gamma} ~\wedgedot~ \bs{\vphi}
    \iff
    \wc{\nabla}\wc{\bs{\vth}} = \vep\ \wc{\bs{\vth}} ~\wedgedot~ \wc{\bs{\Gamma}}
    \quad\text{and}\quad
    \la \wc{\vth}_a \vert \vphi_b \ra = C \delta_{ab},
\end{align}
where $C$ is some constant overall normalization.
That is, there exists a basis of Feynman forms $\{\vphi_a\}$ that shares the same differential equations (up to a sign and possible transpose, depending on conventions). 
Even though one knows the Feynman differential equations from the dual differential equations, it is still important to obtain the basis of Feynman forms so that the proper initial conditions for the differential equations can be computed.

Starting with an arbitrary basis of Feynman forms $\{\phi_a\}$, the sought-after basis $\{\vphi_a\}$ is determined from the intersection matrix $\la \wc{\vth}_a\vert\phi_b\ra$
\begin{align}
    \vphi_a\ \propto\  \phi_b\ \la\wc{\vth}\vert\phi\ra^{-1}_{ba}
    \quad \implies \quad
    \la \wc{\vth}_a \vert \vphi_b \ra = C \delta_{ab}
    \quad \implies \quad
    \nabla \vphi_a = -\vep \wc{\Gamma}_{ab} \vphi_{b}
    .
\end{align}
For example, a natural starting basis of Feynman forms for the sunrise family is
\begin{gather}
    \phi_1  = \frac{
        \ed \ell_{2,\perp}^2 \wedge 
        \ed x_2 \wedge 
        \ed \ell_{1,\perp}^2 \wedge 
        \ed x_1
    }{
        \Dsf_1~\Dsf_2
    }, \quad
    \phi_4  = \frac{
        \ed \ell_{2,\perp}^2 \wedge 
        \ed x_2 \wedge 
        \ed \ell_{1,\perp}^2 \wedge 
        \ed x_1
    }{
        \Dsf_1~\Dsf_2~\Dsf_3 
    }, \label{eq:basisFI}
    \\
    \phi_5  = \frac{
        \ed \ell_{2,\perp}^2 \wedge 
        \ed x_2 \wedge 
        \ed \ell_{1,\perp}^2 \wedge 
        \ed x_1
    }{
        \Dsf_1^2~\Dsf_2~\Dsf_3 
    }, \notag
\end{gather}
where the remaining elements are given by permuting the $\Dsf_a$ (given in \eqref{eq:3mass propagators})
\begin{align}
    \phi_2 = \phi_1\vert_{\Dsf_2 \leftrightarrow \Dsf_3},
    \qquad
    \phi_3 = \phi_1\vert_{\Dsf_1 \leftrightarrow \Dsf_3},
    \qquad
    \phi_6 = \phi_5\vert_{\Dsf_1 \leftrightarrow \Dsf_2},
    \qquad
    \phi_7 = \phi_5\vert_{\Dsf_1 \leftrightarrow \Dsf_3}.
\end{align}
To compute the basis $\vphi$ dual to $\wc\vth$, we must compute intersection numbers $\la\wc\vth_a\vert\phi_b\ra$. 
In particular, the intersection number factorizes \emph{loop-by-loop} 
\begin{align}
\label{eq:lblIntersection}
    \la \wc\eta_a \vert \eta_b \ra
    = \la \wc\eta_{F,c} \wedge \wc\eta_{B,ca} \vert \eta_b \ra
    = \Big\la 
        \wc\eta_{B,ca} 
    \Big\vert 
        \underset{\eta_{B,cb}}{\underbrace{
        \la \wc\eta_{F,c} \vert \eta_b \ra_F
        }}
    \Big\ra_B
    = \la \wc\eta_{B,ca} \vert \eta_{B,cb} \ra_B.
\end{align}
Here, the explicit subscript $F$ (and similarly, $B$) signify that the intersection number $\la\bullet\vert\bullet\ra_F$ is computed while holding all but the fibre (base) variables constant (i.e., we compute the $\ell_2$ intersection number, then the $\ell_1$ intersection number).  
Since the main focus of this work is not about the computation of intersection numbers, we only provide one simple (yet illustrative) example. We direct interested readers to \cite{Caron-Huot:p2} for more details.

As a simple example, we consider the intersection number of the maximal-cut dual form $\wc\eta=\wc{\vth}_7 = \widecheck{\boldsymbol{\vphi}}_{F} ~\wedgedot~ \widecheck{\boldsymbol{\vth}}_{B,7}$ (near \emph{four}-dimension) and a Feynman form $\eta_b=\phi_b$ (near \emph{two}-dimension).\footnote{We consider Feynman forms near 2-dimension in order to compare with the basis constructed in \cite{Bogner:2019lfa}.}
Since any tadpole contributions are killed by the dot product with the factor of $(0,0,1)^\top$ in $\wc{\bs{\vartheta}}_{B,7}$, it is clear that we only need to compute the fibre intersection number with \textcolor{red}{maximal-cut support}. That is, we only need the $\eta_{B,3b}$ component of
\begin{align}
    \bs\eta_{B,b} = 
    \begin{pmatrix}    
        \la \wc\vphi_{F,1} \vert \eta_b \ra_F
        \\
        \la \wc\vphi_{F,2} \vert \eta_b \ra_F
        \\
        {\color{red}
        \la \wc\vphi_{F,3} \vert \eta_b \ra_F
        }
    \end{pmatrix}.
\end{align}
The $\delta_{23}$-function in $\wc\vphi_{F,3}$ (given in \eqref{fibBasis}) simply takes a residue on the $23$-boundary
\begin{align}
\label{eq:fibreIntFormula}
    \eta_{B,3b}
    = \frac{1}{ 
        \sqrt{p^2 \ell_{1,\perp}^2}
        \sqrt{q^2 \ell_{2,\perp}^{2} \vert_{_{23}}}
    } 
    \text{Res}_{23}\left[
        \frac{\wc{u}_F\vert_{23}}{\wc{u}_F}
        \eta_b
    \right].
\end{align}
As a sanity check, we have verified with \texttt{FIRE6} that equation \eqref{eq:fibreIntFormula} correctly extracts the coefficient of the \emph{pure} two-dimensional bubble times the additional normalization $(p^2 \ell_{1,\perp}^2)^{-1/2}$
\begin{align}
    \sqrt{p^2 \ell_{1,\perp}^2}
    \bigg\la 
        \wc\vphi_{F,3} 
    \bigg\vert 
        \sqrt{\ell_{2,\perp}^{2} \vert_{_{23}}}\
        \frac{\ed\ell_{\perp,2}^2 \wedge \ed x_2}{\Dsf_2~\Dsf_3} 
    \bigg\ra_F 
    = 1.
\end{align}
Here, $\ell_{2,\perp}^{2} \vert_{_{23}} = (12)^2\vert_{p \to q, m_1 \to m_2, m_2 \to m_3} = \frac{1}{4}(q^2 + (m_2-m_3)^2) (q^2 + (m_2+m_3)^2)$ is the minor defined in \cite{Caron-Huot:p1}, which ensures that the bubble has unit leading singularity in 2-dimension.
For example, using equation \eqref{eq:fibreIntFormula}, the fibre intersection for $\eta_{4}$ is
\begin{gather}
   \sqrt{p^2 \ell_{1,\perp}^2} 
   \eta_{B,34}
    = 
    \frac{1}{
        \sqrt{ 4q^2 \ell_{2,\perp}^{2} \vert_{_{23}} }
    }
    \frac{
        ( \wc{u}_B u_{B} )\
        \ed \ell_{\perp,1}^2 \wedge \ed x_1 
    }{
        \Dsf_1\vert_{23}
    }.
\end{gather}

Next, \eqref{eq:lblIntersection} instructs us to compute the intersection number of the $\eta_{B,3b}$'s with the base basis. 
Again, we use the $\delta$-function in the definition of $\widecheck{\boldsymbol{\vartheta}}_{B,7}$ to take a residue on the $1$-boundary\footnote{When the twist is matrix-valued, the c-map of $(a,b,c)^\top\delta_1$ is modified to $\wc{u}_B^{-1}\cdot(\wc{u}_B\vert_1) \cdot (a,b,c)^\top \ \ed\th_1$. In the case where $(a,b,c)^\top = (0,0,1)^\top$, the c-map of $(0,0,1)^\top\delta_1$ is simply $ \wc{u}^{-1}_{B,33}\cdot(\wc{u}_B\vert_1) \cdot (0,0,1)^\top\ \ed\th_1= (0,0,\frac{\wc{u}_{B,33}^{-1}}{ \wc{u}_{B,33}\vert_1} )^\top\ \ed\th_1$.}
\begin{align}
\label{eq:baseIntFormula1}
    \la \widecheck{\boldsymbol{\vartheta}}_{B,7} \vert \boldsymbol{\eta}_{B,b} \ra_B
    & = \bigg\la
        \frac{m_1^{-4\vep} \pi c_4}{\psi_1}
        \frac{\ed x_1}{Y_\textnormal{r}}  
    \bigg\vert 
       \text{Res}_1\left[  
            \frac{\wc{u}_{B,33}\vert_1}{\wc{u}_{B,33}}
            \eta_{B,3b} 
       \right]
    \bigg\ra_B,
\end{align}
where $\wc{u}_{B,33}=(\ell_{1,\perp}^2 q_+^2 q_-^2 / q^{2})^\vep$ is the dual twist on the maximal cut (see \eqref{eq:nabla base}). 
Explicitly for $\boldsymbol{\eta}_{B,4}$, we find
\begin{align}
    \text{Res}_1\left[  
            \frac{\wc{u}_{B,33}\vert_1}{\wc{u}_{B,33}}
            \eta_{B,34} 
    \right]
    = \frac{
        \ed x_1
    }{
        2\sqrt{
            [q^2~\ell_{\perp,1}~
            \ell_{\perp,2}^2/p^2]\vert_{123}
        }
    }
    = \frac{1}{2ip^2} \frac{\ed x_1}{Y_\textnormal{r}}.
\end{align}
It is important to note that both our dual and Feynman forms on the base (\eqref{2loopBaseBasis} and \eqref{eq:fibreIntFormula}, respectively) contain a  square root $Y_\textnormal{r}$ of the loop variables. This is, in principle, an obstruction to the construction of the c-map. To sidestep this difficulty, we absorb a factor of $Y_\textnormal{r}$ into the dual twist $\wc{u}\vert_1 \to Y_\textnormal{r}~ \wc{u}\vert_1$. This adds a factor of $Y_\textnormal{r}^{-1}$ into the dual form
\begin{align}
    \frac{m_1^{-4\vep} \pi c_4}{\psi_1}
        \frac{\ed x_1}{Y_\textnormal{r}}
    \to \frac{m_1^{-4\vep} \pi c_4}{\psi_1}
        \frac{\ed x_1}{Y_\textnormal{r}^2},
\end{align}
as well as a factor of $Y_\textnormal{r}$ into the form
\begin{align}
    \text{Res}_1\left[  
            \frac{\wc{u}_{B,33}\vert_1}{\wc{u}_{B,33}}
            \eta_{B,34} 
    \right]
    \to Y_\textnormal{r}~\text{Res}_1\left[  
            \frac{\wc{u}_{B,33}\vert_1}{\wc{u}_{B,33}}
            \eta_{B,34} 
    \right]=\frac{\ed x_1}{2ip^2}.
\end{align}
The last step requires  the c-map of $\ed x_1/Y_r^2$ with respect to the rescaled connection $\wc{\omega}_{B,33}\vert_1 \to \wc{\omega}_{B,33}\vert_1 + \ed \log Y_r$.
To build it, we need the one-dimensional primitives, $\psi_a$,  of $\ed x_1/Y_r^2$ near the six twisted points 
\begin{equation}
    \textsf{TP}=\{r_1,r_2,r_3,r_4,c,\infty\}.
\end{equation}
Equipped with the primitives, the c-map is given by
\begin{equation}
\label{eq:c-map}
    \frac{\ed x_1}{Y_\textnormal{r}^2}\xrightarrow{\textnormal{c-map}}\left(\prod_{\alpha\in \textsf{TP}}\theta_\alpha\right)\frac{\ed x_1}{Y_\textnormal{r}^2}+\sum_{\alpha\in \textsf{TP}}\left(\left(\prod_{\beta\neq\alpha}\theta_\beta\right)\textnormal{d}\theta_\alpha~\psi_\alpha\right),
\end{equation}
where  
\begin{equation}
    (\textnormal{d}+\textnormal{d}\log(Y_\textnormal{r} ~\wc{u}_{B,33}\vert_1)\wedge)\psi_\alpha= \frac{\textnormal{d}x_1}{Y_\textnormal{r}^2} \quad \textnormal{near} \ x_1=\alpha.
\end{equation}
Inserting \eqref{eq:c-map} into the definition of the intersection pairing, yields the following residue formula
\begin{align}
\label{eq:baseIntFormula2}
    \la \widecheck{\boldsymbol{\vartheta}}_{B,7} \vert \boldsymbol{\eta}_{B,4} \ra_B
    & = \bigg\la
        \frac{m_1^{-4\vep} \pi c_4}{\psi_1}
        \frac{\ed x_1}{Y_\textnormal{r}^2}  
    \bigg\vert 
       Y_\textnormal{r}~\text{Res}_1\left[  
            \frac{\wc{u}_{B,33}\vert_1}{\wc{u}_{B,33}}
            \eta_{B,34} 
       \right]
    \bigg\ra_B
    \nn\\&=\frac{m_1^{-4\vep} \pi c_4}{\psi_1}\sum_{\alpha\in \textsf{TP}}\textnormal{Res}_{\alpha}\left[\psi_\alpha~Y_\textnormal{r}~\text{Res}_1\left[  
            \frac{\wc{u}_{B,33}\vert_1}{\wc{u}_{B,33}}
            \eta_{B,34} 
       \right]\right]
    \nn\\&=\frac{m_1^{-4\vep} \pi c_4}{2i\psi_1p^2}\sum_{\alpha\in \textsf{TP}}\textnormal{Res}_{\alpha}\left[\psi_\alpha~\textnormal{d}x_1\right]
    \nn\\&=0. 
\end{align} 

To cross-check the loop-by-loop intersection computation outlined above, we tested the procedure on IBPs generated in \texttt{FIRE6}.\footnote{These include IBPs with numerators (like $\texttt{G[1,\{1,1,1,-1,0\}]-F[1,\{1,1,1,-1,0\}]=0}$) and IBPs with higher power propagators (like $\texttt{G[1,\{2,3,1,0,0\}]-F[1,\{2,3,1,0,0\}]=0}$).} More explicitly, we verified the above steps produce the obvious identity 
\begin{equation}
\label{eq:IntExample1}
    \la \wc{\vth}_7\vert \textnormal{IBP} \ra=0,
\end{equation}
where $\textnormal{IBP}$ is an \emph{exact} form and thus cohomologous to zero.

Summarizing, either by using loop-by-loop intersections or a direct differential equations comparison (c.f., \eqref{eq:dualFromDE}), we can find a \emph{two}-dimensional Feynman integrand basis
\begin{equation}
    \boldsymbol{\varphi}=\textnormal{diag}\left(1 \ 1 \ \frac{i}{2} \ \frac{1}{4} \ \frac{i}{2} \ \frac{1}{2i} \ \frac{-1}{16} \right)\cdot \bs{U} \cdot \boldsymbol{\phi},
\end{equation}
satisfying
\begin{equation}
    \la \widecheck{\boldsymbol{\vartheta}}\vert\boldsymbol{\varphi}\ra~\propto~ \textbf{Id}.
\end{equation}
The basis $\bs{U} \cdot \boldsymbol{\phi}$ is the Feynman integrand basis given in \cite{Bogner:2019lfa}. It is related to \eqref{eq:basisFI} via a gauge transformation $\bs{U}$, which is too complicated to be displayed here. We refer the reader to  \cite{Bogner:2019lfa} where it is explicitly given.
\section{An elliptically fibered K3-surface in momentum space at three-loop \label{SEC:K3banana}}

In this section, we analyse the underlying geometry of the four-mass three-loop sunrise integral directly in momentum space. To uncover the geometry, we examine the twist on the maximal-cut  in four-dimension. 
This yields a quintic in two complex variables whose zero set describes an elliptically fibered K3-surface (section \ref{sec:max cut quintic}). 
Then, in section \ref{SEC:periodsK3}, we give an interpretation of the K3-surface as the configuration space of a (degenerate) conic and four lines in general position on the projective plane $\mathbb{CP}^2$. 
We anticipate this geometric perspective to be  practical in future studies of analytic expressions for periods and differential forms on K3-surfaces. 
While we leave the construction of the full three-loop basis for future work, we describe how the loop-by-loop method constrains the choice of basis elements and construct modular invariant basis elements corresponding to the triple-tadpoles in section \ref{SEC:3loopBasis}. 

This section will be expository on the mathematical side -- several facts are stated without formal proofs (but with references as much as possible), which are too technical for this paper.

\subsection{The maximal-cut quintic surface \label{sec:max cut quintic}}

\begin{figure}
    \centering
    \includegraphics[scale=0.3]{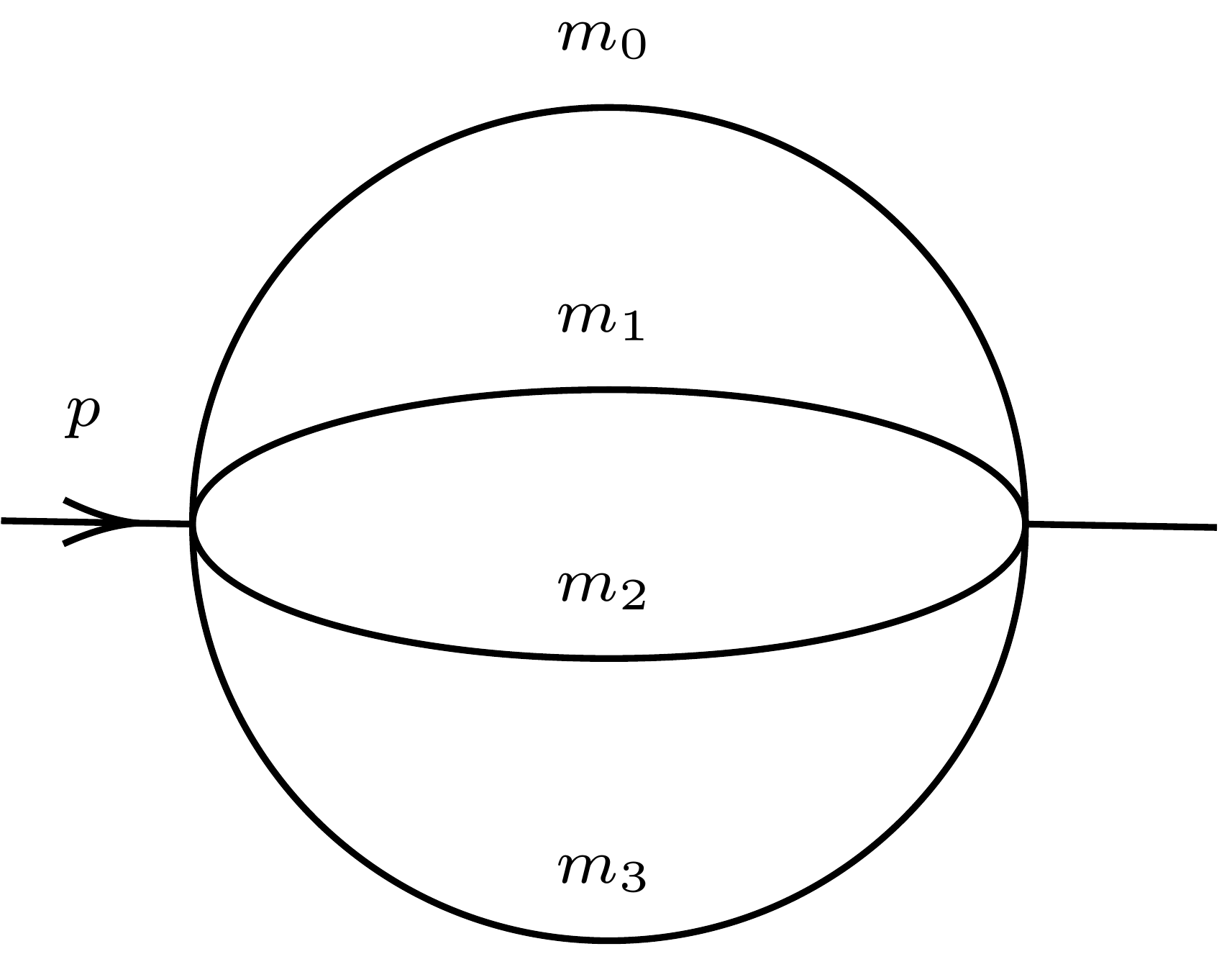}
    \caption{The three-loop unequal mass sunrise diagram.}
    \label{fig:banana}
\end{figure}


In this section, we consider the maximal-cut of the four-mass three-loop sunrise family (see figure \ref{fig:banana}). 

We can recycle and incorporate the two-loop data into the three-loop problem by promoting the external momentum $p$ to a loop-momentum $p \to p+\ell_0$. 
Making this replacement in the two-loop propagators \eqref{eq:2loop props} and adjoining the additional propagator 
\begin{align}
    \textsf{D}_0=\ell_0^2+m_0^2,
\end{align}
we form the set of three-loop propagators. 
The three-loop twist is obtained by the replacement 
\vspace{-0.5cm}
\begin{align}
    \wc{u} 
    \to \wc{u}_\text{3-loop} 
    \equiv 
    \overset{\wc{u}_\mathcal{F}}{\overbrace{
        \ (\wc{u}\vert_{p\to p+\ell_0}) \
    }}
    \overset{\wc{u}_\mathcal{B}}{\overbrace{
       \ \sqrt{p^2} (\ell_{0,\perp}^2)^{\vep+\frac12} \
    }},
\end{align}
where the two-loop twist $\wc{u}$ is given in equation \eqref{eq:lbl udual}.
Here, we have also defined a loop-by-loop parameterization for $\ell_0$\footnote{More explicitly, we have the following three-loop loop-by-loop parameterization for the loop momenta 
\begin{eqnarray}
    \ell_2^\mu=\ell_{2,\|}^\mu+\ell_{2,\perp}^\mu:=x_2 q_2^\mu+\ell_{2,\perp}^\mu, 
    \qquad& 
    q_2\cdot \ell_{2,\perp}=0,
    \qquad&
    q_2^\mu=\ell_1^\mu+q_1^\mu,
\notag \\ 
    \ell_1^\mu=\ell_{1,\|}^\mu+\ell_{1,\perp}^\mu:=x_1 q_2^\mu+\ell_{1,\perp}^\mu, 
    \qquad& 
    q_1\cdot \ell_{1,\perp}=0,
    \qquad&
    q_1^\mu=\ell_0+q_0,
\notag \\
    \ell_0^\mu=\ell_{0,\|}^\mu+\ell_{0,\perp}^\mu:=x_0 q_0^\mu+\ell_{0,\perp}^\mu, 
    \qquad& 
    q_0\cdot \ell_{0,\perp}=0,
    \qquad& 
    q_0=p. 
\notag
\end{eqnarray}
}
\begin{align}
    \ell_0^\mu = \ell_{0,\perp}^\mu + x_0 p^\mu. 
\end{align}
Thus, the three-loop integration measure becomes a $6$-form
\begin{equation}
     \frac{\textnormal{d}^D\ell_0}{\pi^{D/2}}\wedge\frac{\textnormal{d}^D\ell_1}{\pi^{D/2}}\wedge \frac{\textnormal{d}^D\ell_2}{\pi^{D/2}}\sim u_{3\textnormal{-loop}}~ \textnormal{d}\ell_{2,\perp}^2\wedge \textnormal{d}x_2\wedge \textnormal{d}\ell_{1,\perp}^2\wedge \textnormal{d}x_1\wedge \textnormal{d}\ell_{0,\perp}^2\wedge \textnormal{d}x_0,
\end{equation}
multiplied by the three-loop twist. 
As before, the exterior powers of $\textnormal{d}\Omega_{D-2} / 2\pi^{D/2}$ are irrelevant for the purposes of computing differential equations and have been dropped.

To uncover the underlying geometry, we examine the restriction of the twist $\widecheck{u}_{3\textnormal{-loop}}$ to the maximal-cut boundary 
\begin{equation}
    \textsf{D}_0=\textsf{D}_1=\textsf{D}_2=\textsf{D}_3=0.
\end{equation}
in integer dimension ($\vep=0$)
\begin{align} \label{eq:integrand1}
    \lim_{\vep\to0}
    \widecheck{u}_{3\textnormal{-loop}}\vert_{0123}
    &=\sqrt{p^2}\sqrt{F_1}  
    \sqrt{p^2x_0^2+m_0^2} 
    \sqrt{x_1^2 F_1+m_1^2}
    \nn\\&\qquad\times
    \sqrt{F_2^2+2(m_2^2+m_3^2)F_2 +(m_2^2-m_3^2)^2},
\end{align}
where $ F_1=(2 x_0+1) p^2-m_0^2$ and $F_2=(2 x_0+1) (2 x_1+1) p^2-m_0^2(2x_1+1)-m_1^2$. 
To reduce the length of the expressions below, we introduce the dimensionless variables
\begin{equation}
    X_0=\frac{p}{m_1}, 
    \quad X_2=\frac{m_2}{m_1}, 
    \quad X_3=\frac{m_3}{m_1}, 
    \quad X_4=\frac{m_0}{m_1},
\end{equation}
and consider the following rational change of variable
\begin{equation}
\label{reparam2}
    x_1\to\frac{X_0^{-2}~x_1}{F_1/m_1^2}.
\end{equation}
After this change of variable, \eqref{eq:integrand1} can be rewritten in a particularly convenient form 
\begin{equation} \label{4Dintegrand}
       \widecheck{u}_{3\textnormal{-loop}}|_{0123,\vep\to0}
       = \sqrt{x_0^2+X_0^{-2}X_4^2} ~ \mathcal{Y}_{x_0}(x_1),
\end{equation}
where $\mathcal{Y}_{x_0}(x_1)$ is a root of an irreducible quartic polynomial in $x_1$
\begin{equation}
    \mathcal{Y}_{x_0}^2(x_1)-(x_1-\mathcal{R}_1(x_0))(x_1-\mathcal{R}_2(x_0))(x_1-\mathcal{R}_3(x_0))(x_1-\mathcal{R}_4(x_0))=0.
\end{equation}
The set 
\begin{equation}
\label{K3ellCurve}
    E_{\mathcal{Y}_{x_0}}(\mathbb{C})=\{(x_1,\pm \mathcal{Y}_{x_0})|x_1\in\mathbb{C}\},
\end{equation}
defines an elliptic curve. 
The roots of $\mathcal{Y}_{x_0}^2$ depends of the external kinematics and $x_0$ (hence the subscript). 
Explicitly, they are
\begin{align}
\label{3loopRoots}
    \mathcal{R}_1
    = X_0^{4} \frac{2 x_0+\left((X_2+X_3)^2-X_4^2-1\right)X_0^{-2}+1}{2}, 
    & \quad \mathcal{R}_2=X_0^{3}\sqrt{X_0^{-2}X_4^2-1-2 x_0},
    \\
    \mathcal{R}_4 = X_0^{4}\frac{2x_0+\left((X_2-X_3)^2-X_4^2-1\right) X_0^{-2}+1}{2}, 
    & \quad \mathcal{R}_3= -X_0^{3}\sqrt{X_0^{-2}X_4^2-1-2 x_0}.
    \nn
\end{align}
Note that \eqref{3loopRoots} contains a square root in $x_0$. In principle, it can be rationalized by a quadratic change of variables. However, this procedure causes higher powers of $x_0$ to proliferate. Since it is practical to keep the degree of the polynomial as low as possible, this is not what we do.

The \emph{K3-surface} \cite{griffiths2014principles,huybrechts2016lectures,husemoller2004families,schutt2019elliptic,shimada2000elliptic,friedman2013smooth} (named after Kummer, Kähler, and Kodaira) associated to the four-mass three-loop sunrise family manifests itself in two ways.

On one hand, by integrating out $x_1$ first (fixing $x_0$), it is clear from \eqref{4Dintegrand} and \eqref{K3ellCurve} that the underlying space $G$ admits an elliptic fibration over the Riemann sphere $\mathbb{CP}^1$. Indeed, for each generic base point $x_0\in \mathbb{CP}^1$, there is an elliptic curve $E_{\mathcal{Y}_{x_0}}(\mathbb{C})$ such that
\begin{equation}
\label{ellFibration}
	        E_{\mathcal{Y}_{x_0}}(\mathbb{C})\hookrightarrow G \twoheadrightarrow \mathbb{CP}^1.
	    \end{equation}
It is well known by geometers that such space $G$ is a special case of (algebraic) K3-surfaces, called an \emph{elliptically fibered K3-surface} \cite{husemoller2004families}. We will provide quantitative arguments from which this can be seen more directly below. 

\begin{figure}
    \centering
    \includegraphics[scale=0.38]{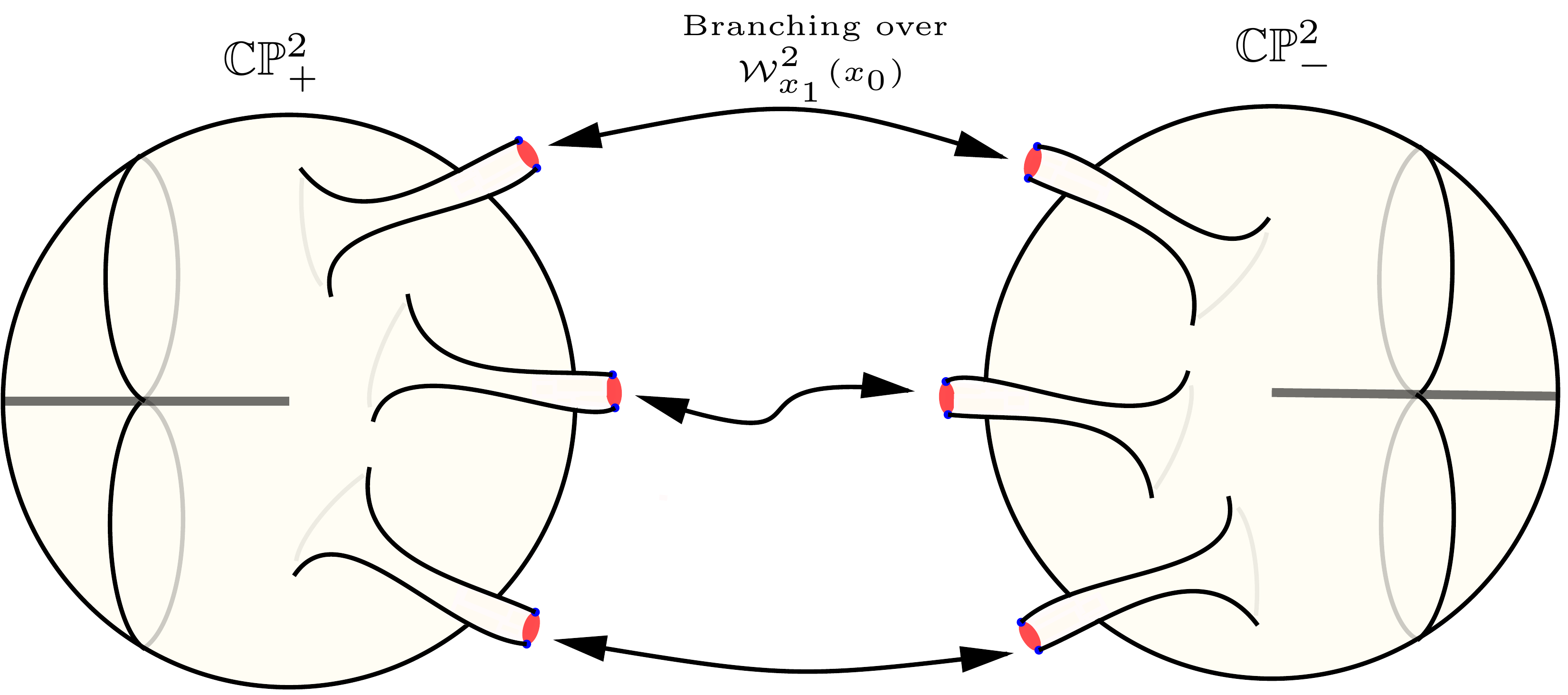}
    \caption{Real section of a K3-surface as the double cover $\mathcal{X}$ of $\mathbb{CP}^2$, branched over the branch locus \eqref{Wcurve}.}
    \label{fig:K3picture}
\end{figure}

On the other hand, integrating out $x_0$ first (fixing $x_1$), yields a square root of an \emph{irreducible quintic} curve
\begin{equation}
\label{Wcurve}
   G:\mathcal{W}_{x_1}^2(x_0)-(x_0^2+X_0^{-2}X_4^2)~\mathcal{Y}_{x_0}^2(x_1)=0.
\end{equation}
Since equation \eqref{Wcurve} defines six branch points  (including infinity) and three branch cuts in $\mathbb{CP}^2$, the zero set $G$ is immediately interpreted as the \emph{double} cover $\mathcal{X}$ of the projective plane branched over the three branch cuts. Figure \ref{fig:K3picture} sketches the real section of this geometry.

 For a non-singular choice of $x_1$, a codimension-1 submanifold of $\mathcal{X}$ is naturally identified with a curve of genus-2.
Indeed, in such cases equation \eqref{Wcurve} is viewed as an \emph{hyperelliptic curve}. The degree of an hyperelliptic polynomial determines the genus of the corresponding curve \cite{griffiths2014principles}. It is well known that a hyperelliptic polynomial of degree $(2g+1)$ or $(2g+2)$ is a curve of arithmetic genus-$g$. For quintics and sextics, $g=2$. This is consistent with the naive perturbative expectation that, for each loop added, the genus of the underlying geometry should increase by one. 

In the remaining part of this section, we provide two independent quantitative checks confirming the underlying maximal-cut geometry corresponds to a K3-surface. 

On one hand, it can be shown \cite[Ch. 11, \S 2.1-2.2]{huybrechts2016lectures} that a Weierstra\ss \ equation defines an elliptically fibered K3-surface \emph{if and only if} 
\begin{equation}
	\deg_{x_0}(f)\leq 8  \quad \textnormal{and} \quad \deg_{x_0}(g)\leq 12,
\end{equation}
and at least one of the bounds
\begin{equation}
	\deg_{x_0}(f)\geq 5  \quad \textnormal{or} \quad \deg_{x_0}(g)\geq 7,
\end{equation}
is  fulfilled. The first task is then to find a Weierstra\ss \ form for $G$ as given in \eqref{Wcurve}. Using the \texttt{Maple} command
\begin{equation}
	\texttt{Weierstrassform}(G,x_1,\mathcal{W}_{x_1},\tilde{x}_1,\tilde{\mathcal{W}}_{\tilde{x}_1}),
\end{equation}
we find  a surface
\begin{equation}
	\label{eq:WformSurface}
	\tilde{G}:~\tilde{\mathcal{W}}_{\tilde{x}_1}^2(x_0)-\tilde{x}_1^3-f(x_0)\tilde{x}_1+g(x_0)=0,
\end{equation}
birationally equivalent to \eqref{Wcurve} such that
\begin{equation}
    \textnormal{deg}_{x_0}(f)=8 \quad \textnormal{and} \quad \textnormal{deg}_{x_0}(g)=12.
\end{equation}
Hence, $G$ is indeed a K3-surface. The poles of the $j$-invariant
\begin{equation}
    j(\tau(x_0))=55296\frac{f^3}{\Delta_j} \quad \textnormal{where} \quad   \Delta_j=27 g^2+4f^3,
\end{equation}
corresponds to the $3\times \textnormal{deg}_{x_0}(f)=2\times  \textnormal{deg}_{x_0}(g)=24$ nodal curves on the K3. These are the so-called singular fibres. To find their locations, one needs to solve the degree 24 polynomial
\begin{equation}
\label{eq:singFib}
    \Delta_j=0.
\end{equation}
In general, there is no way to express roots of fifth (or higher) order polynomials in terms of radicals.
In particular, we find analytically only 12 solutions to \eqref{eq:singFib}, namely $x_0=\pm iX_4X_0^{-1}$ (both with multiplicity six). It is not hard to see that these solutions describe points on \eqref{Wcurve}. In addition, contrastingly to what happens in the elliptic case (c.f., the elliptic $j$-invariant in \eqref{eq:jInv}), a direct computation indicates that roots of \eqref{eq:WformSurface} colliding do not yield further solutions to \eqref{eq:singFib}. 
 
On the other hand, we can compute the Hodge numbers for the (homogenized) zero set $G$. With the help of \texttt{Macaulay2} \cite{M2}, we find
\begin{equation}
	\label{hodgeD}
\tikzset{every picture/.style={line width=0.75pt}} 
\begin{tikzpicture}[x=0.75pt,y=0.75pt,yscale=-1,xscale=1]
\draw (164,132.41) node [anchor=north west][inner sep=0.5pt]    {$\begin{matrix}
    &  & h^{2,2} &  & \\
    & h^{2,1} &  & h^{1,2} & \\
    h^{2,0} &  & h^{1,1} &  & h^{0,2}\\
    & h^{1,0} &  & h^{0,1} & \\
    &  & h^{0,0} &  & 
\end{matrix} ~=~ \begin{matrix}
      &   & 1  &   &   \\
      & 0 &    & 0 &   \\
    1 &   & 20 &   & 1 \\
      & 0 &    & 0 &   \\
      &   & 1  &   & 
\end{matrix}$,};
\end{tikzpicture}
\end{equation}
where $h^{a,b}$ is the dimension of the corresponding Dolbeault cohomology $h^{a,b} = \text{dim}H^{a,b}$.
It is well known that every K3-surface has Hodge diamond of this form \cite{griffiths2014principles}.

It is worth noting the Hodge diamond contains a lot of useful topological information about a given space.  For example, the condition $h^{2,0}=1$ says that $G$ has a unique and everywhere nonzero holomorphic two-form, while the condition $h^{1,0}=0$ makes it clear that $G$ is \emph{not} a higher genus torus.  

We can also triple-check that $G$ is a K3-surface by computing its Euler characteristic $\mathcal{X}_{\textnormal{E}}$. 
First, we compute the Betti numbers\footnote{The dimensions of the corresponding de Rham cohomology groups.} by summing each row of the Hodge diamond. 
Then the Euler characteristic is given by the alternating sum of the Betti numbers. From \eqref{hodgeD}, we find
\begin{equation}
\label{eq:ECk3}
    \mathcal{X}_{\textnormal{E}}(G)=1-0+22-0+1=24.
\end{equation}
It is well known that a non-singular surface is a K3 if it admits a nowhere-vanishing holomorphic two-form (which is guaranteed here by $h^{2,0}=1$) and its Euler characteristic is 24 (see for example \cite[p.230]{yoshida2013hypergeometric}).  

We quickly mention that, historically, elliptically fibered K3-surfaces played a central role in the development of F-theory on Calabi-Yau 2-folds \cite{Vafa:1996xn,Sen:1996vd,Banks:1996nj,Aspinwall:1994rg,Aspinwall:1996mn}. It is therefore not the first time they show up in the physics literature.
\subsection{K3-surfaces as configuration spaces on the projective plane \label{SEC:periodsK3}}

Given the crucial role that an analytic understanding of (co)cycles and periods played in the construction of \eqref{epsCon}, 
 it is clear that an analytical understanding of K3-cycles and periods is a valuable input in determining a basis at three-loop. 

That said, we already know that a new set of
transcendental functions (going beyond elliptic) should appear there. Indeed, since the corresponding Picard-Fuchs operator \cite{Muller-Stach:2012tgj} fails to factorize into a symmetric square of the degree two operator \cite{Bloch:2014qca,Vanhove:2014wqa,Lairez:2022zkj}, it cannot annihilate powers and products of $\psi_1$ and $\psi_2$ as it does in the equal mass case (see for example \cite{Primo:2016ebd,Primo:2017ipr,Broedel:2021zij,Pogel:2022yat}).

The goal of this subsection is to review an interesting equivalence between certain K3-surfaces and configuration spaces of curves on the projective plane. The hope is that this geometric perspective will provide additional tools to directly analyze the periods of and differential forms on K3-surfaces in a near future.

We start by reviewing a useful construction from the existing mathematics literature, where a certain kind of K3-surface (closely related to the one showing up in \eqref{Wcurve}) is interpreted as the blow-up of the configuration space of six generic lines on the projective plane. 
Then, we discuss the blown-up configuration space of four generic lines and a degenerate conic on the projective plane associated to the K3-surface in \eqref{Wcurve}. 
  
\subsubsection{Configuration space of six lines on the projective plane}

In the first few sections of this paper, we considered the double cover of the Riemann sphere $\mathbb{CP}^1$ branched along four points in general positions -- i.e., a torus.  This covering defined families of elliptic curves over the moduli space of the configurations of four points on $\mathbb{CP}^1$, which are naturally parametrized by a unique variable: the conformal cross-ratio/elliptic modulus $k$ of the four branch points in \eqref{eq:ellQUantities}. 

\begin{figure}
    \centering
    \includegraphics[scale=0.3]{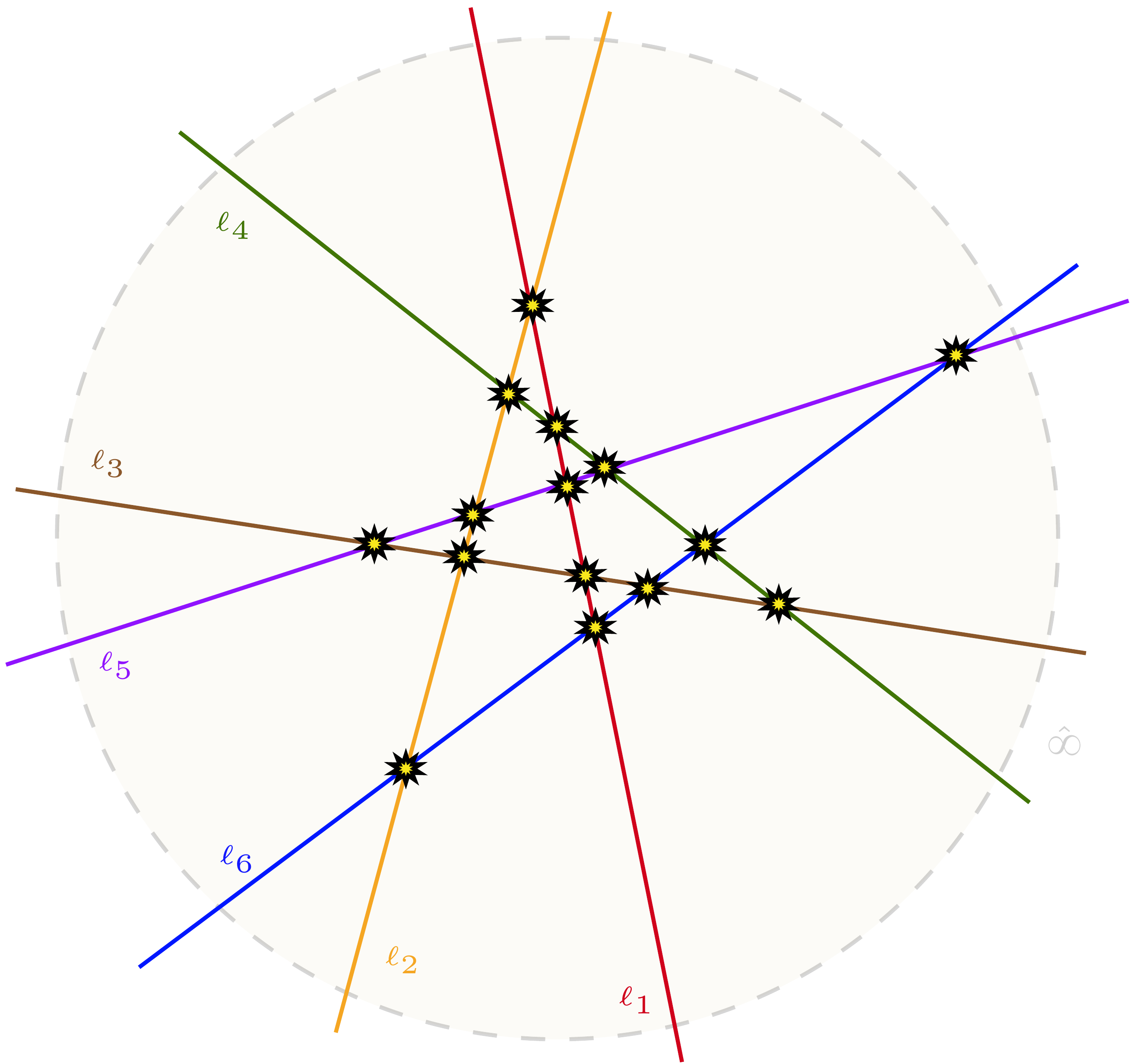}
    \caption{Six lines (one for each branch point) in general positions on the projective plane. Their $\binom{6}{2}=15$ points $P_{ij}$ of normal crossing are illustrated by the yellow stars. There are 10 bounded regions and 12 regions going to infinity. The blown-up moduli space describes a smooth K3-surface.}
    \label{fig:6lines}
\end{figure}

Higher dimensional analogues have been studied in many different contexts of mathematics (a modest list of references would be  \cite{shiga1979one,shiga1981one,aomoto2011theory,Matsumoto1992THEMO,Yoshida1998TheRL,Matsumoto1993ConfigurationSO,Sekiguchi1997WO,Matsumoto1989ThePM,Sasaki1997OnTR,Sasaki1992MonodromyOT,matsumoto1992monodromy,Matsumoto1993ThetaFO,Matsumoto1993MonodromyOT,yoshida2013hypergeometric,jorgK3}).
Among others, Matsumoto, Sasaki and Yoshida \cite{matsumoto1992monodromy} have extensively studied the two-dimensional generalization in the 90's, namely the double cover of the projective plane branched along six lines $\{l_i\}_{i=1}^6$ in general positions (see figure \ref{fig:6lines}).
After making suitable resolutions (blow-ups) at the fifteen intersection points, the double cover defines a family of smooth K3-surface for each non-degenerate configuration of the six lines. 

In \cite{matsumoto1992monodromy}, the authors studied the periods of the family and fully determined the associated monodromy group. 
Furthermore, they derived the set of differential equations satisfied by the periods in terms of the so-called Aomoto-Gel'fand system \cite{aomotostructure,gel1990hypergeometric,gel1986generalized} on the Gra\ss mannian $\textnormal{Gr}(3, 6)$.

The six lines are said to be in \say{general position} or \say{non-degenerate} on	 $\mathbb{CP}^2$ when the coefficients defining them are sufficiently generic (i.e., no more than one line crosses at each other at the same point).  
Degenerate arrangements and their compactification are discussed in \cite[\S 0.9-10]{matsumoto1992monodromy}.

Let $[x_1:x_0:X]$ symbolize the coordinate of $\mathbb{CP}^2$. Then, we denote the system formed by the six (homogenized) lines by 
\begin{equation}
    \boldsymbol{l}=\{l_i(x_1,x_0,X):=\alpha_{1,i}~x_1+\alpha_{2,i}~x_0+\alpha_{3,i}~X=0\}_{i=1}^6\subset \mathbb{CP}^2,
\end{equation}
where the $\alpha_{\bullet,\bullet}$ are generic complex coefficients. 
The configuration space of six lines on the projective plane has exactly four parameters. 
To see this, we represent $\boldsymbol{l}$ by a $(3\times6)$-matrix $M_{3,6}$ with non-vanishing minors, which ensure the lines are in generic positions. 
We then \say{gauge away} the obvious $\textnormal{GL}(3,\mathbb{C})$- and $(\mathbb{C}^\times)^6$-redundancies of this space by a suitable choice of left $ \textnormal{GL}(3,\mathbb{C})$-action and right $(\mathbb{C}^\times)^6$-action such that\footnote{Here, $\mathbb{C}^\times$ denotes the punctured complex plane.}
\begin{equation}
    M_{3,6}^{\textnormal{gauged}}= g\cdot M_{3,6}\cdot h=\begin{pmatrix}
        1 &0 & 0& 1 & 1 & 1\\
        0 & 1 & 0 & 1 & a_{52} & a_{62}\\
        0 & 0 & 1 & 1 & a_{53} & a_{63}\\
    \end{pmatrix}.
\end{equation}
This shows the configuration space is an open subset of $\mathbb{C}^4$, which is what we wanted to show.

When the lines are in general position, the double cover branched along the six lines $\boldsymbol{l}$ (call it $\mathcal{G}'(\boldsymbol{l})$) defines a \emph{singular} K3-surface with singularities at each of the fifteen intersection points
\begin{equation}
\label{eq:poinSpace}
    P_{ij}=l_i\cap l_j, \quad i\neq j.
\end{equation}
Blowing-up these singularities along algebraic curves results in a \emph{smooth} K3-surface $\mathcal{G}(\boldsymbol{l})$. In particular, each blown-up configuration defines a four-dimensional K3-surface. The set of all such configurations is referred to as the double cover \emph{family} of K3-surfaces. 

It is instructive to cross-check quantitatively the above statements by direct computation of geometric invariants -- e.g., the Euler characteristic of $\mathcal{G}(\boldsymbol{l})$. In particular, if we denote by \say{$\bullet$} the zero-dimensional point space \eqref{eq:poinSpace}, the punctured arrangement of six lines has Euler characteristic
\begin{equation}
\chi_{\textnormal{E}}(\boldsymbol{l})=6\chi_{\textnormal{E}}(\mathbb{CP}^1)-15\chi_{\textnormal{E}}(\bullet)=12-15=-3.    
\end{equation}
Going back to the smooth covering definition of $\mathcal{G}(\boldsymbol{l})$, we immediately find
\begin{equation}
    \begin{split}
        \chi_{\textnormal{E}}(\mathcal{G}(\boldsymbol{l}))&=[\underbrace{2(\chi_{\textnormal{E}}(\mathbb{CP}^2)-\chi_{\textnormal{E}}(\boldsymbol{l}))}_{\textnormal{double cover}}+\underbrace{ \chi_{\textnormal{E}}(\boldsymbol{l})}_{\textnormal{branching}}]+\underbrace{15 [\chi_{\textnormal{E}}(\mathbb{CP}^1)- \chi_{\textnormal{E}}(\bullet)]}_{\textnormal{blow-up}}\\&=[2(3-(-3))-3]+15\\&=24.
    \end{split}
\end{equation}
 
 Analogous to the definitions of the elliptic curve's periods (c.f., \eqref{periodsEllCurve}), the periods for the K3 family $\mathcal{G}(\boldsymbol{l})$ are obtained by integrating the \emph{unique} holomorphic two-form
\begin{equation}
\label{holom2form}
    \rho(\boldsymbol{l}):=\frac{x_1~\textnormal{d}x_0\wedge \textnormal{d}X-x_0~\textnormal{d}x_1\wedge \textnormal{d}X+X~\textnormal{d}x_1\wedge \textnormal{d}x_0}{\sqrt{\prod_{i=1}^6 l_i(x_1,x_0,X)}},
\end{equation}
over independent two-cycles $\{\Gamma_i(\boldsymbol{l})\}$ depending continuously on the linear system $\boldsymbol{l}$  
\begin{equation}
\label{perK3abstract}
    \psi_{\mathcal{G}(\boldsymbol{l}),i}:=\int_{\Gamma_i(\boldsymbol{l})}\rho(\boldsymbol{l}).
\end{equation}

The relevant two-dimensional contours are the so-called \emph{transcendental cycles} of the integral homology group $H_2(\mathcal{G}(\boldsymbol{l}),\mathbb{Z})$. Here, transcendental means non-algebraic, signifying these cycles are not represented by algebraic equations of the form $\mathcal{C}(x_1,x_0,X)=0$ where $\mathcal{C}$ is some polynomial in its arguments. In particular, transcendental cycles form the subset of cycles in $H_2(\mathcal{G}(\boldsymbol{l}),\mathbb{Z})$ such that \eqref{perK3abstract} does not vanish.

In order to be able to define the transcendental cycles, we first need to know how many of them there are on $\mathcal{G}(\boldsymbol{l})$. From the discussion above, it is not hard to see there are in total 16 independent algebraic curves on $\mathcal{G}(\boldsymbol{l})$: 15 from the desingularization at the $P_{ij}$'s and one from the pullback of a generic line $L$ in the projective plane, namely the hyperplane class
\begin{equation}
\label{eq:Hclass}
    H=\{L(x_1,x_0,X)=0\}.
\end{equation}

Since summing the middle row of the Hodge diamond gives a total of $22$ independent two-cycles, this implies there are $22-16=6$ transcendental (non-algebraic) cycles.\footnote{The number 16 also naturally shows up in the study of F-theory dual singular heterotic K3 models. They are associated to Abelian
toroidal orbifolds $\mathbb{T}^2\times\mathbb{T}^2/\mathbb{Z}^2$. In particular, see figure 1 in \cite{Ludeling:2014oba} sketching a K3-surface dual to the orbifold which has 16  $\mathbb{Z}^2$-singularities. Blowing these up gives 16 algebraic cycles.}\footnote{It is important to note that the number six is in agreement with the size $\lambda$ of the so-called Frobenius basis of the four-mass three-loop sunrise \cite{Bonisch:2021yfw,Forum:2022lpz,Duhr:2022pch}.}
An explicit, yet rather complicated, realization of these cycles is detailed in \cite[\S 2.1]{matsumoto1992monodromy}. For alternative realizations, see \cite{jorgK3,yoshida2013hypergeometric}.  
\begin{table}[]
\centering
{\renewcommand{\arraystretch}{1.5}
\begin{tabular}{ccc}
{\color[HTML]{656565} } 
    \emph{\textbf{K3-surfaces}} &  & \emph{\textbf{Elliptic curves}}  
    \\
        System $\boldsymbol{l}$ of 6 lines in $\mathbb{CP}^2$ & $\leftrightarrow$ & System $\boldsymbol{p}$ of 4 points in $\mathbb{CP}^1$ 
    \\ \hline
        Space $\mathcal{G}(\boldsymbol{l})$ & $\leftrightarrow$ &  Space $\mathbb{C}-\boldsymbol{p}$
    \\ \hline 
        $M_{3,6}^{\textnormal{gauged}}=\begin{scriptsize}
        \begin{pmatrix}
            1 &0 & 0& 1 & 1 & 1\\
            0 & 1 & 0 & 1 & a_{52} & a_{62}\\
            0 & 0 & 1 & 1 & a_{53} & a_{63}\\
        \end{pmatrix}\end{scriptsize}$ 
        & $\leftrightarrow$ 
        & $M_{2,4}^{\textnormal{gauged}} = \begin{scriptsize}\begin{pmatrix}
            1 &0 & 1& 1 &\\
            0 & 1 & 1 & k\\
        \end{pmatrix}\end{scriptsize}$ 
    \\ \hline
        Holomorphic two-form $\rho$   & $\leftrightarrow$ & Holomorphic one-form $\textnormal{d}x/y$ 
    \\ \hline
        Transcendental cycles $\{\Gamma_{i}\}_{i=1}^6$   &  $\leftrightarrow$ & $A$ and $B$ cycles  
    \\ \hline
        Intersection form $\mathcal{A}$  & $\leftrightarrow$ & Intersection form \eqref{ellCurveInForm} 
    \\ \hline
        Period vector $\boldsymbol{\psi}_{\mathcal{G}(\boldsymbol{l})}$ & $\leftrightarrow$ & Period vector $\boldsymbol{\psi}=\begin{scriptsize}\begin{pmatrix}
            \psi_2\\ \psi_1
        \end{pmatrix}\end{scriptsize}$\\ \hline
        Riemann inequality \eqref{reimRels}  & $\leftrightarrow$ &  $\textnormal{Im}(\tau)>0$ 
    \\ \hline
        Period map $\boldsymbol{l}\to \boldsymbol{\psi}_{\mathcal{G}(\boldsymbol{l})}$  & $\leftrightarrow$ & Period map $\boldsymbol{p}\to \boldsymbol{\psi}(\boldsymbol{p})$
    \\ \hline
        Group $\Gamma_{\mathcal{A}}$ &  $\leftrightarrow$ & Goup $\textnormal{SL}(2,\mathbb{Z})$ 
    \\ \hline
     $\mathcal{D}\simeq \mathbb{H}_2$ & $\leftrightarrow$ & $\Lambda\simeq\mathbb{H}=\left\{\left. \tau\in \mathbb{C}\right|\textnormal{Im}(\tau)= \frac{\tau-\overline{\tau}}{2i}>0\right\}$
        \\ \hline Space $\mathcal{K}$  & $\leftrightarrow$ & Fundamental cell $\mathbb{C}/\Lambda$\\ \hline $f(Y)>0$ & $\leftrightarrow$ &  $ad-cb=1$ 
\end{tabular}
}
\caption{Similarities between the modular interpretations of K3-surfaces and elliptic curves.}
\label{tab:K3ellCurve}
\end{table}

We now discuss some properties satisfied by the transcendental cycles and associated quantities. The transcendental cycles constructed in \cite{yoshida2013hypergeometric} and \cite{matsumoto1992monodromy} have a self-intersection matrix
\begin{equation}
\label{eq:Amat}
    \mathcal{A}(\boldsymbol{l})=[\Gamma(\boldsymbol{l}),\Gamma(\boldsymbol{l})]=-\frac{1}{2}\begin{pmatrix}
        U &0 & 0\\ 0&U & 0\\ 0&0&-\textbf{Id}
    \end{pmatrix} \quad \textnormal{where} \ U=\begin{pmatrix}
        0&1\\1&0
    \end{pmatrix},
\end{equation}
of signature $(-2,+4)$. We can think of $(-\mathcal{A})^{-1}$ as the analogue of the intersection form 
\begin{equation}
\label{ellCurveInForm}
    \begin{pmatrix}
        [\widecheck{\gamma}_5|\widecheck{\gamma}_5] & [\widecheck{\gamma}_5|\widecheck{\gamma}_4]\\
        [\widecheck{\gamma}_4|\widecheck{\gamma}_5] & [\widecheck{\gamma}_4|\widecheck{\gamma}_4]
    \end{pmatrix}=\begin{pmatrix}
        0&1\\-1&0
    \end{pmatrix},
\end{equation}
satisfied by the integral homology basis of the bare\footnote{By bare, we mean without marked points.} elliptic curve (c.f., figure \ref{fig:homologyCycles}).\footnote{{In both \eqref{eq:Amat} and \eqref{ellCurveInForm}, the square bracket of two cycles $C_1$ and $C_2$ is an integer obtained by summing over all intersection points $$[C_1|C_2]=\sum_{p\in C_1\cap C_2}\pm1.$$ This sum comes with a relative choice of orientation. Above, the prescription of  \cite{matsumoto1992monodromy} is understood.}} Moreover, it can be checked from the obvious equality and inequality 
\begin{equation}
    0=\langle \rho(\boldsymbol{l})|\rho(\boldsymbol{l})\rangle=\int_{\mathcal{G}(\boldsymbol{l})}\rho(\boldsymbol{l})\wedge\rho(\boldsymbol{l}) \quad \textnormal{and} \quad 0<\langle \rho(\boldsymbol{l})|\bar{\rho}(\boldsymbol{l})\rangle=\int_{\mathcal{G}(\boldsymbol{l})}\rho(\boldsymbol{l})\wedge\bar{\rho}(\boldsymbol{l}),
\end{equation}
{where $\bar{\rho}$ is Hodge-star conjugate to the differential $\rho$}, that the periods $\boldsymbol{\psi}_{\mathcal{G}(\boldsymbol{l})}$ obey the so-called \emph{Riemann relations} and \emph{Riemann inequality} \cite{springer2008introduction}
\begin{equation}
\label{reimRels}
  \boldsymbol{\psi}_{\mathcal{G}(\boldsymbol{l})}^\top \cdot \mathcal{A}^{-1} \cdot \boldsymbol{\psi}_{\mathcal{G}(\boldsymbol{l})}=0 \quad \textnormal{and} \quad  -\boldsymbol{\psi}_{\mathcal{G}(\boldsymbol{l})}^\top \cdot \mathcal{A}^{-1} \cdot \overline{\boldsymbol{\psi}}_{\mathcal{G}(\boldsymbol{l})}>0,
\end{equation}
respectively. The Riemann inequality is the K3 analogue of the condition $\textnormal{Im}(\tau)>0$ on the torus. Indeed, an explicit computation shows
\begin{equation}
    (\psi_2,\psi_1)\cdot\begin{pmatrix}
        [\widecheck{\gamma}_5|\widecheck{\gamma}_5] & [\widecheck{\gamma}_5|\widecheck{\gamma}_4]\\
        [\widecheck{\gamma}_4|\widecheck{\gamma}_5] & [\widecheck{\gamma}_4|\widecheck{\gamma}_4]
    \end{pmatrix}\cdot \begin{pmatrix}
        \bar{\psi}_2\\ \bar{\psi}_1\end{pmatrix}>0\iff \tau>\bar{\tau}\iff \textnormal{Im}(\tau)>0.
\end{equation}
For notational simplicity, let us from now on, write $(a,b)\in(\mathbb{C}^6)^2$ instead of $a^\top\cdot\mathcal{A}^{-1}\cdot b$. Since the cycles $\Gamma_{\bullet}$ depend continuously on $\boldsymbol{l}$, the period map
\begin{equation}
    \phi:\boldsymbol{l}\to \boldsymbol{\psi}_{\mathcal{G}(\boldsymbol{l})}=[\psi_{\mathcal{G}(\boldsymbol{l}),1}:\psi_{\mathcal{G}(\boldsymbol{l}),2}:...:\psi_{\mathcal{G}(\boldsymbol{l}),6}]\in\mathbb{CP}^5,
\end{equation}
is a multi-valued function. As dictated by the Riemann relations in \eqref{reimRels}, the image $\phi(\boldsymbol{l})$ is contained in the set
\begin{equation}
\label{domD0}
    \mathcal{D}_0=\left\{\boldsymbol{\xi}\in\mathbb{CP}^5~|~(\boldsymbol{\xi},\boldsymbol{\xi})=0,-(\boldsymbol{\xi},\overline{\boldsymbol{\xi}})> 0\right\}.
\end{equation}
However, $\mathcal{D}_0$ does not define a proper period domain: \emph{it is disconnected}. To see this, we plug the quadric equality
\begin{equation}
\label{eq:quadHyp}
    0=(\boldsymbol{\xi},\boldsymbol{\xi})=2\xi_1\xi_2+2\xi_3\xi_4-(\xi_5)^2-(\xi_6)^2,
\end{equation}
any $\boldsymbol{\xi}\in\mathcal{D}_0$ satisfies into the quadric inequality $-(\boldsymbol{\xi},\overline{\boldsymbol{\xi}})> 0$ and get
\begin{equation}
    0<4\textnormal{Im}(\xi_3)\textnormal{Im}(\xi_4)-2(\textnormal{Im}(\xi_5))^2-2(\textnormal{Im}(\xi_6))^2.
\end{equation}
By positivity of $(\textnormal{Im}(\xi_i))^2$, this implies both $\textnormal{Im}(\xi_3)/\xi_1$ and $\textnormal{Im}(\xi_4)/\xi_1$ are non-zero and share the same sign. The two disconnected components of $\mathcal{D}_0$ are then
\begin{equation}
\label{domD}
    \mathcal{D}_+=\left\{\boldsymbol{\xi}\in\mathbb{CP}^5~|~(\boldsymbol{\xi},\boldsymbol{\xi})=0,-(\boldsymbol{\xi},\overline{\boldsymbol{\xi}})> 0,\textnormal{Im}\left(\xi_3/\xi_1\right)>0\right\},
\end{equation}
and 
\begin{equation}
    \mathcal{D}_-=\left\{\boldsymbol{\xi}\in\mathbb{CP}^5~|~(\boldsymbol{\xi},\boldsymbol{\xi})=0,-(\boldsymbol{\xi},\overline{\boldsymbol{\xi}})> 0,\textnormal{Im}\left(\xi_3/\xi_1\right)<0\right\}.
\end{equation}
Without loss of generality, we restrict the image $\phi(\boldsymbol{l})$ of the period map to the \emph{connected} component $\mathcal{D}:=\mathcal{D}_+$. On the torus, this is analogous to the restriction $\textnormal{Im}(\tau)>0$.

Working on the physical chart $X=1$, let $\boldsymbol{x}=(x_1,x_0)\in\mathcal{G}(\boldsymbol{l}|_{X=1})$. Then, we define the bounded domain $\mathcal{K}$ as the set of points $\boldsymbol{\xi}(\boldsymbol{x})$ such that
\begin{equation}
\label{eq:genAM}
    \boldsymbol{\xi}(\boldsymbol{x})=\int_{\boldsymbol{x}_{\textnormal{ref}}}^{\boldsymbol{x}}\rho(\boldsymbol{x}') \quad \textnormal{mod} \ \boldsymbol{\psi}_{\mathcal{G}(\boldsymbol{l}|_{X=1})},
\end{equation}
for some reference point $\boldsymbol{x}_{\textnormal{ref}}\in\mathcal{G}(\boldsymbol{l}|_{X=1})$. In particular, \eqref{eq:genAM} is such that each $\boldsymbol{\xi}(\boldsymbol{x})$ is defined up to \emph{integer} multiples of periods. In many respects, $\mathcal{K}$ is the natural higher dimensional analogue to the fundamental parallelogram $\mathbb{C}/\Lambda_{(1,\tau)}$. Among other things, its definition comes with a \say{build-in} ambiguity similar to that of fixing $\tau$ to define $\mathbb{C}/\Lambda_{(1,\tau)}$. Indeed, while for $\mathbb{C}/\Lambda_{(1,\tau)}$ this ambiguity is given by the action of the modular group $\textnormal{SL}(2,\mathbb{Z})$ on the lattice $\Lambda_{(1,\tau)}$, for $\mathcal{K}$ it is given by the action on $\mathcal{D}$ of a subgroup of the automorphism group of \eqref{eq:quadHyp}. This subgroup (see \cite{matsumoto1992monodromy}) is
\begin{equation}
\label{congSubK3}
 \Gamma_{\mathcal{A}}:=\left\{Y\in\textnormal{GL}(6,\mathbb{Z})|(Y,Y)=\mathcal{A}^{-1},~f(Y)>0\right\},
\end{equation}
where the additional condition on $Y$
\begin{equation}
    f(Y):=(Y_{11}+Y_{21})(Y_{33}+Y_{43})-(Y_{31}+Y_{41})(Y_{13}+Y_{23})>0,
\end{equation}
is such that $\boldsymbol{\xi}\to Y\cdot \boldsymbol{\xi}$ preserves $\textnormal{Im}\left(\xi_3/\xi_1\right)>0$. It is therefore analogous to the condition $ad-cb=1$ on the torus.

It is worth mentioning that the inverse map $\mathcal{D}\to \mathcal{G}(\boldsymbol{l})$ can be written exclusively in terms of Siegel theta functions \cite{Matsumoto1993ThetaFO,matsumoto1992monodromy,yoshida2013hypergeometric,igusa2012theta}. This statement can be traced back to the isomorphism between $\mathcal{D}$ and the four-dimensional upper-half space of $\mathbb{H}_2$
\begin{equation}
    \mathbb{H}_2=\left\{\left. W=\begin{pmatrix}
        \tau_1&\tau_2\\ \tau_3& \tau_4
    \end{pmatrix}\in M(2,2)~\right| ~\frac{W-W^\dagger}{2i}>0\right\},
\end{equation}
where $M(p,q)$ denotes the space of complex $(p\times q)$-matrices and where \say{$\textnormal{matrix}>0$} means the matrix is hermitian and positive definite. For the explicit realization, see \cite[Ch. VIII, \S 9]{yoshida2013hypergeometric}. Incidentally, there is a well-defined generalization of modular forms on $\mathbb{H}_2$ \cite[Ch. VIII, \S 12.6]{yoshida2013hypergeometric}. It would be interesting to see if differential forms defined from generalized modular forms are enough to fully describe the function space of K3 Feynman integrals with four kinematic parameters (e.g., the three-loop four-mass sunrise). We make this more precise at the end of next subsection.

The various relations between the modular interpretations of K3-surfaces and elliptic curves discussed throughout this section are summarized in table \ref{tab:K3ellCurve}.

\subsubsection{Configuration space of four lines and a conic on the projective plane}

\begin{figure}
    \centering
\includegraphics[scale=0.3]{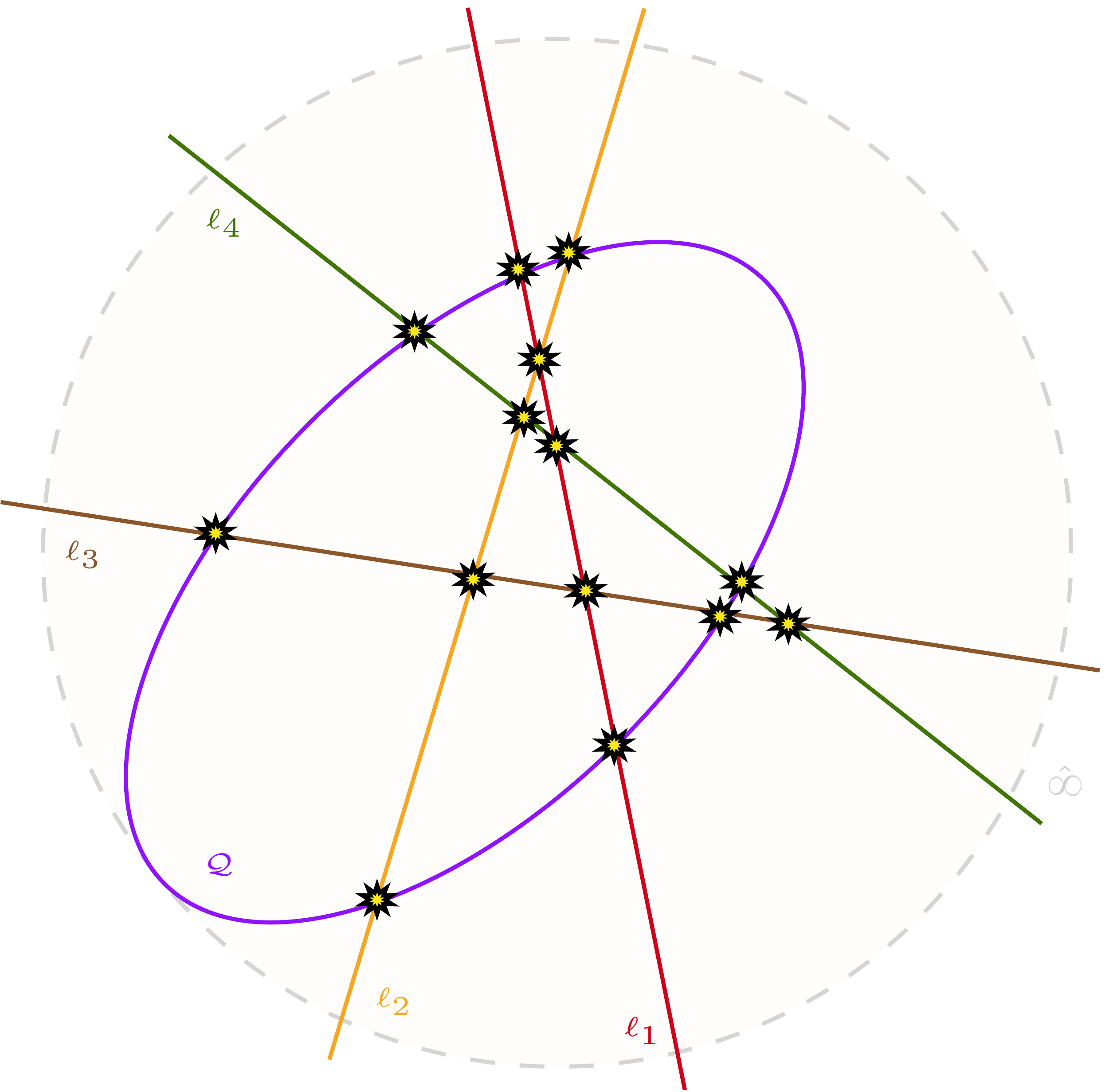}
    \caption{The configuration space of four lines and a quadric on the projective plane. There are 14 normal crossing points $P_{ij}$ (labelled by yellow stars). The shaded dashed circle represents the line at infinity, which is not in our configuration space. Non-degenerate configurations always seem to come with 11 bounded regions.}
    \label{fig:Our6Lines2}
\end{figure}

The focus of this subsection is the configuration space associated with the three-loop sunrise K3-surface \eqref{Wcurve}. This surface is described by an irreducible quintic polynomial \eqref{Wcurve} in $x_0$, with coefficients depending on $x_1$ as well as on the external kinematics. Equation \eqref{Wcurve} factorizes into the product of four \emph{generic} lines $\ell_{1\le i\le 4}$ and a \emph{degenerate} quadric $\mathcal{Q}$. Explicitly, we have 
\begin{equation}
\label{Wcurve2}
   G:\mathcal{W}_{x_1}^2(x_0)-\ell_1\ell_2\ell_3\ell_4 \mathcal{Q}=0.
\end{equation}
The four lines $\ell_{1\le i\le4}= 
(x_1 \ x_0 \ X) \cdot \alpha_{1\le i\le4}=0$ are given by
\begin{align}
    \alpha_1=\begin{pmatrix}
        0\\1\\-i X_3/X_0
    \end{pmatrix}, &\quad \alpha_2=\begin{pmatrix}
        0\\1\\i X_3/X_0
    \end{pmatrix},  \\ \alpha_3=-\begin{pmatrix}
        -2 X_0^{-4}\\2\\1+\left((X_2+X_3)^2-X_3^2-1\right) X_0^{-2}
    \end{pmatrix}, &\quad \alpha_4=-\begin{pmatrix}-2 X_0^{-4}\\2\\1+\left((X_2-X_3)^2-X_3^2-1\right) X_0^{-2}   \end{pmatrix}, \notag
\end{align}
and the quadric $\mathcal{Q}=(x_1 \ x_0 \ X)\cdot \boldsymbol{\beta}\cdot(x_1 \ x_0 \ X)^\top=0$ describes a parabola (see figure \ref{fig:Our6Lines2})
\begin{equation}
\label{eq:betaMat}
    \boldsymbol{\beta}=-\left(
\begin{array}{ccc}
 X_0^{-6} & 0 & 0 \\
 0 & 0 & 0 \\
 0 & 2 & 1-X_0^{-2} X_4^2 \\
\end{array} 
\right).
\end{equation}

A very important exercise is to make sure the configuration space
\begin{equation}
\label{eq: cfSpace}
    \ell_1\cup\ell_2\cup \ell_3\cup\ell_4\cup\mathcal{Q}\subset\mathbb{CP}^2.
\end{equation}
 has the same number of degrees of freedom than the three-loop \emph{four}-mass mass sunrise integral. Since we can rescale each line and the degenerate quadric individually, we have $4\times(3-1)=8$ parameters for the former plus $1\times(6-1-1)=4$ parameters for the latter and minus $8$ parameters coming from the finite subgroup of $\textnormal{GL}(3,\mathbb{C})$ preserving the lines. To see this in action, it suffices to find a $g\in \textnormal{GL}(3,\mathbb{C})$ and a $h_1\in (\mathbb{C}^\times)^4$ such that
\begin{equation}
    \boldsymbol{\alpha}'=g\cdot \boldsymbol{\alpha}\cdot h_1=g\cdot \begin{pmatrix}
        \alpha_1 & \alpha_2 & \alpha_3 & \alpha_4 
    \end{pmatrix}\cdot h_1=\begin{pmatrix}
        1&0&0&1\\0&1&0&1\\0&0&1&1
    \end{pmatrix},
\end{equation}
as well as a $h_2\in \mathbb{C}^\times$ such that
\begin{equation}
\label{eq:betaPrime}
    \boldsymbol{\beta}'=g\cdot\boldsymbol{\beta}\cdot g^\top\cdot h_2= \begin{pmatrix}\beta_{11}'&\beta_{12}'&\beta_{13}'\\\beta_{12}'&\beta_{22}'&\beta_{23}'\\\beta_{13}'&\beta_{23}'&1\end{pmatrix}.
\end{equation}

The non-trivial transformations are 
\begin{equation}
    g=\left(
\begin{array}{ccc}
 \frac{\left((X_2+X_3)^2-X_4^2-1\right) X_0^{4}-2 i X_4 X_0^{5}+X_0^{6}}{8 X_2 X_3} & -\frac{i
   X_0 X_4}{4 X_2 X_3} & \frac{X_0^{2}}{4 X_2 X_3} \\
 \frac{\left((X_2+X_3)^2-X_4^2-1\right) X_0^{4}+2 i X_4 X_0^{5}+X_0^{6}}{8 X_2 X_3} & \frac{i
   X_0 X_4}{4  X_2 X_3} & \frac{X_0^{2}}{4 X_2 X_3} \\
 \frac{X_0^{4}}{2} & 0 & 0 \\
\end{array}
\right),
\end{equation}
and 
\begin{equation}
     h_1=\textnormal{diag}\left(\frac{2 i  X_2 X_3}{X_0X_4} \ \frac{2X_2 X_3}{iX_0X_4} \ 1 \ 1 \right).
\end{equation}
Moreover, it is clear from \eqref{eq:betaMat} and \eqref{eq:betaPrime} that $\boldsymbol{\beta}'$ is constrained by $\det(\boldsymbol{\beta}')=0$. Thus, the \say{gauge fixed} configuration space is an open subset of $\mathbb{C}^4$ parameterized by five parameters $\beta_{11}', \beta_{12}', \beta_{13}', \beta_{22}'$ and $\beta_{23}'$ subject to that constraint. 

We checked explicitly
that the \say{gauged fixed} configuration space 
\begin{equation}
\label{eq: cfSpacePrime}
    \ell_1'\cup\ell_2'\cup \ell_3'\cup\ell_4'\cup\mathcal{Q}'\subset\mathbb{CP}^2,
\end{equation}
is in general position in the  three charts $\mathcal{U}_{1\le i \le 3}$ covering $\mathbb{CP}^2$ 
\begin{equation}
    \mathbb{CP}^2=\bigcup_{i=1}^3\mathcal{U}_i, \quad \textnormal{where} \quad \mathcal{U}_i=\{\boldsymbol{X}=(X,x_1,x_0)| \boldsymbol{X}_i\neq 0\},
\end{equation}
for most kinematic configurations.\footnote{One has to choose specific kinematics (such as vanishing masses) for the configuration space to degenerate.} Indeed, each line intersects each other once (giving $\binom{4}{2}=6$ intersection points), while each line intersects the quadric twice (giving $4\times 2=8$ intersection points). Thus, there are in total $6+8=14$ points of normal crossing. This is illustrated in figure \ref{fig:Our6Lines2} in the case where the lines and the quadric are real, which happens for purely imaginary $X_0$.

As a sanity check, we compute the Euler characteristic of $G$ interpreted as the double-cover of $\mathbb{CP}^2$ branched over \eqref{eq: cfSpacePrime}. If we denote union of lines by $\boldsymbol{\ell}'=\bigcup_{i=1}^4\ell_i'$, then the Euler characteristic is given by
\begin{equation}
\label{confSpaceEulCar}
\begin{split}
    \chi_{\textnormal{E}}(G)&=[\underbrace{2(\chi_{\textnormal{E}}(\mathbb{CP}^2)-\chi_{\textnormal{E}}(\boldsymbol{\ell}'\cup \mathcal{Q}'))}_{\textnormal{double cover}}+\underbrace{ \chi_{\textnormal{E}}(\boldsymbol{\ell}'\cup \mathcal{Q}')}_{\textnormal{branching}}]+\underbrace{14 (\chi_{\textnormal{E}}(\mathbb{CP}^1)- \chi_{\textnormal{E}}(\bullet))}_{\textnormal{blow-up}}\\&=[2(3-(-4))-4]+14\\&=24.
\end{split}
\end{equation}
which matches \eqref{eq:ECk3}. In deriving \eqref{confSpaceEulCar}, we used
\begin{equation}
    \chi_{\textnormal{E}}(\boldsymbol{\ell}'\cup \mathcal{Q}')=4\chi_{\textnormal{E}}(\mathbb{CP}^1)+\chi_{\textnormal{E}}(\mathcal{Q}')-14\chi_{\textnormal{E}}(\bullet)=4\times 2+2-14=-4,
\end{equation}
The equality $\chi_{\textnormal{E}}(\mathcal{Q})=2$ follows from two observations: (\emph{i}) Every irreducible quadric in $\mathbb{CP}^n$ is birational to $\mathbb{CP}^{n-1}$ \cite{harris2013algebraic} and, (\emph{ii}) the Euler characteristic is a birational invariant, which can be shown in a variety of ways \cite{hirzebruch1966topological, maakestad2019birational,van1983birationale,chatzistamatiou2012higher}.

Next, we count the number of non-algebraic and independent cycles on \eqref{Wcurve2}. Out of the 22 closed cycles, there are $14+1+1=16$ algebraic ones coming, respectively, from the blow-ups of the points of normal-crossing $P_{ij}$, the birational transformation (blow-up) $\mathcal{Q}\dashrightarrow\mathbb{CP}^1$ and the pullback of the hyperplane class $H$ as defined in \eqref{eq:Hclass}.

Naively, since the K3-surface in \eqref{Wcurve2} admits an elliptic fibration \eqref{eq:WformSurface}, the construction of these six non-algebraic cycles should go along the same lines than in \cite{matsumoto1992monodromy} (see also \cite{jorgK3} as well as the construction of \say{loaded cycles} in \cite{yoshida2013hypergeometric}). It would be interesting to investigate relations between the period domain automorphism group for $G$ and \eqref{congSubK3}. In the case they match, this would hint generalized modular forms on $\mathbb{H}_2$ \cite[Ch. VIII, \S 12.6]{yoshida2013hypergeometric} are the appropriate set of differential forms to consider for the three-loop four-mass sunrise.

\subsection{Towards a (modular invariant) three-loop basis} \label{SEC:3loopBasis}


In this section, we initiate the construction of the four-mass three-loop basis. 
Within this section, the three-loop fibre space is denoted by $\mathcal{F}$, while the three-loop base space is denoted by $\mathcal{B}$ (notice the use of \emph{calligraphic} font). 
From the loop-by-loop construction, it follows that $\mathcal{F}$ is the base space $B$ of the two-loop sunrise.

It is useful to recall the noteworthy requirements for the construction of a \say{good} two-loop basis. 
First we required the fibre basis to satisfy a differential equation \eqref{omegaF} such that \eqref{eq:nabla base} is in $\vep$-form  (c.f., section \ref{sec:2loopfibBasis}). This forced us to  renormalize the one-loop basis presented in \cite{Caron-Huot:p1} by a factor of the dual base twist $\widecheck{u}_B^{-1}|_{\varepsilon\to 0}$. The resulting basis is found in \eqref{fibBasis}.
Secondly, we required the base basis to be such that the sv-constraint was satisfied (c.f., section \ref{SEC:lblDE}). Since the fibre forms are fixed at this point, this is a constraint purely on the base forms. In particular, it forced \eqref{baseTads} and \eqref{mcBasis2New} to contain square root normalizations designed to make the fibre-base wedge product single-valued. Although this constraint alone was not sufficient to uniquely fix the base basis, it was a necessary one.

At three-loop, the first step is essentially the same: we require the fibre basis $\{ \wc\vphi_{\mathcal{F},a} \}$ to satisfy a differential equation such that the resulting covariant derivative on the base is in $\vep$-form (i.e., $\wc{\boldsymbol{\nabla}}_{\mathcal{B}}\vert_{\vep=0} = \boldsymbol{d}$).
If we do not renormalize the two-loop basis $\{\wc\vth_a\}$ in \eqref{eq:2loopSunBasis}, the resulting covariant derivative on the base would not be in $\varepsilon$-form
\begin{equation}
\label{eq:badCovDer}
\wc{\boldsymbol{\nabla}}_{\mathcal{B}}=\bs{d}+\left(\varepsilon\wc{\boldsymbol{\Gamma}}+\textbf{Id}~\ed\log\left[\wc{u}_\mathcal{B} \right] \right)=\boldsymbol{d}+\left( 
        \vep \wc{\bs{\Gamma}}
        + \mathbf{Id}\ \ed\log\left[
            \sqrt{p^2}(\ell_{0,\perp}^2)^{\vep+\frac12}
        \right] 
   \right).
\end{equation}
To bring \eqref{eq:badCovDer} into $\vep$-form, we define a new two-loop basis 
\begin{align} \label{eq:3loopfibre}
    \wc{\bs{\vphi}}_{\mathcal{F}} 
    = \frac{ \wc{\bs{\vth}} }{ (\wc{u}_\mathcal{B}\vert_{\vep\to0}) }
    = \frac{ \wc{\bs{\vth}} }{ \sqrt{p^2 \ \ell_{0,\perp}^2} },
\end{align}
that satisfies a differential equation
\begin{align}
\varepsilon\wc{\boldsymbol{\Gamma}}\to \varepsilon\wc{\boldsymbol{\Gamma}}- \mathbf{Id}\ \ed\log\left[
            \sqrt{p^2}(\ell_{0,\perp}^2)^{\frac12}
        \right],
\end{align}
such that \eqref{eq:badCovDer} becomes
\begin{equation}
    \widecheck{\boldsymbol{\nabla}}_{\mathcal{B}}\to \widecheck{\boldsymbol{\nabla}}_{\mathcal{B}}
    = \boldsymbol{d}+\widecheck{\boldsymbol{\varpi}}~\wedgedot,
    \qquad
    \widecheck{\boldsymbol{\varpi}}
    := 
    \vep \left(
        \wc{\bs{\Gamma}} 
        + \ed\log(\ell_{0,\perp}^2) \textbf{Id}
    \right)\xrightarrow{\varepsilon\to 0}0.
\end{equation}
The second step is to construct a 15-dimensional three-loop basis 
(4 triple-tadpoles and 11 maximal-cut elements)\footnote{On the third loop, the two-loop modular parameters $\tau, z_1, z_2$ in \eqref{epsCon} become functions of the loop variables $x_0$ and $\ell_{0,\perp}^2$. The number of critical points is therefore obtained by counting, on each boundary, the number of elements in the zero set of $\widecheck{\boldsymbol{\varpi}}$ diagonal block determinants. While this exercise was particularly easy at two-loop, it isn't at three-loop for two main reasons. First, due to the matrix nature of the maximal-cut block and, second, because the function space is much more complicated. Following \cite{Mizera:2019vvs}, we can however easily compute the number of critical points on each cuts and find that there are four tadpoles and 11 maximal-cut  elements.}
\begin{align}
\label{eq:lbl3loopform}
    \wc{\vphi}_{3\textnormal{-loop},a} 
    = \wc{\vphi}_{\mathcal{F},b}
    \wedge \widecheck{\vphi}_{\mathcal{B},b a}
    = \widecheck{u}_\mathcal{B}^{-1}|_{\varepsilon\to 0}~\wc{\vth}_{b}
    \wedge \widecheck{\vphi}_{\mathcal{B},ba}
    = \widecheck{u}_\mathcal{B}^{-1}|_{\varepsilon\to 0}~\wc{\vphi}_{F,c} 
    \wedge \wc{\vth}_{B,cb}
    \wedge \widecheck{\vphi}_{\mathcal{B},ba},
\end{align}
 such that the square root normalizations and the non-trivial modular properties on the fibre cancel by wedging with the $\{\widecheck{\vphi}_{\mathcal{F},a}\}$.

 As a proof-of-concept, we discuss the schematic form of $\{\widecheck{\boldsymbol{\varphi}}_{\mathcal{B},i}\}_{i=1}^4$ producing algebraic and modular invariant triple-tadpoles. The first three of these forms on the base space $\mathcal{B}$ have a simple structure
 \begin{gather}
    \widecheck{\boldsymbol{\vphi}}_{\mathcal{B},1}
    =\delta_0\wedge\textnormal{d}\log(h_1)~(1~0~0~0~0~0~0)^\top,\\
    \widecheck{\boldsymbol{\vphi}}_{\mathcal{B},2}
    =\delta_0\wedge\textnormal{d}\log(h_1)~(0~1~0~0~0~0~0)^\top,\\
    \widecheck{\boldsymbol{\vphi}}_{\mathcal{B},3}
    =\delta_0\wedge\textnormal{d}\log(h_1)~(0~0~1~0~0~0~0)^\top,
\end{gather}
where the dlog form is modular invariant and cancels the square root normalization of $\wc{\bs{\vphi}}_\mathcal{F}$.
With \eqref{eq:3loopfibre} in mind, a simple choice for $h_1$ is
\begin{equation}
 h_1 = \frac{1-\frac{ix_0}{\sqrt{\ell_{0,\perp}^2\vert_0/m_1^2}}}{1+\frac{ix_0}{\sqrt{\ell_{0,\perp}^2\vert_0/m_1^2}}}.
\end{equation}
The fourth triple-tadpole is produced by wedging a bulk form on $\mathcal{B}$ with the maximal-cut  forms $\wc{\vphi}_{\mathcal{F},4\le a \le 7}$ on the fibre. Since the $\wc{\vphi}_{\mathcal{F},4\le a \le 7}$ have non-trivial modular properties, $  \wc{\bs{\vphi}}_{\mathcal{B},4}$ is constrained by modular invariance of \eqref{eq:lbl3loopform}. To see this, first set
\begin{equation}
    \boldsymbol{C}=(0~0~0~c_4C_{44}~C_{45}~C_{46}~c_4C_{47})^\top,
\end{equation}
where the $C_{\bullet \bullet}$'s contain no square root and $c_4$ is defined in \eqref{eq:ellQUantities}. Then, define the bulk form to be
\begin{equation}
\label{bulkFormOn3rdloop}
      \widecheck{\boldsymbol{\vphi}}_{\mathcal{B},4}
      = \theta_0 ~ \textnormal{d} \log(h_2) \wedge \textnormal{d} \log(h_3) ~ \boldsymbol{C},
\end{equation}
 where $h_{3}$ and $h_{4}$ are defined to be modular functions such that $\textnormal{d}\log(h_2)\wedge\textnormal{d}\log(h_3)$ cancels the square root normalization of $\widecheck{\vphi}_{\mathcal{F},4\le i\le 7}$. A 
straightforward choice would be
 \begin{equation}
     h_2=\frac{1-\sqrt{\ell_{0,\perp}^2/m_1^2}}{1+\sqrt{\ell_{0,\perp}^2/m_1^2}} \quad \textnormal{and} \quad h_3=x_0.
 \end{equation}
 Furthermore, since $\widecheck{\boldsymbol{\vphi}}_{\mathcal{F}}$ contains \textcolor{blue}{factors} that are manifestly modular invariant, it is clear that the triple-tadpole 
 \begin{equation}
 \label{eq:bulkTad3loop}
 \begin{split}
      \widecheck{\vphi}_{3\textnormal{-loop},4}
      &= \widecheck{\boldsymbol{\vphi}}_{\mathcal{F}}~\wedgedot~\widecheck{\boldsymbol{\vphi}}_{\mathcal{B},4}
      \\
      &= \widecheck{u}_\mathcal{B}^{-1}|_{\varepsilon\to 0}~\widecheck{\bs{\vphi}}_F 
    ~\wedgedot~ \widecheck{\bs{\vartheta}}_{B}~\wedgedot~\widecheck{\boldsymbol{\vphi}}_{\mathcal{B},4}
    \\
    &= \textcolor{blue}{\theta_0~\widecheck{u}_\mathcal{B}^{-1}|_{\varepsilon\to 0}}~\textcolor{blue}{\textnormal{d}\log(h_2)}\wedge\textcolor{blue}{\textnormal{d}\log(h_3)}\wedge\textcolor{blue}{\widecheck{\bs{\vphi}}_{F}}~\wedgedot~\Lambda,
 \end{split}
 \end{equation}
 is modular invariant if and only if  
\begin{equation}
\label{eq:Lambda}
    \begin{split}
        \Lambda&:=\widecheck{\boldsymbol{\vartheta}}_B\cdot\boldsymbol{C}\\&=c_4C_{44}\left[\widecheck{\boldsymbol{\varphi}}_{B,4}+v_{54}~\widecheck{\boldsymbol{\varphi}}_{B,5}+v_{64}~\widecheck{\boldsymbol{\varphi}}_{B,6}+\left(v_{74}^{\textnormal{(I)}}+v_{74}^{\textnormal{(II)}}+u_{75}v_{54}+u_{76}v_{64}\right)\widecheck{\boldsymbol{\varphi}}_{B,7}\right]\\&\quad +C_{45}\left[\widecheck{\boldsymbol{\varphi}}_{B,5}+(u_{75}+v_{75})\widecheck{\boldsymbol{\varphi}}_{B,7}\right]+C_{46}\left[\widecheck{\boldsymbol{\varphi}}_{B,6}+(u_{76}+v_{76})\widecheck{\boldsymbol{\varphi}}_{B,7}\right]+c_4C_{47}\widecheck{\boldsymbol{\varphi}}_{B,7}.
    \end{split}
\end{equation}
is. It is also apparent from \eqref{eq:Lambda} and appendix \ref{SEC:details} that $\Lambda$ is multi-valued/non-algebraic for \emph{generic} choices of $C$'s through its dependence on $\partial_\tau X_i$. Below, we will look for $C_{\bullet\bullet}$'s such that this dependence drops out.

Before proceeding with the derivation of the constraints, it is useful to observe that the right-hand side (and so the
left-hand side) of
\begin{equation}
\label{eq:rescaleShift}
    \psi_1\widecheck{\nabla}_0\widecheck{\boldsymbol{\varphi}}_{B,7}+(\partial_0\psi_1)\widecheck{\boldsymbol{\varphi}}_{B,7}=\widecheck{\nabla}_0(\psi_1~\widecheck{\boldsymbol{\varphi}}_{B,7}),
\end{equation}
is modular invariant. Applying to \eqref{eq:rescaleShift} the rescaling
\begin{equation}
    C_{44}\to \frac{\pi\varepsilon W_0}{\psi_1}C_{44} \quad \textnormal{followed by the shift} \quad C_{47}\to C_{47}+(\partial_0\psi_1)C_{44},
\end{equation}
the term $ C_{44}\widecheck{\boldsymbol{\varphi}}_{B,4}+C_{47}\widecheck{\boldsymbol{\varphi}}_{B,7}\subset \Lambda$ becomes $C_{44}\widecheck{\nabla}_0(\psi_1~\widecheck{\boldsymbol{\varphi}}_{B,7})+C_{47}~\widecheck{\boldsymbol{\varphi}}_{B,7}$.
Fixing the constraints to be
\begin{equation}
    \begin{cases}
      C_{47}=-C_{44}\frac{\pi\varepsilon W_0}{\psi_1}\left(v_{74}^{\textnormal{(I)}}+v_{74}^{\textnormal{(II)}}+u_{75}v_{54}+u_{76}v_{64}\right)\\ \quad \quad \quad \quad -c_4^{-1}C_{45}(u_{75}+v_{75})-c_4^{-1}C_{46}(u_{76}+v_{76}),\\
       C_{45}=-c_4C_{44}\frac{\pi\varepsilon W_0}{\psi_1}v_{54},
       \\
       C_{46}=-c_4C_{44}\frac{\pi\varepsilon W_0}{\psi_1}v_{64},
    \end{cases}
\end{equation}
we have that
\begin{equation}
    \Lambda=c_4C_{44}\widecheck{\nabla}_0(\psi_1~\widecheck{\boldsymbol{\varphi}}_{B,7}).
\end{equation}
Since $\Lambda$ is manifestly modular invariant for $C_{44}$ a modular function, so is the algebraic  three-loop tadpole form in \eqref{eq:bulkTad3loop}.

In principle, the procedure outlined in this section can be used as a guideline to produce constraints for an algebraic and modular invariant basis on the maximal-cut. We leave the resolution of the maximal-cut constraints for future work.
\section{Conclusion}

Motivated by the ``naive'' expectation that multi-loop Feynman integrals are an iteration of simple one-loop problems, we introduced a loop-by-loop technique for computing the differential equation for families of multi-loop \emph{dual}
Feynman integrands.  
Through an inner product called the intersection number, dual Feynman integrands are in one-to-one correspondence with ``regular'' Feynman integrands. 
In particular, dual Feynman integrands are simpler since they only have support on generalized unitarity cuts. 
Moreover, the differential equation for a basis of Feynman integrands is simply related to the associated dual differential equation (they are equivalent up to a sign and possible transpose depending on conventions). 
Thus, studying dual differential equations is equivalent to studying the differential equations of Feynman integrands. 

The main advantage of working with dual Feynman integrands is two-fold. 
First, dual integration-by-parts is simpler since dual Feynman integrands are localized to generalized unitary cuts and propagators cannot be squared. 
Secondly, but more importantly, the localization to generalized unitarity cuts helps inform a good choice of basis. 
Since the geometry associated to a Feynman integrand is hidden in its cuts, dual integrands help make the associated geometry apparent.
Then, since dual integrands do not have to ``look like'' Feynman integrands, we can choose a basis purely motivated by the underlying geometry.

The loop-by-loop fibration introduced here is applicable when the space of all loop-momentum $X$ locally looks like the Cartesian product of two subspaces: a fibre $F$ characterized by the loop momentum of a sub-diagram and a base $B$ defined as the orthogonal complement of $F$ in $X$. 
At the level of the differential forms, this fibration induces a splitting of $(p+q)$-forms on $X$ into a wedge product of $q$-forms defined on $F$ together with $p$-forms defined on $B$. 
This splitting can be used to compute the differential equation on $X$ \say{one-loop-at-a-time}. 
One first computes the differential equation for a basis on $F$, which in turn, gives rise to a covariant derivative on the base space $B$.
If the base space contains only one loop momenta, one can simply compute the differential equation for a basis on $B$ to get the differential equation on $X$.  
On the other hand, if the base space contains more than one loop momenta, one divides $B$ into a new base and fibre. 
Each step is repeated until there are no remaining loop-momenta left. 

As a simple yet non-trivial example, we constructed an $\vep$-form dual basis for the three-mass two-loop \emph{elliptic} sunrise family and the associated differential equation \eqref{epsCon}.

Breaking up the problem into simpler one-loop problems yields several advantages. 
First, we can reuse the known $\vep$-form basis and differential equation at one-loop to construct the two-loop basis and differential equation. 
Second, at each step only a small subset of variables are active on the fibre simplifying algebraic manipulations as well as integration-by-parts.

It was important that the fibre basis had a differential equation such that the resulting covariant derivative on the base is in $\vep$-form (i.e., $\boldsymbol{\wc{\nabla}}_B\vert_{\vep=0} =\boldsymbol{d}$).
This fixed the normalization of the fibre forms and helped motivate a good basis on the base. 
Since the wedge product of the fibre and base bases must be single-valued and the $\vep$-form basis on the fibre has normalizations that are multi-valued on the base, we were forced to include  monodromy canceling factors in the base basis. 
These monodromy canceling factors actually help geometrically motivate a good starting choice of base basis. 

While the starting base basis \eqref{mcBasis2New} is geometrically well motivated, it does not have an $\vep$-form differential equation. 
Two gauge transformations are needed to bring the system into $\vep$-form. 
These gauge transformations are entirely fixed by the modular properties of the differential equation and the resulting $\vep$-form basis is presented in \eqref{2loopBaseBasis}. 
Similarly, we bootstrap the $\vep$-form differential equation on the torus using only its modular properties. 
The result is a \say{ready-to-integrate} $\varepsilon$-form differential equation written in terms of modular and Kronnecker forms on the torus \eqref{epsCon}.

Importantly, our analysis sidesteps the need for $q$-series analysis. 
We believe that the gauge transformation procedure and modular bootstrap developed here will help streamline and systematize the process of constructing $\vep$-form differential equations with elliptic geometry. 

Since the differential equation \eqref{epsCon} is written in terms of modular and Kronnecker forms, the elliptic symbol letters are easily read off. 
However, a direct comparison with \cite{Wilhelm:2022wow}, requires integrating \eqref{epsCon} with the proper initial conditions. 
We leave such analysis for future work. 

Surprisingly, there is also a simple relation between our $\vep$-form basis of dual forms and the $\vep$-form basis of Feynman integrands constructed in \cite{Bogner:2019lfa}. Up to the trivial rescaling discussed in section \ref{SEC:feynmanBasis}, our dual basis is dual to their basis of Feynman integrands.
We interpret this unexpected relation as a sign that there are probably not many different \say{good} basis for the sunrise topology (it is highly constrained).

Then, the underlying K3-surface associated to the four-mass three-loop sunrise topology is identified in section \ref{SEC:K3banana}. 
We also give an interpretation of this K3-surface as the configuration space of a conic and four lines in generic positions on the projective plane. 
While we leave any attempt to properly define the relevant non-algebraic cycles as future work, we think that the configuration space interpretation of the K3-surface will make the analysis more tractable. 
Proper answers to these questions are necessary to fully understand the function space at three-loop. 

Since this work serves as a proof of concept, it would be interesting to test our methods on either three-loop examples or two-loop higher multiplicity examples. 
It would also be interesting to understand how the choice of fibre effects the difficulty of intermediate steps when studying less symmetric examples.
When the underlying function space is well understood (i.e., elliptic or polylogarithmic) we expect our methods to be quite efficient. 
Some natural next examples are the generic mass kite and ice cone diagrams, the two-loop non-planar EW form factor, or the massive double box in Bhabha scattering. 

\acknowledgments
We first thank our supervisor Simon Caron-Huot for his encouragement and support during different stages of this project. We would also like to thank David Broadhurst, Vsevolod Chestnov, Claude Duhr, Albrecht Klemm, Hjalte Frellesvig, Christoph Nega, Sebastian Pögel, Franziska Porkert, Lorenzo Tancredi, Christian Vergu, Stefan Weinzierl, Matthias Wilhelm, Chi Zhang, Anastasia Volovich and Marcus Spradlin for valuable discussions.
M.G. is grateful for the hospitality offered at Bonn University and the Niels Bohr Institute during the final stages of this work. M.G. would also like to express special thanks to the organizers of the conference \say{\emph{Elliptic Integrals in Fundamental Physics}} held at the Mainz Institute for Theoretical Physics (MITP) of the Cluster of Excellence PRISMA+ (Project ID 39083149), where part of this work was presented. M.G.’s work is supported by the National Science and Engineering
Council of Canada (NSERC) and the Canada Research Chair program.
A.P. is grateful for support provided by NSERC, the Fonds de Recherche du Québec - Nature et Technologies
and by the Simons Investigator Award $\#$376208 of A. Volovich.

\appendix

\section{Review of classical elliptic functions}
\label{SEC:revClassicalFunctions}

In this appendix, we list our conventions for the classical elliptic functions. 
  
Theta functions are periodic functions that can be represented by series whose convergence is extraordinarily rapid. 
They provide an important means for numeric evaluation of elliptic functions. 
Moreover, theta functions are useful for the solving algebraic equations involving elliptic functions. 
For theta functions, we use the conventions of \cite{rademacher2012topics}. 
We list, for $q=\exp(i\pi \tau)$ and $\tau\in\mathbb{H}$, all the Jacobi theta functions
\begin{gather}
       \theta_{1}\!\left(\zeta , \tau\right) = 2 {e}^{\pi i \tau / 4} \sum_{n=0}^{\infty} {\left(-1\right)}^{n} {q}^{n \left(n + 1\right)} \sin\!\left(\left(2 n + 1\right) \pi\zeta \right),\\
        \theta_{2}\!\left(\zeta , \tau\right) = 2 {e}^{\pi i \tau / 4} \sum_{n=0}^{\infty} {q}^{n \left(n + 1\right)} \cos\!\left(\left(2 n + 1\right) \pi\zeta \right),\\
        \theta_{3}\!\left(\zeta , \tau\right) = 1 + 2 \sum_{n=1}^{\infty} {q}^{{n}^{2}} \cos\!\left(2 n  \pi\zeta \right),\\
           \theta_{4}\!\left(\zeta  , \tau\right) = 1 + 2 \sum_{n=1}^{\infty} {\left(-1\right)}^{n} {q}^{{n}^{2}} \cos\!\left(2 n \pi\zeta \right).
\end{gather}
The factor $q^{n^2}$ usually leads to rapid convergence of these series. 
  
Equally important are Weierstra\ss\ functions. 
The so-called  Weierstra\ss\ $\wp$-function is doubly periodic and defined as
\begin{equation}
\label{wpFunct}
\begin{split}
      \wp\!\left(z, \tau\right) &= \frac{1}{{z}^{2}} + \sum_{\left(m, n\right) \in {\mathbb{Z}}^{2} \setminus \left\{\left(0, 0\right)\right\}} \frac{1}{{\left(z + m + n \tau\right)}^{2}} - \frac{1}{{\left(m + n \tau\right)}^{2}} \\&= {\left(\pi \theta_{2}\!\left(0 , \tau\right) \theta_{3}\!\left(0 , \tau\right) \frac{\theta_{4}\!\left(z , \tau\right)}{\theta_{1}\!\left(z , \tau\right)}\right)}^{2} - \frac{{\pi}^{2}}{3} \left(\theta_{2}^{4}\!\left(0, \tau\right) + \theta_{3}^{4}\!\left(0, \tau\right)\right),
\end{split}
\end{equation}
for $ z \in \mathds{C}, \tau \in \mathbb{H}$. 
Then, the Weierstra\ss\ $\zeta$-function is a primitive for the $\wp$-function 
\begin{equation}
    \frac{\textnormal{d}}{\textnormal{d} z}\, \zeta\!\left(z, \tau\right) = -\wp\!\left(z, \tau\right) \iff \zeta(z)=z^{-1}-\int_{0}^{z}\textnormal{d}w(\wp(w)-w^{-2}).
\end{equation}
This gives a convenient way to express the integral of the $\wp$-function
\begin{equation}
\label{intwp}
    \int_{0}^{z}\textnormal{d}w\wp(w)=-\zeta(z)+z^{-1}+\int_{0}^{z}\textnormal{d}ww^{-2}.
\end{equation}
With a bit of work, one can work out the following relation to the Jacobi theta functions
\begin{equation}
    \zeta\!\left(z, \tau\right) = -\frac{z}{3} \frac{\theta'''_{1}\!\left(0 , \tau\right)}{\theta'_{1}\!\left(0 , \tau\right)} + \frac{\theta'_{1}\!\left(z , \tau\right)}{\theta_{1}\!\left(z , \tau\right)}.
\end{equation}
Moreover, the series representation is given by
\begin{equation}
    \zeta\!\left(z, \tau\right) = \frac{1}{z} + \sum_{\left(m, n\right) \in {\mathbb{Z}}^{2} \setminus \left\{\left(0, 0\right)\right\}} \frac{1}{z - m - n \tau} + \frac{1}{m + n \tau} + \frac{z}{{\left(m + n \tau\right)}^{2}}.
\end{equation}
The $\wp$- and $\zeta$-functions have the following useful analytic properties:
\begin{enumerate}
    \item $\mathop{\operatorname{poles}\,}\limits_{z \in \mathbb{C}} \wp\!\left(z, \tau\right) = \Lambda_{(1, \tau)}$ (double pole at zero),
    \item $\mathop{\operatorname{poles}\,}\limits_{z \in \mathbb{C}} \zeta\!\left(z, \tau\right) = \Lambda_{(1, \tau)}$ (only simple poles),
    \item $\mathop{\operatorname{zeros}\,}\limits_{z \in \mathbb{C}} \wp\!\left(z, i\right) = \left\{ \left(m + \frac{1}{2}\right) + \left(n + \frac{1}{2}\right) i : m \in \mathbb{Z} \;\mathbin{\operatorname{and}}\; n \in \mathbb{Z} \right\}$,
    \item $\wp\!\left(z, \tau\right) \text{ is holomorphic on } z \in \mathbb{C} - \Lambda_{(1, \tau)}$,
    \item $\zeta\!\left(z, \tau\right) \text{ is holomorphic on } z \in \mathbb{C}- \Lambda_{(1, \tau)}$.
\end{enumerate}
Clearly, $\zeta(z)$ is parity-odd -- i.e., $\zeta(-z)=-\zeta(z)$.

\section{Properties of generic quartic elliptic curve}
\label{SEC:GenericEllCurve}


In this appendix, we review some basic facts about elliptic curves. 
Much of the material covered here is borrowed from \cite{Broedel:2017kkb, Broedel:2018iwv, farkas1992riemann, mckean_moll_1997}. Conventions in this appendix are independent to those in the body of this paper.

To start, consider a family $E(\mathbb{C})$ of quartic elliptic curves defined by the irreducible polynomial equation with unspecified roots $e_i$
\begin{equation}
\label{qec}
    E(\mathbb{C}): y^2-(x-e_1)(x-e_2)(x-e_3)(x-e_4)=0.
\end{equation}
Assuming the canonical ordering $-\infty<e_{1}<e_{2}<e_{3}<e_{4}<\infty$ and  choosing the positive branch, we set
\begin{equation}
    \sqrt{y^2}=+\sqrt{|y^2|}\times\begin{cases}
      1& x\le e_1, \\ i& e_1<x\le e_2,\\ 1& e_2<x\le e_3,\\ i& e_3<x\le e_4,\\ 1& e_4<x.
    \end{cases}
\end{equation}
With this convention, the periods of the (framed) \emph{elliptic curve} can be written as
\begin{equation}
\label{periods1}
    \psi_1=2c_4\int_{e_2}^{e_3}\frac{\textnormal{d}x}{y}=2K(k) 
    \quad \textnormal{and} \quad 
    \psi_2=2c_4\int_{e_2}^{e_1}\frac{\textnormal{d}x}{y}=2iK(1-k),
\end{equation}
where $K$ stands for the complete elliptic integrals of $1^\textnormal{st}$ kind and
\begin{equation}
    k:=\frac{e_{14}e_{23}}{e_{13}e_{24}}\in(0,1) 
    \quad \textnormal{and} \quad 
    c_4:=\frac{1}{2}\sqrt{e_{13}e_{24}}>0,
\end{equation}
such that $\tau:=\psi_2/\psi_1\in\mathbb{H}$. In particular, given $k\in(0,1)$, we have $\tau\in i\mathbb{R}_+$. Working with $\{1,\tau\}$ instead of $\{\psi_1,\psi_2\}$ gets rid of the $\textnormal{GL}(1,\mathbb{C})$ normalization ambiguity. 

For completeness, we may also introduce the quasi-periods of the differential form 
\begin{equation}
\label{phiHat}
    \textnormal{d}x\hat{\Phi}_4(x,y):=-\frac{\psi_1^2 \textnormal{d}x}{4c_4^2 y}\left(x^2-\frac{s_1}{2}x+\frac{s_2}{6}\right),
\end{equation}
which has zero residue by design and where $s_n:=s_n(e_1,e_2,e_3,e_4)$ and (later used) $\Bar{s}_n:=s_n(e_2,e_3,e_4)$ are two elementary symmetric polynomials in four and three variables respectively. They are given by combinations of complete elliptic integrals of $1^\textnormal{st}$ and $2^\textnormal{nd}$ kinds
\begin{equation} \label{eq:normalized quasi-period 1}
    \phi_1':=\frac{c_4}{\psi_1}\int_{e_2}^{e_3}\textnormal{d}x \hat{\Phi}_4(x,y)=\psi_1\times \left(E(k)-\frac{1}{3} (2-k)
   K(k)\right),
\end{equation}
and
\begin{equation}
\label{quasiPer}
     \phi_2':=\frac{c_4}{\psi_1}\int_{e_2}^{e_1}\textnormal{d}x \hat{\Phi}_4(x,y)=\psi_1\times \left(\frac{1}{3} i (k +1) K(1-k
   )-i E(1-k )\right).
\end{equation}
The periods and quasi-periods are not independent, but they are related through the \emph{Legendre's relation}
\begin{equation}
\label{LegendreRelTau}
    \phi_2'-\tau\phi_1'=-i\pi.
\end{equation}
  The periods and quasi-periods are strictly speaking not invariants of $E(\mathbb{C})$, since there may be different values of $\{1,\tau\}$ and $\{\phi_1',\phi_2'\}$ that correspond to the same elliptic curve.\footnote{For example, we can perform a global rescaling of the periods without changing the geometry.}
  
  There is however an invariant, called the \emph{$j$-invariant}, that uniquely characterizes an elliptic curve. For quartic elliptic curve, it is given by
\begin{equation}
\label{eq:jInv}
    j=256\frac{(1-k(1-k))^3}{k^2(1-k)^2}.
\end{equation}
This invariant diverges where the elliptic discriminant $\Delta$ vanishes. Mathematically, the elliptic curve becomes singular (nodal) since at least two roots of the quartic need to collide. Physically, this happens as a Landau surface is approached in the kinematic space.

The map from the elliptic curve to its \emph{torus} is provided by the \emph{Abel's map}
\begin{equation}
\label{abelv2}
    E(\mathbb{C})\to \mathbb{C}/\Lambda_{(1,\tau)}: \quad (x,y)\to Z\equiv \frac{c_4}{\psi_1}\int_{\textcolor{black}{x=\lambda(Z)}}\frac{\textnormal{d}x'}{y} \quad \textnormal{mod} \ \Lambda_{(1,\tau)},
\end{equation}
where the formal inverse $\lambda$ will be defined soon. Since the lattice $\Lambda_{(1,\tau)}$ is invariant under the action of the \emph{modular group} $\textnormal{SL}(2,\mathbb{Z})$, any two framed elliptic curves of the form \eqref{qec} with respective periods $\{\psi_1,\psi_2\}$ and $\{\psi_1',\psi_2'\}$ are equivalent provided 
\begin{equation}
    \begin{pmatrix}\psi_2' \\ \psi_1'\end{pmatrix}=\gamma\cdot 
    \begin{pmatrix} \psi_2 \\ \psi_1\end{pmatrix},
\end{equation}
for a $(2\times 2)$ integral matrix in $\gamma\in\textnormal{SL}(2,\mathbb{Z})$.

Conversely, the map from the torus to the elliptic curve is given by
\begin{equation}
\label{lambMap}
    \mathbb{C}/\Lambda_{(1,\tau)}\to E(\mathbb{C}): \quad Z\to [\lambda(Z):c_4\psi_1^{-1}\lambda'(Z):1].
\end{equation}
where $\lambda$ denotes the formal inverse of \eqref{abelv2} and is given by
\begin{equation}
\label{elllambv2}
    \lambda(Z):=\frac{-3e_1e_{13}e_{24}\psi_1^{-2}~\wp(Z)+e_1^2\Bar{s}_1-2e_1\Bar{s}_2+3\Bar{s_3}}{-3e_{13}e_{24}\psi_1^{-2}~\wp(Z)+3e_1^2-2e_1\Bar{s}_1+\Bar{s}_2},
\end{equation}
From its dependence in the \emph{Weierstra\ss \ $\wp$-function}, $\lambda$ is a doubly periodic and meromorphic function. Therefore, $\lambda$ is an \emph{elliptic function}. Furthermore $\lambda(\pm Z_a)=a$. Solving for $\wp$ in \eqref{elllambv2}, we find
\begin{equation}
\label{lambInv}
\small
   \wp (Z)=
   \frac{{-}(e_3 e_4{+}e_2 (e_3{+}e_4)) (2
   e_1{+}\lambda (Z)){+}e_1
   ((e_2{+}e_3{+}e_4) (e_1{+}2 \lambda
   (Z)){-}3 e_1 \lambda (Z)){+}3 e_2 e_3
   e_4}{3\psi_1^{{-}2}~(e_1{-}e_3) (e_2{-}e_4)
   (e_1{-}\lambda (Z))}.
\end{equation}
Since we are working with a \emph{quartic} elliptic curve, let's clarify what we mean by $\wp(Z)$. To do this, we need to find the Weierstra\ss \ form of the quartic elliptic curve in \eqref{qec}. 

The canonical way to proceed is as follows: Starting with \eqref{qec}, we send a finite root, say $e_1$, at (complex) infinity using the rational map\begin{equation}
    x\to \frac{1}{x-e_1}\implies e_{1}\to e'_1=\infty, e_{2\le i\le 4}\to e'_i=\frac{1}{e_i-e_1}.
\end{equation}
Then, we translate the coordinate $x$ to put a root at zero using 
\begin{equation}
    x\to x-e_2'\implies e_{1}'\to e''_1=\infty, e_2'\to e_2''=0, e_{3\le i\le 4}\to e_i''=e_i'-e_2'.
\end{equation}
Finally, we scale by 
\begin{align}
    &x\to x\frac{(e_1-e_2) (e_1-e_4)}{\psi_1^{-2}~(e_2-e_4)} ,
    \\
    &\implies e''_1\to e_1'''=\infty, e_2''\to e_2'''=0, e_3'' \to e_3'''=\psi_1^2k, e_4''=\psi_1^2.
\end{align}
The effect of these changes of variables on $y$ is 
\begin{equation}
    y^2\to Y^2=x(x-\psi_1^2)(x-\psi_1^2k).
\end{equation}
This is a Legendre-like normal form. To get the Weierstra\ss \ normal form we seek $R_1,R_2$ and $R_3$ such that 
\begin{equation}
    \psi_1^2k=R_3-R_2, \ \psi_1^2=R_1-R_2, \ R_2-R_2=0 \ : R_1+R_2+R_3=0 \iff R_2:=-\frac{1+k}{3\psi_1^{-2}}.
\end{equation}
In terms of the original roots, 
\begin{align}
    R_1 &=\frac{-2 e_2 e_3+e_1
   (e_2+e_3-2 e_4)+(e_2+e_3) e_4}{3
   \psi_1^{-2}~(e_1-e_3) (e_2-e_4)},
   \nn\\
   R_2 &=
   \frac{-2 e_3 e_4+e_2
   (e_3+e_4)+e_1 (-2
   e_2+e_3+e_4)}{3\psi_1^{-2}~(e_1-e_3)
   (e_2-e_4)},
   \\ 
   R_3 &=\frac{e_2
   (e_3-2 e_4)+e_3 e_4+e_1 (e_2-2
   e_3+e_4)}{3\psi_1^{-2}~(e_1-e_3)
   (e_2-e_4)}.
   \nn
\end{align}
Then, $\wp$ is the solution to the following differential equation 
\begin{equation}
\label{weirDE}
    (\wp'(Z))^2=4(\wp(Z)-R_1)(\wp(Z)-R_2)(\wp(Z)-R_3).
\end{equation}
From this as well as \eqref{lambInv} it is not hard to check that
\begin{equation}
\begin{split}
     (c_4\psi_1^{-1}\lambda'(Z))^2
   =\prod_{i=1}^4(\lambda(Z)-e_i),
\end{split}
\end{equation}
thus justifying \eqref{abelv2}. 

Note that, by the negation property of elliptic curves, if $Z_a=(a,y(a))$ then $-Z_a=(a,-y(a))$. From this set-up, we can easily deduce that
\begin{equation}
    \lambda(0)=e_1, \ \lambda(\pm\tau/2)=e_2, \ \lambda(\pm(1+\tau)/2)=e_3, \ \ \lambda(\pm1/2)=e_4.
\end{equation}
Moreover, $\lambda(Z)$ has poles for $Z_\infty$ such that $\lambda(\pm Z_\infty)=\infty$. These are given by \eqref{abelv2}
\begin{equation}
    \pm Z_\infty:=\frac{c_4}{\psi_1}\int_{\infty}\frac{\textnormal{d}x'}{y}.
\end{equation}

\section{The isomorphism between the elliptic curve and the torus}
\label{SEC:torusVar}
In this section, we detail the change of variable from $(X_0,X_2,X_3)$ to $(\tau,z_1,z_2)$  \cite{Bogner:2019lfa}. 

Recall, that in section \ref{sec:2loopfibBasis} we defined the distance 
\begin{align} \label{ourAbel}
    z_{x_1} 
    = Z_{x_1} - (-Z_{x_1})
    = 2 Z_{x_1}
    \equiv\frac{F(\sin^{-1}(u_{x_1})|k)}{K(k)} \quad \textnormal{mod} \ \Lambda_{(1,\tau)},
\end{align}
where $k=\frac{r_{23}r_{14}}{r_{13}r_{24}}\in(0,1)$ and 
\begin{equation}
    u_{x_1}=\sqrt{\frac{(x_1-r_1)(r_2-r_4)}{(x_1-r_2)(r_1-r_4)}}.
\end{equation}

In terms of the dimensionless variables kinematic variables, it is easy to check that 
\begin{equation}
    k=-\frac{16 X_0X_2X_3}{(++-)(+-+)(-++)(---)},
\end{equation}
and we define
\begin{equation}
\label{eq:mp1}
    u_{1,\infty}^2:=\lim_{x_1\to\infty}u_{x_1}^2|_{m_1\leftrightarrow m_2}=-\frac{(++-)(---)}{4 X_3},
\end{equation}
as well as 
\begin{equation}
\label{eq:mp2}
    u_{2,\infty}^2:=\lim_{x_1\to\infty}u_{x_1}^2|_{m_1\leftrightarrow m_3}=-\frac{(+-+)(---)}{4 X_2}.
\end{equation}
Since the point $u_{c}$ is tied to the particularities of the loop-by-loop fibration, we prefer use $u_{2,\infty}$ instead. 
We are now ready to give the change of variable $\{X_0,X_2,X_3\}\leftrightarrow\{\tau,z_1,z_2\}$. On one hand we have 
\begin{equation} \label{eq:torus var}
    \{\tau,z_1,z_2\}=\left\{\frac{\psi_2}{\psi_1},\frac{F(\sin^{-1}(u_{1,\infty})|k)}{K(k)},\frac{F(\sin^{-1}(u_{2,\infty})|k)}{K(k)}\right\},
\end{equation}
while on the other the introduction of an intermediate set of variables is needed. To motivate this, we need to refer to the Jacobian elliptic sn-function defined as the formal inverse of $F=F(\sin^{-1}(x)|k)$ -- i.e., 
\begin{equation}
    F=\int_0^{x=\textnormal{sn}(F|k)}\frac{\textnormal{d}x}{\sqrt{(1-x^2)(1-k^2x^2)}}.
\end{equation}
In view of \eqref{ourAbel}, this definition implies 
\begin{equation}
    K(k)\ z =\int_0^{x=\textnormal{sn}(K(k)z|k)}\frac{\textnormal{d}x}{\sqrt{(1-x^2)(1-k^2x^2)}}.
\end{equation}
The connection to Jacobi's $\theta$-functions comes from the identities (given $q=\exp(i\pi \tau)$)
\begin{align}
  &\sqrt{k}=\frac{\theta_2^2(0|q)}{\theta_3^2(0|q)} \quad \textnormal{and} \quad  \textnormal{sn}(F|k)=\frac{1}{k^{1/4}}\frac{\theta_1\left(\frac{\pi F}{2K(k)}|q\right)}{\theta_4\left(\frac{\pi F}{2K(k)}|q\right)}
  \\
  &\implies \textnormal{sn}(K(k)z|k)=\frac{\theta_3(0|q)}{\theta_2(0|q)}\frac{\theta_1\left(\pi \frac{z}{2}|q\right)}{\theta_4\left(\pi \frac{z}{2}|q\right)}.
\end{align}
In particular, these give raise to the following intermediate variables
\begin{align}
\label{eq:cvIso0}
    \lambda_0:= k &= \frac{\theta_2^4(0|q)}{\theta_3^4(0|q)},
    \nn\\
    \kappa_1:=u_{1,\infty}^2=\frac{\theta_3^2(0|q)}{\theta_2^2(0|q)}\frac{\theta_1^2\left(\pi \frac{z_1}{2}|q\right)}{\theta_4^2\left(\pi \frac{z_1}{2}|q\right)}, 
    & \quad 
    \kappa_2:=u_{2,\infty}^2=\frac{\theta_3^2(0|q)}{\theta_2^2(0|q)}\frac{\theta_1^2\left(\pi \frac{z_2}{2}|q\right)}{\theta_4^2\left(\pi \frac{z_2}{2}|q\right)}.
\end{align}
Solving for $\{X_0,X_2,X_3\}$ in the above, one finds (specific to our root ordering)
\begin{equation*}
    X_0^2=\frac{(\kappa_1-1)(\kappa_2-1)
   \kappa_1 \kappa_2 \lambda_0^2}{(\kappa_1
   \lambda_0-1) (\kappa_2 \lambda_0-1)},
\end{equation*}
\begin{equation}
\label{eq:cvIso}
    X_2^2=\frac{(\kappa_1-1) \kappa_1}{(\kappa_1-\kappa_2)^2 (\kappa_1 \lambda_0-1) (\kappa_2 \kappa_1 \lambda_0-\kappa_1-\kappa_2+1)^2}R,
\end{equation}
\begin{equation*}
    X_3^2=\frac{(\kappa_2-1) \kappa_2}{(\kappa_1-\kappa_2)^2 (\kappa_2 \lambda_0-1) (\kappa_2 \kappa_1 \lambda_0-\kappa_1-\kappa_2+1)^2}R,
\end{equation*}
where
\begin{equation}
\begin{split}
      R&=(\kappa_1{+}\kappa_2)^3 \lambda_0 (\kappa_1 \kappa_2 \lambda_0{+}1){-}8 \kappa_1
   \kappa_2 (\kappa_1{+}\kappa_2)^2 \lambda_0{+}8 \kappa_1 \kappa_2 (\kappa_1{+}\kappa_2) (\kappa_1 \kappa_2 \lambda_0{+}1)\\&{+}3 \kappa_1 \kappa_2 (\kappa_1{+}\kappa_2) \lambda_0 (\kappa_1 \kappa_2 \lambda_0{+}1){+}(\kappa_1{+}\kappa_2) \left(\kappa_1^3 \kappa_2^3 \lambda_0^3{+}1\right){+}4 \kappa_1^2 \kappa_2^2
   (1{-}\lambda_0) \lambda_0\\&{-}8
   \kappa_1 \kappa_2 \left(\kappa_1^2 \kappa_2^2 \lambda_0^2{+}1\right){-}\left(\kappa_1^2{+}\kappa_2^2\right) (\lambda_0{+}1) \left(\kappa_1^2
   \kappa_2^2 \lambda_0^2{+}1\right){-}8 \kappa_1^2 \kappa_2^2\\&{+}2 \left(\kappa_1^2 \lambda_0{-}2 \kappa_1{+}1\right)
   \sqrt{(\kappa_1{-}1) \kappa_1 (\kappa_2{-}1) \kappa_2 (\kappa_1 \lambda_0{-}1)
   (\kappa_2 \lambda_0{-}1)} \left(\kappa_2^2 \lambda_0{-}2 \kappa_2{+}1\right).
\end{split}
\end{equation}
Using, for example, sum of squares identities we can expand the kinematic variables in terms of the torus variables
\begin{equation}
    X_0^2=\frac{\theta_1^2(\pi  \frac{z_1}{2}|q)
   \theta_1^2(\pi  \frac{z_2}{2}|q)
   \theta_2^2(\pi \frac{z_1}{2}|q)
   \theta_2^2(\pi 
   \frac{z_2}{2}|q)}{\theta_3^2(\pi 
   \frac{z_1}{2}|q) \theta_3^2(\pi 
   \frac{z_2}{2}|q) \theta_4^2(\pi 
   \frac{z_1}{2}|q) \theta_4^2(\pi 
  \frac{z_2}{2}|q)}.
\end{equation}

\section{Differential forms on the torus and relevant modular transformations}
\label{SEC:DiffFormOnTorus}
\emph{Kronnecker forms} \cite{Weinzierl:2020fyx,Weinzierl:2020kyq} are closed, quasi-periodic and meromorphic differential forms on the punctured torus. They are explicitly given by
\begin{equation}
\label{kronForm}
    \omega_{k}^{\textnormal{Kr}}(L(z),K\tau):=(2\pi i)^{2{-}k}\left(g^{(k{-}1)}(L(z),K\tau)\textnormal{d}L(z){+}K (k{-}1)g^{(k)}(L(z),K\tau)\frac{\textnormal{d}\tau}{2\pi i}\right),
\end{equation}
where $K\in\mathbb{N}$ and $L(z)$ is a linear function
\begin{equation}
    L(z):=\sum_{m=1}^{n-1}\alpha_mz_m+\beta.
\end{equation}
The functions $g^{(\bullet)}$ are the Laurent coefficients of the so-called \emph{Eisenstein-Kronnecker function}
\begin{equation}
   F(x,y,\tau):=\frac{\pi  \theta _1^{\prime
   }\left(e^{i \pi  \tau }\right)
   \theta _1\left(\pi  (x+y),e^{i \pi
    \tau }\right)}{\theta _1\left(\pi
    x,e^{i \pi  \tau }\right) \theta
   _1\left(\pi  y,e^{i \pi  \tau
   }\right)}=\sum_{k\ge 0}g^{(k)}(x,y)y^{k-1},
\end{equation}
A natural question is: \say{\emph{Why Kronnecker forms}}? The answer to this question lies in the fact that the generating function $F$ is the \emph{unique} quasi-periodic meromorphic function of $\tau$ \emph{and} $z_i$ on the torus \cite[Thm. 3]{zagier1991periods}. It follows that $\omega_{k,n}^{\textnormal{Kr}}$ is the unique closed and quasi-periodic form in $\textnormal{d}\tau$ \emph{and} $\textnormal{d}z_i$ with \emph{simple poles}. 

Since $g$-functions and Kronnecker forms have been studied extensively in the past few years  and we refer the reader to  \cite{weil1999elliptic,Weinzierl:2020fyx,Weinzierl:2020kyq,Broedel:2017kkb,brown2013multiple,zagier1991periods} for more details. In particular, the modular transformation rule for $\omega_\bullet^\textnormal{Kr}$ is \cite{Weinzierl:2020fyx} 
\begin{equation}
     \omega_k^{\textnormal{Kr}}\left(L'\left(z'\right),\tau'\right) = 
 \left(c\tau +d \right)^{k-2}
 \sum\limits_{j=0}^k
 \frac{1}{j!}
 \left( \frac{c L\left(z\right)}{c\tau+d} \right)^j
 \omega_{k-j}^{\textnormal{Kr}}\left(L\left(z\right),\tau\right).
\end{equation}
The other important class of differential forms on the torus are  \emph{modular forms}. These are differential forms in $\tau$ exclusively. In particular, the modular forms showing up in \eqref{epsCon} are given by
\begin{equation}
    \eta_2(\tau)=[e_2(\tau)-2e_2(2\tau)]\frac{\textnormal{d}\tau}{2\pi i},
\end{equation}
and 
\begin{equation}
    \eta_4(\tau)
    =e_4(\tau)\frac{\textnormal{d}\tau}{(2\pi i)^3},
\end{equation}
where the $e_\bullet$'s are \emph{Eisenstein series}. In terms of standard elliptic functions, the relevant Eisenstein series are given by
\begin{equation}
    e_2(\tau)=2\zeta\left(\frac{1}{2},\tau\right) \quad \textnormal{and} \quad  e_4(\tau)=\frac{1}{2} \left(\theta_{2}^{8}\!\left(0,\tau\right) + \theta_{3}^{8}\!\left(0,\tau\right) + \theta_{4}^{8}\!\left(0,\tau\right)\right).
\end{equation}
It is not demanding to show that $e_2(\tau)-2e_2(2\tau)\in\mathcal{M}_2(\Gamma_0(2))$ and $e_4(\tau)\in\mathcal{M}_4(\textnormal{SL}(2,\mathds{Z}))$, where $\mathcal{M}_{k}(\Gamma)$ denotes the set of modular forms of weight $k$ in the subgroup $\Gamma$ of $\textnormal{SL}(2,\mathds{Z})$. For more on modular form, we refer the reader to the nice review in \cite{Adams:2017ejb}.

An understanding of the modular properties of the special functions appearing in the base basis was a crucial piece of knowledge in section \ref{SEC:lblDE}. In particular, we took all the algebraic functions of the kinematic to be modular invariant. Below we describe the modular transformation rules of the elliptic functions relevant to our derivation of \eqref{epsCon}.

First, it is easy to show that under a $\textnormal{SL}(2,\mathds{Z})$ transformation we have (up to a $\textnormal{GL}(1,\mathbb{C})$-ambiguity)
\begin{equation}
    \begin{cases}
      \psi_2\to(a \tau+b)\psi_1,\\
      \psi_1\to(c \tau+d)\psi_1.
    \end{cases}
\end{equation}
Among other useful modular transformation rules are those for the partial derivatives of the standard kinematic variables with respect to the (punctured) torus parameters -- e.g., 
    \begin{equation}
        \frac{\partial X_i}{\partial \tau}, \quad \frac{\partial X_i}{\partial z_1} \quad \textnormal{and} \quad \frac{\partial X_i}{\partial z_2}.
    \end{equation}
    These are found by applying the inverse function theorem. For instance, under the action of the modular group
    \begin{equation}
        \frac{\partial X_i}{\partial \tau}=\frac{\partial X_i(\tau,z_1,z_2)}{\partial \tau}\to \frac{\partial X_i(\tau',z_1',z_2')}{\partial \tau'},
    \end{equation}
    where 
    \begin{equation}
        \tau'=\frac{a \tau+b}{c\tau+d} \quad \textnormal{and} \quad z_i'=\frac{z_i}{c\tau+d},
    \end{equation}
    the inverse function theorem tells us that 
\begin{equation}
\label{partialDerMod}
\begin{split}
    \begin{pmatrix}
        \frac{\partial X_i}{\partial \tau'}\\\frac{\partial X_i}{\partial z_1'}\\ \frac{\partial X_i}{\partial z_2'}
    \end{pmatrix}&=\begin{pmatrix}
        \frac{\partial\tau}{\partial \tau'}&\frac{\partial z_1}{\partial \tau'}&\frac{\partial z_2}{\partial \tau'}\\ \frac{\partial\tau}{\partial z_1'}&\frac{\partial z_1}{\partial z_1'}&\frac{\partial z_2}{\partial z_1'}\\ \frac{\partial\tau}{\partial z_2'}&\frac{\partial z_1}{\partial z_2'}&\frac{\partial z_2}{\partial z_2'}
    \end{pmatrix}\cdot  \begin{pmatrix}
        \frac{\partial X_i}{\partial \tau}\\\frac{\partial X_i}{\partial z_1}\\ \frac{\partial X_i}{\partial z_2}
    \end{pmatrix}
    =\begin{pmatrix}
        \frac{\partial\tau'}{\partial \tau}&\frac{\partial z_1'}{\partial \tau}&\frac{\partial z_2'}{\partial \tau}\\ \frac{\partial\tau'}{\partial z_1}&\frac{\partial z_1'}{\partial z_1}&\frac{\partial z_2'}{\partial z_1}\\ \frac{\partial\tau'}{\partial z_2}&\frac{\partial z_1'}{\partial z_2}&\frac{\partial z_2'}{\partial z_2}
    \end{pmatrix}^{-1}\cdot  \begin{pmatrix}
        \frac{\partial X_i}{\partial \tau}\\\frac{\partial X_i}{\partial z_1}\\ \frac{\partial X_i}{\partial z_2}
    \end{pmatrix}
    \\&=\begin{pmatrix}
        (c\tau+d)^2& c(c\tau+d)z_1& c(c\tau+d)z_2\\ 0& c\tau+d & 0\\ 0&0&  c\tau+d
    \end{pmatrix}\cdot  \begin{pmatrix}
        \frac{\partial X_i}{\partial \tau}\\\frac{\partial X_i}{\partial z_1}\\ \frac{\partial X_i}{\partial z_2}
    \end{pmatrix}.
\end{split}
\end{equation}

\section{Details on the gauge transformations}
\label{SEC:details}
In this appendix, we enumerate explicitly the expressions for matrix components of $\boldsymbol{\mathcal{U}}$ and $\boldsymbol{\mathcal{V}}$ defined, respectively, in \eqref{eq:Ugauge} and \eqref{eq:Vgauge}. The former are given by 
\begin{equation}
\small
    u_{75}=\frac{\psi_1}{\pi}\frac{\left(X_0^2+(X_2-X_3)^2-1\right)(-++)(+--)(---)(+++)}{\mathcal{D}},
\end{equation}
and
\begin{equation}
\small
    u_{76}=\frac{\psi_1}{\pi}\frac{\sqrt{\left(X_0^2{-}1\right)^2 \left(X_2^2{-}X_3^2\right)^2}}{\left(X_0^2{-}1\right)}\frac{\left(5 X_0^4{-}2 \left(3 X_2^2{+}3 X_3^2{-}5\right) X_0^2{+}X_3^4{+}\left(X_2^2{-}1\right)^2{-}2 \left(X_2^2{+}1\right) X_3^2\right)}{\mathcal{D}},
\end{equation}
where the common denominator is given by
\begin{equation}
    \mathcal{D}:=2~c_4~X_0^2~\left(-3 X_0^4+2 \left(X_2^2+X_3^2+1\right) X_0^2+X_3^4+\left(X_2^2-1\right)^2-2 \left(X_2^2+1\right) X_3^2\right).
\end{equation}
The latter are given by
\begin{equation}
\begin{split}
    v_{54}&=\frac{3  N}{2i~\psi_1~\mathcal{D}}\left[\frac{ \left(X_0^2+X_2^2-X_3^2-1\right)}{X_3}\frac{\partial X_3}{\partial \tau}+\frac{
   \left(X_0^2-X_2^2+X_3^2-1\right)}{X_2}\frac{\partial X_2}{\partial \tau}\right],
\end{split}
\end{equation}
\begin{equation}
\small
\begin{split}
    v_{64}&=\frac{S~N}{2i~\psi_1~\mathcal{D}}\left[\frac{\left(3 X_0^2-3 X_2^2-X_3^2+1\right)}{X_2}\frac{\partial X_2}{\partial \tau}-\frac{\left(3 X_0^2-X_2^2-3 X_3^2+1\right) }{X_3}\frac{\partial X_3}{\partial \tau}\right],
\end{split}
\end{equation}
\begin{equation}
\begin{split}
    v_{75}&=\frac{i N}{\psi_1~\mathcal{D}}\left[\frac{ \left(X_0^2+X_2^2-X_3^2-1\right)}{X_3}\frac{\partial X_3}{\partial \tau}+\frac{
   \left(X_0^2-X_2^2+X_3^2-1\right)}{X_2}\frac{\partial X_2}{\partial \tau}\right],
\end{split}
\end{equation}
\begin{equation}
\begin{split}
    v_{76}&=\frac{S~N}{i\psi_1~\mathcal{D}}\left[\frac{~ \left(-3 X_0^2+3 X_2^2+X_3^2-1\right)}{X_2}\frac{\partial X_2}{\partial \tau}-\frac{\left(-3 X_0^2+X_2^2+3 X_3^2-1\right)}{X_3}\frac{\partial X_3}{\partial \tau}\right],
\end{split}
\end{equation}
and
\begin{align}
    v_{74}^{(\textnormal{I})}
    &=\frac{\psi_1^2}{2\pi^2\mathcal{D}^2}
    \bigg[
        9 X_0^{12} 
        {-} 22 X_0^{10} \left(X_2^2{+}X_3^2{+}1\right) 
        {-} X_0^8 \left(X_2^4{-}50 \left(X_3^2{+}1\right) X_2^2{+}X_3^4{-}50 X_3^2{+}1\right) 
    \nn\\&\qquad
        {+} 4  X_0^6 \bigg(
            11 X_2^6{-}19 \left(X_3^2{+}1\right) X_2^4
            {+}\left({-}19 X_3^4{+}54X_3^2{-}19\right) X_2^2
        \nn\\&\qquad\qquad\qquad
            {+}\left(X_3^2{+}1\right) \left(11 X_3^4{-}30 X_3^2{+}11\right)
        \bigg)
    \nn\\&\qquad
        {-} X_0^4 \bigg(
            41 X_2^8{-}84 \left(X_3^2{+}1\right) X_2^6
            {+} \left(86 X_3^4{+}52 X_3^2{+}86\right) X_2^4
        \nn\\&\qquad\qquad\qquad
            {+} \left({-}84 X_3^6{+}52X_3^4{+}52 X_3^2{-}84\right) X_2^2
            {+} \left(X_3^2{-}1\right)^2 \left(41 X_3^4{-}2 X_3^2{+}41\right)            
        \bigg) 
    \nn\\&\qquad
        {+} 2 X_0^2 (X_2{-}X_3{-}1) (X_2{-}X_3{+}1) 
        (X_2{+}X_3{-}1) (X_2{+}X_3{+}1)
        \nn\\&\qquad\qquad\qquad
        \times \left(
            5 X_2^6{-}5 \left(X_3^2{+}1\right)
            X_2^4{+}\left({-}5 X_3^4{+}2 X_3^2{-}5\right) X_2^2
            {+}5 \left(X_3^2{-}1\right)^2 
            \left(X_3^2{+}1\right)
        \right) 
    \nn\\&\qquad
        {+}\left(X_2^4{-}2 \left(X_3^2{+}1\right) X_2^2{+}\left(X_3^2{-}1\right)^2\right)^3
   \bigg],
\end{align}
where the common numerator is given by
\begin{equation}
    N=(-++)(++-)(+-+) (---),
\end{equation}
while the sign function is given by
\begin{equation}
    S=\textnormal{sign}\left(\frac{\left(X_0^2-1\right) \left(X_2^2-X_3^2\right)}{X_0^4}\right).
\end{equation}
Additionally, it can be checked that 
\begin{equation}
    v_{75}=-\frac{2}{3}v_{54} \quad \textnormal{and} \quad v_{76}=-2v_{64}.
\end{equation}
Finally, since $v_{74}^{\textnormal{(II)}}$ is determined by the other $v$'s, we may write
\begin{equation}
    \begin{split}
       \hspace{-4cm} v_{74}^{\textnormal{(II)}}&=\frac{N^2 \left(3 \left(X_0^2-X_2^2+X_3^2-1\right)^2+\left(-3 X_0^2+3 X_2^2+X_3^2-1\right)^2\right)}{4 \psi_1^2 \mathcal{D}^2 X_2^2}\left(\frac{\partial X_2}{\partial \tau}\right)^2\\&+\frac{N^2 \left(3 \left(X_0^2+X_2^2-X_3^2-1\right)^2+\left(-3
   X_0^2+X_2^2+3 X_3^2-1\right)^2\right) }{4 \psi_1^2 \mathcal{D}^2 X_3^2}\left(\frac{\partial X_3}{\partial \tau}\right)^2\\&+\frac{N^2 ~F_{74}}{2 \psi_1^2\mathcal{D}^2 X_2 X_3} \frac{\partial X_2}{\partial \tau} \frac{\partial X_3}{\partial \tau},
    \end{split}
\end{equation}
where
\begin{equation}
\begin{split}
      F_{74}&=\left(-6 X_0^2-\left(3 X_0^2-X_2^2-3 X_3^2+1\right) \left(3 X_0^2-3 X_2^2-X_3^2+1\right)\right.\\& \hspace{3cm} \left.+3
   \left(X_0^4-\left(X_2^2-X_3^2\right)^2\right)+3\right).
\end{split}
\end{equation}


\bibliographystyle{JHEP}
\bibliography{main}
\end{document}